\documentclass[twocolumn,prl,aps,superscriptaddress, amsmath,amssymb,10pt]{revtex4-2}
\usepackage{amsmath}
\usepackage{amssymb}
\usepackage{amsfonts}
\usepackage{bm}
\usepackage{amssymb}
\usepackage{graphicx}
\usepackage{dcolumn}
\usepackage{txfonts}
\usepackage{makeidx}
\usepackage{color}
\usepackage{mathtools}
\usepackage{threeparttable}
\usepackage[linkcolor=blue,anchorcolor=black,citecolor=blue,colorlinks=true]{hyperref}
\usepackage{breakurl}
\usepackage{xr}  
\begin{document}

\title{Stable Atomic Magnetometer in Parity-Time Symmetry Broken Phase}
\author{Xiangdong Zhang}
\author{Jinbo Hu}
\author{Nan Zhao}
\email{nzhao@csrc.ac.cn}
\affiliation{ Beijing Computational Science Research Center }%

\date{\today}

\begin{abstract}
  Random motion of spins is usually detrimental in magnetic resonance experiments. 
  The spin diffusion in non-uniform magnetic fields causes broadening of the resonance and limits the sensitivity and the spectral resolution in applications like magnetic resonance spectroscopy. 
  Here, by observation of the parity-time ($PT$) phase transition of diffusive spins in gradient magnetic fields, 
  we show that the spatial degrees of freedom of atoms could become a resource, 
  rather than harmfulness, for high-precision measurement of weak signals. 
  In the normal phase with zero or low gradient fields, the diffusion results in dissipation of spin precession. 
  However, by increasing the field gradient, the spin system undergoes a $PT$ transition, and enters the $PT$ symmetry broken phase. 
  In this novel phase, the spin precession frequency splits due to spatial localization of the eigenmodes. 
  We demonstrate that, using these spatial-motion-induced split frequencies, 
  the spin system can serve as a stable magnetometer, whose output is insensitive to the inevitable long-term drift of control parameters. 
  This opens a door to detect extremely weak signals in imperfectly controlled environment.
\end{abstract}

\maketitle

\textit{Introduction.}—Measurement of extremely weak signals requires sensors with high sensitivity and high stability. 
High sensitivity allows the sensor to generate large enough signal against background noise. 
However, the signal-to-noise ratio is ultimately limited by the measurement time. 
During a long measurement, even if the detected signal is actually unchanged, 
the sensor output is prone to vary over time due to the imperfect control of 
measurement conditions (e.g., the low frequency drift of electronic devices). 
In this sense, the ability of rejection or compensation of long-term drift, 
i.e. the stability, of a sensor is essential for measuring extremely weak signals.

Atomic spins are useful in the sensing of weak magnetic fields~\cite{Budker2007} 
or signals which are regarded as effective magnetic fields, 
such as inertial rotations~\cite{Walker2016} and extraordinary interactions of fundamental 
physics~\cite{Jiang2021a, Su2021, Yan2015, Bulatowicz2013, Terrano2022}. 
These fields to be detected will manifest themselves by shifting the precession 
frequency of atomic spins. For atoms in liquid or gas phases, their spatial motion 
is usually governed by the diffusion law. With inevitable magnetic field inhomogeneity, 
the diffusion causes spin relaxation and decoherence, which increase the uncertainty of 
the spin precession frequency and degrade the weak field sensing.

The spin precession with spatial motion has been extensively studied decades 
ago~\cite{Hahn1950, Carr1954, Torrey1956}. The dynamics of diffusive atomic spins 
is governed by the Torrey equation~\cite{Torrey1956}. When confined in a finite 
volume, the atomic motion is described by a series of eigenmodes with 
complex eigenvalues. Stoller, Happer and Dyson~\cite{Stoller1991} gave the exact 
solution to the Torrey equation with a linear magnetic field gradient and demonstrated 
the branch behavior of the eigenvalue spectrum due to the non-Hermitian nature 
of the Torrey equation. 

The spin diffusion in non-uniform magnetic field is an ideal platform for studying 
the non-Hermitian physics. The branch spectrum of the Torrey equation proposed in 
ref.~\cite{Stoller1991} is indeed the signature of the $PT$ transition~\cite{Bender1998, Bender2007}. 
Among a number of experimental demonstrations of the $PT$ transition in various 
physical systems~\cite{Makris2008, Ruter2010, Chang2014, Peng2014, Feng2014, Zhen2015, Doppler2016, Xu2016, Zhang2016, Li2019}, 
Zhao, Schaden and Wu~\cite{Schaden2007, Zhao2008b, Zhao2008a, Zhao2010, Wu2021} observed 
the $PT$ transition in system of diffusive electron spin of Rb atoms using 
ultra-thin vapor cells. Here, we study the $PT$ transition process of diffusive 
nuclear spins. The full eigenvalue spectrum in both $PT$-symmetric and $PT$-broken 
phases and, particularly, the mode localization (also known as edge-enhancement~\cite{Saam1996}) 
behavior in the $PT$-broken phase are observed.

\begin{figure}[hbpt]
  \includegraphics[scale=1]{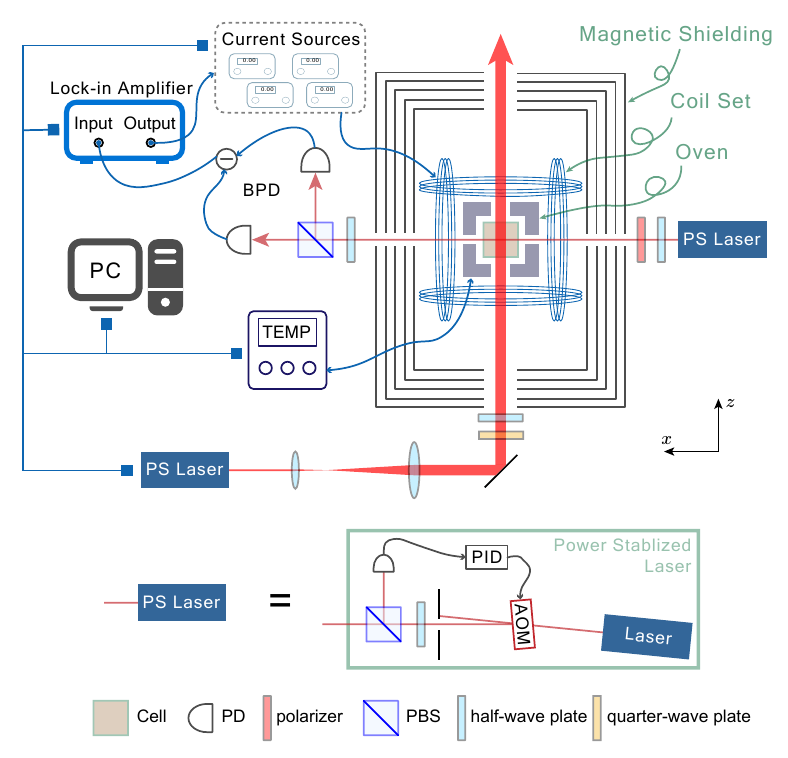}
  \caption{\label{fig:expSetup} {Experimental Setup.} 
  A cubic glass cell containing Xe gas and Rb metal is placed inside a magnetic 
  shielding and heated. 
  A parametric magnetometer~\cite{Cohen-Tannoudji1970, [][{, Section 3.1.}]Eklund2008, Tang2019, Zhang2020, Song2021} 
  is used to detect the nuclear spin signals. 
  See Section~\ref{SMsec:expSetup} of SM for more details. 
  }
\end{figure}

We further demonstrate the application of the $PT$ transition of diffusive spins 
in magnetometry. In contrast to previous studies on the improvement of sensitivity 
near the exceptional points (EPs)~\cite{Wiersig2014, Wiersig2016, Liu2016a, Chen2017, Hodaei2017, Lai2019, Hokmabadi2019}, 
we show that, the spatial motion of spins in 
the $PT$-broken phase could be a resource for improving measurement stability. 

\begin{figure*}[hbpt]
  \includegraphics[]{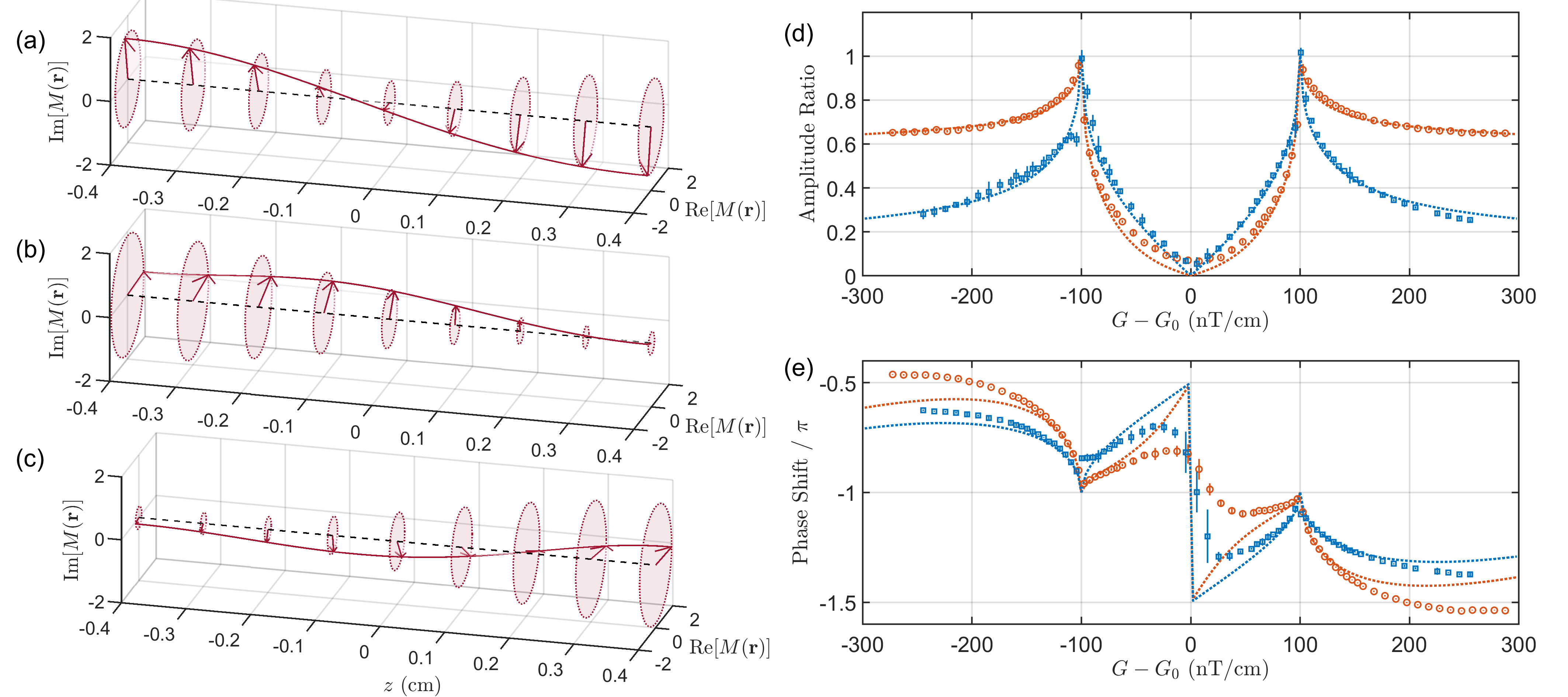}
  \caption{\label{fig:eigenModeDist} {Spatial distribution of the eigenmodes of Torrey equation.} 
  (a), The distribution of $M_+(z)$ in $PT$-symmetric phase with $G=80~{\rm nT/cm}$. 
  The eigenmode $M_-(z)$ is nearly uniform along $z$ direction. 
  (b) and (c), Similar to (a) but for $M_{\pm}(z)$ in 
  $PT$-broken phase with $G=250~{\rm nT/cm}$. 
  (d) and (e), The amplitude ratio $\eta(G; z_{\rm p})$ and the phase shift 
  $\delta \theta(G; z_{\rm p})$ at different probe beam positions. 
  $\delta \theta \equiv \theta_+ - \theta_-$ for the $PT$-symmetric phase and the $PT$-broken 
  phase with $G<0$; $\delta \theta \equiv \theta_- - \theta_+$ for the $PT$-broken phase 
  with $G>0$. The symbols are measured data points extracted from the FID spectrum. 
  The displacement of the probe beam in our experiment is 
  $z_{\rm p,red} - z_{\rm p,blue}=1.18\pm0.02~{\rm mm}$. The dashed lines are theoretical 
  results using the same parameters as in Fig.~\ref{fig:lambdaSpliting} and with probe beam 
  positions $z_{\rm p,red} = -0.56~{\rm mm}$ and $z_{\rm p,blue} = -1.74~{\rm mm}$. 
  The red (blue) data points are the mean value of five (three) repeated measurements. 
  The error bars \footnote{The error bars in all figures of this work represent the $95\%$ confident intervals of experiment data.} 
  in the $PT$-symmetric phase may be underestimated, see Section~\ref{SMsec:fittingBenchmark} of SM. 
  }
\end{figure*}

\textit{$PT$ Transition of Diffusive Spins.}—We observe the $PT$ transition of diffusive nuclear spins by using the experimental setup 
shown in Fig.~\ref{fig:expSetup}. Two isotopes of noble gas ($\rm ^{129}Xe$ and 
$\rm ^{131}Xe$), both carrying nuclear spins, are sealed in a cubic glass cell with inner side 
length $L=0.8~{\rm cm}$. The free-induction decay (FID) signal of the Xe nuclear spins 
is measured to explore their dynamics.

The dynamics of the Xe nuclear spins is governed by the Torrey equation~\cite{Torrey1956}
\begin{equation}
  \frac{\partial  K_{+}(\mathbf{r}, t)}{\partial t} = D \nabla^2 K_{+}(\mathbf{r}, t) - \left( i\gamma B_z +\Gamma_{\rm 2c} \right)  K_{+}(\mathbf{r}, t),
  \label{Eq:BlochTorrey}
\end{equation}
where $K_{+}(\mathbf{r}, t) \equiv K_x + i K_y$ is the transverse 
component of the Xe nuclear spin magnetization $\mathbf{K}(\mathbf{r}, t)$, 
$D$ is the diffusion constant, 
$\gamma$ is the gyromagnetic ratio of Xe nuclear spins, 
$B_z (\mathbf{r})$ is the magnetic field along $z$ direction 
and $\Gamma_{\rm 2c}$ is the intrinsic spin relaxation rate due to inter-atom collisions. 
We present a detailed solution of Eq.~\eqref{Eq:BlochTorrey} in Section~\ref{SMsec:theoreticalModel} 
of Supplemental Material (SM). 

Consider the special case where $B_z(\mathbf{r}) = B_0 + G \cdot z$ and 
the boundary condition is $\hat{\mathbf{n}}\cdot \nabla K_{+}(\mathbf{r}) =0$ on the cell walls. 
The eigen-problem corresponding to Eq.~(\ref{Eq:BlochTorrey}) can simplify to 
\begin{equation}
  \left( D \frac{\rm d^2}{{\rm d}z^2} - i \gamma G \cdot z \right) M_k(z)
  = - s_{k} M_k (z), \quad
  k=0,1,2,\dots
  \label{Eq:EigenProblem}
\end{equation}
Here, we ignore the $x,y$ directions because $B_z$ is uniform along these directions, 
and thus $M_k (\mathbf{r}) = M_k (z)$ for the ground modes. The $i \gamma B_0$ and 
$\Gamma_{\rm 2c}$ terms are dropped because they only contribute a constant shift to 
all eigenvalues $\{s_k\}$. 

The $PT$ operation changes Eq.~\eqref{Eq:EigenProblem} into 
\begin{equation}
  \left( D \frac{\rm d^2}{{\rm d}z^2} - i \gamma G \cdot z \right) M_k^*(-z)
  = - s_{k}^* M_k^* (-z), 
  \label{Eq:EigenProblem_PTtrans}
\end{equation}
meaning that both $\{s_k, M_k(z)\}$ and $\{s_k^*, M_k^*(-z)\}$ solve Eq.~\eqref{Eq:EigenProblem}. 

In the small gradient region, all $\{s_k\}$ are purely real and no degeneracy exists, which means 
the eigenmodes should have $PT$ symmetry, i.e. $M_k(z) = M_k^*(-z)$. Figure.~\ref{fig:eigenModeDist}(a) 
shows an example of $M_1(z)$ in this region. The eigenmodes extend over the whole cell, 
mode localization at the boundary is prevented by the $PT$ symmetry. 

However, predicted by the solution in ref.~\cite{Stoller1991}, there is a critical gradient called 
exceptional point (EP) where $s_0$ and $s_1$ become the same. 
In the region $G>G_{\rm EP}$, the imaginary part of $s_0$ and $s_1$ are non-zero, 
and the $PT$ symmetry of $M_0(z)$ and $M_1(z)$ breaks. 
Instead, $PT$ operation transforms $M_0(z)$ into $M_1^*(-z)$. 
As the gradient gets larger, $M_0(z)$ and $M_1(z)$ start to localize on the opposite ends of the cell. 
This lead to the splitting of resonance frequency of these eigenmodes since they 
``feel'' a different average field. 
Figure.~\ref{fig:eigenModeDist}(b)(c) show an example of $M_0(z)$ and $M_1(z)$ in this region. 
(For more details, see Section~\ref{SMsec:PT_explanation} of SM.) 

Based on the symmetry of eigenmodes, the $G<G_{\rm EP}$ region is named as $PT$-symmetric phase and 
the $G>G_{\rm EP}$ region is $PT$-broken phase. 
The theoretical prediction of EP for ${\rm ^{129}Xe}$ in our experiment is $G_{\rm EP} = 99.4~{\rm nT/cm}$. 
The EP for ${\rm ^{131}Xe}$ ($\sim 335~{\rm nT/cm}$) is larger than the gradient region we can reach. 

\begin{figure}[hbpt]
  \includegraphics[]{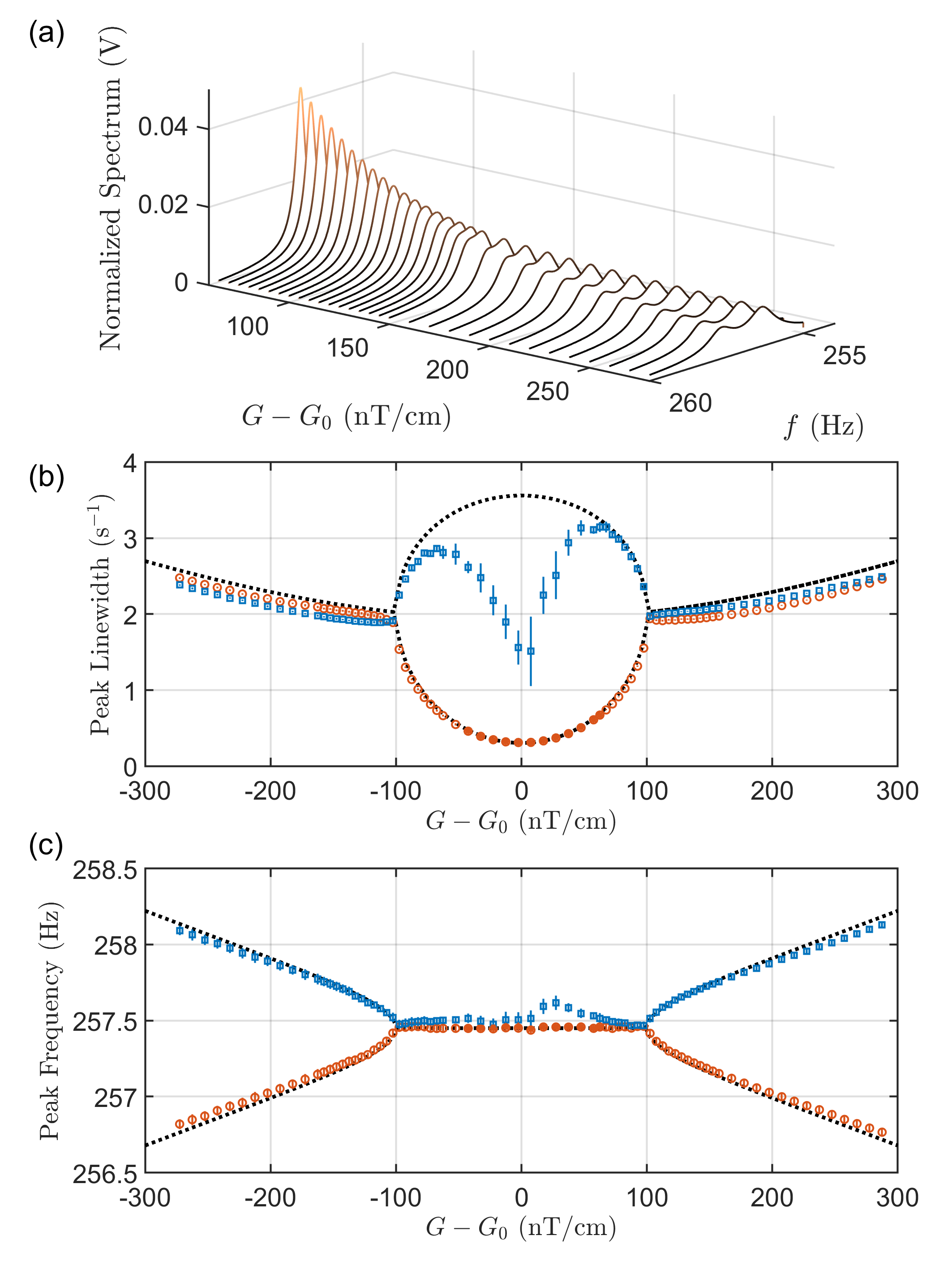}
  \caption{\label{fig:lambdaSpliting} {Eigenvalue spectrum of diffusive spins.}
  (a), Fourier spectrum of measured FID signals of ${\rm ^{129}Xe}$ as a function 
  of $G$ at pump power $250~{\rm mW}$. 
  (b) and (c), The peak linewidth $\Gamma_{\pm}$ and precession frequency $\omega_{\pm}/(2\pi)$ 
  of ${\rm ^{129}Xe}$ spins. The symbols are measured data extracted from the FID spectrum, 
  and the dashed curves are the calculated eigenvalues of 
  Eq.~(\ref{Eq:BlochTorrey}) with $\left| \gamma B_0 \right| /(2\pi)=257.45~{\rm Hz}$, 
  $\Gamma_{\rm 2c}=0.306~{\rm s^{-1}}$ and $D=0.211~{\rm cm^2/s}$. 
  The blue data points in the $|G|<70~{\rm nT/cm}$ region 
  are less reliable due to the fast decay of the excited mode $M_+(z)$
  (See Section~\ref{SMsec:fittingBenchmark} of SM). 
  All data points are the mean value of five repeated measurements. 
  See Section~\ref{SMsec:spectrumFitting} of SM for details of FID spectrum fitting. 
  }
\end{figure}

The evolution of $K_+$ can be expanded using the eigenmodes as 
$K_{+}(\mathbf{r}, t)  = \sum_{k}  c_k M_k(\mathbf{r}) {\rm e}^{-s_k t }$, 
where $\{c_k\}$ are expansion coefficients determined by the initial spin distribution. 
The FID signal is proportional to (see Eq.~\eqref{Eq:FIDSignal_twoModeSolution} of SM): 
\begin{equation}
  K_{x}(z_{\rm p}, t) = 
  \sum_{k=0}^{\infty} A_k(z_{\rm p})\cos[\omega_k t+ \theta_k(z_{\rm p})]{\rm e}^{-\Gamma_k t},
\label{Eq:FIDSignal}
\end{equation}
where $z_{\rm p}$ is the position of probe beam (on $z$ axis), 
$A_k(z_{\rm p}) \equiv \vert c_k M_k(z_{\rm p}) \vert$, 
$\theta_k(z_{\rm p}) \equiv -\arg[c_k M_k(z_{\rm p})]$ and
$s_k \equiv \Gamma_k + i \omega_k$. 
Since higher excited modes decays very fast, only $M_0$ and $M_1$ have experimentally observable effect. 
In the following, the subscript $k=1$ and $0$ is replaced by ``$+$'' and ``$-$'' signs, respectively. 

Figure~\ref{fig:lambdaSpliting}(a) shows the spectrum of FID signals at different gradient. 
The resonance peak splits as gradient gets larger. 
Figures~\ref{fig:lambdaSpliting}(b)(c) compare the measured eigenvalues $s_\pm$ 
with the theoretical values. The behavior of $\Gamma_\pm$ and $\omega_\pm$ 
fits well with theory. 
Figures~\ref{fig:eigenModeDist}(d)(e) show the amplitude ratio 
$\eta(z_{\rm p}) \equiv A_{\rm min}/A_{\rm max}$ and 
phase shift $\delta \theta(z_{\rm p}) \equiv \pm \left( \theta_+ - \theta_- \right)$ 
of the two eigenmodes $M_\pm$, where $A_{\rm max}$ ($A_{\rm min}$) is the larger (smaller) 
amplitude between $A_\pm(z_{\rm p})$. 
Two probe beam positions are used to verify the spatial distribution of eigenmodes. 
The amplitude ratio $\eta$ fits well with theory, 
and is sensitive to $z_{\rm p}$ in the $PT$-broken phase due to the localization 
of eigenmodes. 
The phase shift fits not so good with theory because of the fitting accuracy and 
the difficulty on determining the precise time origin of an FID signal.

\begin{figure*}[hbpt]
  \includegraphics[]{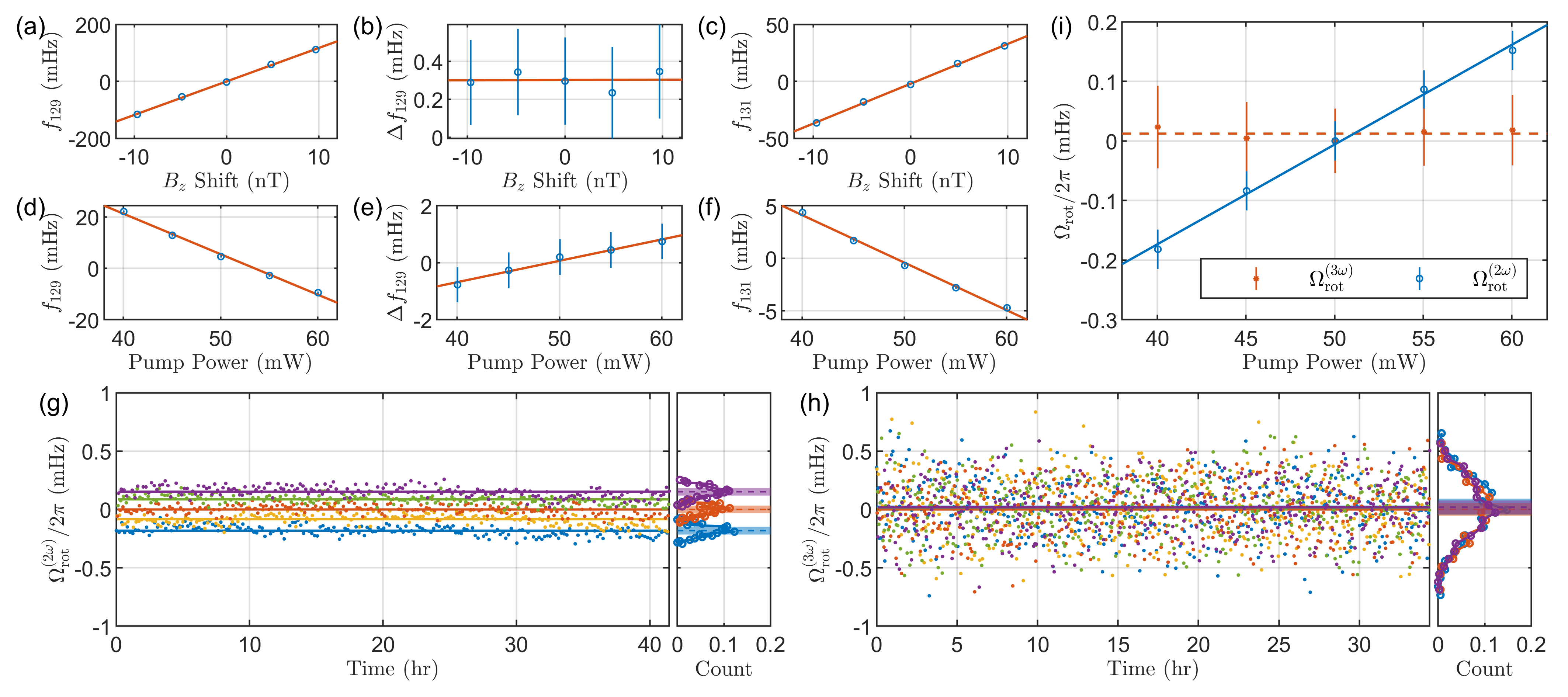}
  \caption{\label{fig:stableMag} {Stable comagnetometer.}
  (a)-(c), Magnetic field dependence of three measured frequencies. 
  $[f_{129}, \Delta f_{129}, f_{131}] \equiv [\omega_{129}, \Delta \omega_{129}, \omega_{131}]/(2\pi)$. 
  These frequencies are shifted by 257.16, 1.17 and 76.24~Hz respectively for clarity reason. 
  The $B_z$ shift is relative to $21.92~{\rm \mu T}$. 
  The fitted slopes of (a)-(c) are $(11.74 \pm 0.64)$, 
  $(0.000 \pm 0.011)$ and $(3.49 \pm 0.19)~\rm mHz/nT$.
  (d)-(f), Pump power dependence of the three measured frequencies with the fitted slopes 
  $(-1.57 \pm 0.19)$, $(0.075 \pm 0.021)$ and $(-0.452 \pm 0.055)~ \rm mHz/mW$ respectively.
  (g), Traces of 1,000 successive FID measurements at zero gradient ($PT$-symmetric phase). The pump power 
  changes among [40, 45, 50, 55, 60]~mW periodically and the color of points represents 
  the value of pump power. The histogram on the right shows the distribution and average 
  value of $\Omega_{\rm rot}^{(2\omega)}$ at 40, 50 and 60 mW.
  (h), The same as (g), but for 2,000 successive FID measurements at $|G|=250~{\rm nT/cm}$ ($PT$-broken phase). 
  (i), The average values in (g) and (h). Solid and dashed lines are 
  linear fitting of the data points. The slope of blue solid line is $(16.7 \pm 2.0)~\rm \mu Hz/mW$, 
  while red dashed line is $(0.0 \pm 2.2)~\rm \mu Hz/mW$. 
  Data points in (a)-(f) are the mean values of 400 repeated measurements.
  }
\end{figure*}

\textit{Stable Comagnetometer in $PT$ Broken Phase.}—One can utilize the split spin 
precession frequencies in the $PT$-broken phase to stabilize 
the output of the atomic magnetometers. In general, the precession frequency $\omega_\alpha$ 
of a given spin species $\alpha$ is usually influenced by a number of input variables and 
is expressed by a multivariate function $\omega_\alpha=\omega_\alpha(\mathbf{x})$ of the 
input vector $\mathbf{x}=[x_1, x_2, …, x_M]^{\rm T}$. Among the $M$ components of $\mathbf{x}$, 
only one is the real signal we want to detect (e.g., the unknown magnetic field). The 
remaining $M-1$ variables will cause systematic error if they are not well controlled. 
Comagnetometers use $M$ precession frequencies $\bm{\omega}=[\omega_1, \omega_2, …, \omega_M]^{\rm T}$ 
of different spin species to determine the $M$ variables in $\mathbf{x}$ unambiguously. 
As long as the Jacobian matrix $J=[\partial \omega_i/\partial x_j]_{i,j=1 \dots M}$ of the 
multidimensional function $\bm{\omega}=\bm{\omega}(\mathbf{x})$ is invertible, 
the comagnetometer is immune to the drift of all the variables in $\mathbf{x}$.

The dual-species nuclear magnetic resonance gyroscope (NMRG)~\cite{Walker2016}, a kind of 
comagnetometer, uses the precession frequencies $\bm{\omega}=[\omega_{129}, \omega_{131}]^{\rm T}$ 
of $\rm ^{129}Xe$ and $\rm ^{131}Xe$ nuclear spin to determine the rotation rate $\Omega_{\rm rot}$ of 
the system. The precession frequencies $\bm{\omega}$ depend on $\mathbf{x}=[B, \Omega_{\rm rot}]^{\rm T}$ 
through the relation $\omega_\alpha=\gamma_\alpha B+ \Omega_{\rm rot}$, where $\gamma_\alpha$ 
is the gyromagnetic ratio of $\rm ^{129}Xe$ or $\rm ^{131}Xe$ nuclear spin and $B$ is the 
magnetic field along $z$ direction. The rotation rate is estimated by (assume $B>0$)
\begin{equation}
  \Omega_{\rm rot}^{\rm (2 \omega)} \equiv \frac{| R \omega_{131} | - |\omega_{129}|}{1 +|R|},
  \label{Eq:OmegaR_2f_def}
\end{equation}
with $R \equiv \gamma_{129}/\gamma_{131} \approx -3.373417$ the ratio of gyromagnetic ratios~\cite{Makulski2015}.

The above relation $\omega_\alpha= \gamma_\alpha B+ \Omega_{\rm rot}$ is only valid when the 
magnetic field $B$ is spatially uniform. Due to the difference of boundary conditions and 
gyromagnetic ratios, the diffusive $\rm ^{129}Xe$ and $\rm ^{131}Xe$ spins can have 
different responses to a non-uniform magnetic field. The spin precession frequencies are actually  
$\omega_\alpha=\gamma_\alpha B_0 + \Omega_{\rm rot}+ \gamma_\alpha \bar{B}_{\rm A,\alpha}$, 
where $\bar{B}_{\rm A,\alpha}$ is an isotope-dependent effective magnetic field originated 
from the inhomogeneity of $B(\mathbf{r})$~\cite{Sheng2014} and $B_0$ is the mean value of 
$B(\mathbf{r})$. The differential part 
$b_{\rm A} \equiv \bar{B}_{\rm A,129} - \bar{B}_{\rm A,131}$ of the isotope-dependent effective 
field produces a systematic error on the estimator Eq.~(\ref{Eq:OmegaR_2f_def}) as
\begin{equation}
  \Omega^{(\rm 2\omega)}_{\rm rot} - \Omega_{\rm rot} = 
     - \frac{| \gamma_{129} \gamma_{131} |}{|\gamma_{129}|+|\gamma_{131}|}b_{\rm A}.
  \label{Eq:OmegaR_2f_err}
\end{equation}

One origin of the inhomogeneity of $B(\mathbf{r})$ is the polarization field generated by the 
spin-exchange collisions between Xe and Rb atoms. In our experiment, 
the $\bar{B}_{\rm A,\alpha}$ from polarization field is in the order of $\sim 10^1~{\rm nT}$, 
and the observed value of $b_{\rm A}$ can be as large as $\sim 10^0~{\rm nT}$. 
More importantly, $b_{\rm A}=b_{\rm A}(P_{\rm pump},f_{\rm pump},T, \dots)$ depends on several control 
parameters such as the laser power $P_{\rm pump}$, laser frequency $f_{\rm pump}$ 
and cell temperature $T$, etc. The drift of these control parameters will eventually 
limit the long-term stability of $\Omega_{\rm rot}$ measurement. Great efforts based on 
pulse control of the alkali-metal atoms have been made to eliminate the 
influence of the polarization field~\cite{Limes2018, Korver2015, Thrasher2019}. Here we 
demonstrate a new method utilizing the $PT$ transition.

The $PT$ transition extends the dual-species NMRG to a 3-component comagnetometer. 
Particularly, we measure three frequencies $\bm{\omega}=[\omega_{129}, \Delta \omega_{129}, \omega_{131}]^{\rm T}$ 
as functions of three input variables $\mathbf{x}=[B_0, \Omega_{\rm rot}, P_{\rm pump}]^{\rm T}$, 
where $\omega_{129} \equiv (\omega_+ + \omega_-)/2$ and $\Delta \omega_{129} \equiv |\omega_+|-|\omega_-|$ 
are the mean frequency and the $PT$ splitting of $\rm ^{129}Xe$. As a proof-of-principle 
experiment, we assumed the pump power $P_{\rm pump}$ to be the dominating parameter which 
affects the non-uniform magnetic field, i.e., $b_{\rm A}=b_{\rm A}(P_{\rm pump})$. The Jacobian matrix is 
experimentally determined and the gyroscope signal $\Omega_{\rm rot}^{\rm (3 \omega)}$ of 
3-component comagnetometer is calculated by solving the following linear equation
\begin{equation}
  \delta 
  \begin{pmatrix}
    | \omega_{129} | \\
    \Delta \omega_{129} \\
    | \omega_{131} | \\
  \end{pmatrix}
   = 
   \begin{pmatrix}
      |\gamma_{129}| & \chi_1 & -1 \\
            0        & \chi_2 &  0 \\
      |\gamma_{131}| & \chi_3 & +1 \\
  \end{pmatrix}
  \cdot \delta 
  \begin{pmatrix}
    B_0 \\
    P_{\rm pump} \\
    \Omega_{\rm rot}^{\rm (3 \omega)} \\
  \end{pmatrix},
  \label{Eq:OmegaR_3f_def}
\end{equation}
where $\chi_1$, $\chi_2$ and $\chi_3$ are the fitted slopes in 
Figs.~\ref{fig:stableMag}(d)-(f), respectively.
The $PT$ splitting $\Delta \omega_{129}$ is insensitive to $B_0$ but proportional to the 
change of $P_{\rm pump}$. This can be understood by 
noticing that the spatial distribution of the polarization field  
relies on $P_{\rm pump}$. The $\rm ^{129}Xe$ 
spins sense the change of the inhomogeneous polarization field and 
manifest it as the splitting $\Delta \omega_{129}$ between the two localized modes in the 
$PT$-broken phase.

Figures~\ref{fig:stableMag}(g)(h)(i) compare the measurement 
stability of $\Omega_{\rm rot}$ against the change of $P_{\rm pump}$. During the whole 
measurement, the actual rotation rate $\Omega_{\rm rot}$ is unchanged. The traditional 
dual-species NMRG estimator $\Omega_{\rm rot}^{(\rm 2 \omega)}$ shows a $17~{\rm \mu Hz/mW}$ 
dependence on $P_{\rm pump}$, while slope almost vanishes for 
our 3-component comagnetometer estimator $\Omega_{\rm rot}^{\rm (3 \omega)}$. 
This result demonstrates the great potential for improving comagnetometer stability.

We have demonstrated the stability of 3-component comagnetometer against the fluctuation of 
pumping power, but the key idea is that we can choose two arbitrary parameters $x_1$ and $x_2$, and then 
configure the 3-component comagnetometer to be stable against the fluctuation of both $x_1$ and $x_2$. 
$x_1$, $x_2$ can be any continuous scalar parameters of the experimental system such as laser 
power, laser wavelength, cell temperature, coil current or even linear combination of them.
Due to the mode localization nature in $PT$-broken phase, $\Delta \omega_{129}$ is directly 
sensitive to the non-uniform distribution of magnetic field. Conversely, $\omega_{129}$ and 
$\omega_{131}$ are mainly determined by the average magnetic field, non-uniformity only 
contributes perturbative corrections. These make the $\Delta \omega_{129}$ a good indicator
for monitoring the change of parameters that can induce non-uniform magnetic field, and
then to suppress their influence on $\omega_{129}$ and $\omega_{131}$.

\textit{Discussion and Outlook.}—In this paper, we report the observation of the $PT$ transition of diffusive nuclear spins. 
Particularly, the spin precession frequency splitting and the mode localization are measured 
in the $PT$-broken phase. In this phase, boundary between coherent and incoherent spin motion 
is blurred. The random spin diffusion in a gradient field behaves like a coherent coupling 
(e.g., spin-orbit coupling) in a Hermitian system, rather than a pure dissipation as in the 
$PT$-symmetric phase. The diffusive nuclear spin system provides an excellent testbed for 
further exploring the non-Hermitian physics.

We also demonstrate the application of $PT$ transition in sensing of weak signals. 
Comagnetometer in the $PT$-broken phase is sensitive to magnetic field gradient, which 
enables the design of gradiometer~\cite{Sheng2017, Zhang2020a} measuring the magnitude 
and gradient of magnetic field in a single atomic cell. Furthermore, the $PT$ transition 
was shown to be useful in improving the sensitivity of parameter estimation near 
the EPs previously~\cite{Wiersig2014, Wiersig2016, Liu2016a, Chen2017, Hodaei2017, Lai2019, Hokmabadi2019}, 
although the signal-to-noise ratio and the fundamental precession limit are still 
under debate~\cite{Langbein2018, Zhang2019, Lau2018, Chen2019}. Our work show that, 
assisted by the $PT$ transition, the spatial motion is engaged in the sensing process and 
the sensor stability, another important aspect of high-precession measurement, is 
significantly enhanced. This paves the way to develop stable comagnetometers for the 
detection of extremely weak signals.

\begin{acknowledgments}
We thank Yanhua Wang for the assistance in establishing the experiment setup and 
Dawu Xiao for the preliminary theoretical calculations. We thank Kang Dai for 
providing the vapor cell. This work is supported by NSAF 
(Grants No. U2030209 and U1930402).

X.Z. and N.Z. designed the experiment. 
X.Z. performed the measurements and analyzed the data. 
J.H., N.Z. and X.Z. carried out the theoretical analysis and numerical simulation. 
N.Z. and X.Z. wrote the manuscript. N.Z. supervised the project.
\end{acknowledgments}

Supplemental Material (SM) is provided on [url], which includes 
Refs.~\cite{Happer1972, Walker1997, Seltzer2008, Grover1978, 
            Zheng2011, Poling2001,  Feng2020}. 

\bibliographystyle{apsrev4-2}
\bibliography{maintext-APS.bib}

\providecommand{\noopsort}[1]{}\providecommand{\singleletter}[1]{#1}%
\begin{thebibliography}{59}%
\makeatletter
\providecommand \@ifxundefined [1]{%
 \@ifx{#1\undefined}
}%
\providecommand \@ifnum [1]{%
 \ifnum #1\expandafter \@firstoftwo
 \else \expandafter \@secondoftwo
 \fi
}%
\providecommand \@ifx [1]{%
 \ifx #1\expandafter \@firstoftwo
 \else \expandafter \@secondoftwo
 \fi
}%
\providecommand \natexlab [1]{#1}%
\providecommand \enquote  [1]{``#1''}%
\providecommand \bibnamefont  [1]{#1}%
\providecommand \bibfnamefont [1]{#1}%
\providecommand \citenamefont [1]{#1}%
\providecommand \href@noop [0]{\@secondoftwo}%
\providecommand \href [0]{\begingroup \@sanitize@url \@href}%
\providecommand \@href[1]{\@@startlink{#1}\@@href}%
\providecommand \@@href[1]{\endgroup#1\@@endlink}%
\providecommand \@sanitize@url [0]{\catcode `\\12\catcode `\$12\catcode
  `\&12\catcode `\#12\catcode `\^12\catcode `\_12\catcode `\%12\relax}%
\providecommand \@@startlink[1]{}%
\providecommand \@@endlink[0]{}%
\providecommand \url  [0]{\begingroup\@sanitize@url \@url }%
\providecommand \@url [1]{\endgroup\@href {#1}{\urlprefix }}%
\providecommand \urlprefix  [0]{URL }%
\providecommand \Eprint [0]{\href }%
\providecommand \doibase [0]{https://doi.org/}%
\providecommand \selectlanguage [0]{\@gobble}%
\providecommand \bibinfo  [0]{\@secondoftwo}%
\providecommand \bibfield  [0]{\@secondoftwo}%
\providecommand \translation [1]{[#1]}%
\providecommand \BibitemOpen [0]{}%
\providecommand \bibitemStop [0]{}%
\providecommand \bibitemNoStop [0]{.\EOS\space}%
\providecommand \EOS [0]{\spacefactor3000\relax}%
\providecommand \BibitemShut  [1]{\csname bibitem#1\endcsname}%
\let\auto@bib@innerbib\@empty
\bibitem [{\citenamefont {Budker}\ and\ \citenamefont
  {Romalis}(2007)}]{Budker2007}%
  \BibitemOpen
  \bibfield  {author} {\bibinfo {author} {\bibfnamefont {D.}~\bibnamefont
  {Budker}}\ and\ \bibinfo {author} {\bibfnamefont {M.}~\bibnamefont
  {Romalis}},\ }\href {https://doi.org/10.1038/nphys566} {\bibfield  {journal}
  {\bibinfo  {journal} {Nature Physics}\ }\textbf {\bibinfo {volume} {3}},\
  \bibinfo {pages} {227} (\bibinfo {year} {2007})}\BibitemShut {NoStop}%
\bibitem [{\citenamefont {Walker}\ and\ \citenamefont
  {Larsen}(2016)}]{Walker2016}%
  \BibitemOpen
  \bibfield  {author} {\bibinfo {author} {\bibfnamefont {T.~G.}\ \bibnamefont
  {Walker}}\ and\ \bibinfo {author} {\bibfnamefont {M.~S.}\ \bibnamefont
  {Larsen}},\ }\href {https://doi.org/10.1016/bs.aamop.2016.04.002} {\bibfield
  {journal} {\bibinfo  {journal} {Advances in Atomic, Molecular and Optical
  Physics}\ }\textbf {\bibinfo {volume} {65}},\ \bibinfo {pages} {373}
  (\bibinfo {year} {2016})}\BibitemShut {NoStop}%
\bibitem [{\citenamefont {Jiang}\ \emph {et~al.}(2021)\citenamefont {Jiang},
  \citenamefont {Su}, \citenamefont {Garcon}, \citenamefont {Peng},\ and\
  \citenamefont {Budker}}]{Jiang2021a}%
  \BibitemOpen
  \bibfield  {author} {\bibinfo {author} {\bibfnamefont {M.}~\bibnamefont
  {Jiang}}, \bibinfo {author} {\bibfnamefont {H.}~\bibnamefont {Su}}, \bibinfo
  {author} {\bibfnamefont {A.}~\bibnamefont {Garcon}}, \bibinfo {author}
  {\bibfnamefont {X.}~\bibnamefont {Peng}},\ and\ \bibinfo {author}
  {\bibfnamefont {D.}~\bibnamefont {Budker}},\ }\href
  {https://doi.org/10.1038/s41567-021-01392-z} {\bibfield  {journal} {\bibinfo
  {journal} {Nature Physics}\ }\textbf {\bibinfo {volume} {17}},\ \bibinfo
  {pages} {1402} (\bibinfo {year} {2021})}\BibitemShut {NoStop}%
\bibitem [{\citenamefont {Su}\ \emph {et~al.}(2021)\citenamefont {Su},
  \citenamefont {Wang}, \citenamefont {Jiang}, \citenamefont {Ji},
  \citenamefont {Fadeev}, \citenamefont {Hu}, \citenamefont {Peng},\ and\
  \citenamefont {Budker}}]{Su2021}%
  \BibitemOpen
  \bibfield  {author} {\bibinfo {author} {\bibfnamefont {H.}~\bibnamefont
  {Su}}, \bibinfo {author} {\bibfnamefont {Y.}~\bibnamefont {Wang}}, \bibinfo
  {author} {\bibfnamefont {M.}~\bibnamefont {Jiang}}, \bibinfo {author}
  {\bibfnamefont {W.}~\bibnamefont {Ji}}, \bibinfo {author} {\bibfnamefont
  {P.}~\bibnamefont {Fadeev}}, \bibinfo {author} {\bibfnamefont
  {D.}~\bibnamefont {Hu}}, \bibinfo {author} {\bibfnamefont {X.}~\bibnamefont
  {Peng}},\ and\ \bibinfo {author} {\bibfnamefont {D.}~\bibnamefont {Budker}},\
  }\href {https://doi.org/10.1126/sciadv.abi9535} {\bibfield  {journal}
  {\bibinfo  {journal} {Science Advances}\ }\textbf {\bibinfo {volume} {7}},\
  \bibinfo {pages} {eabi9535} (\bibinfo {year} {2021})}\BibitemShut {NoStop}%
\bibitem [{\citenamefont {Yan}\ \emph {et~al.}(2015)\citenamefont {Yan},
  \citenamefont {Sun}, \citenamefont {Peng}, \citenamefont {Zhang},
  \citenamefont {Fu}, \citenamefont {Guo},\ and\ \citenamefont
  {Liu}}]{Yan2015}%
  \BibitemOpen
  \bibfield  {author} {\bibinfo {author} {\bibfnamefont {H.}~\bibnamefont
  {Yan}}, \bibinfo {author} {\bibfnamefont {G.~A.}\ \bibnamefont {Sun}},
  \bibinfo {author} {\bibfnamefont {S.~M.}\ \bibnamefont {Peng}}, \bibinfo
  {author} {\bibfnamefont {Y.}~\bibnamefont {Zhang}}, \bibinfo {author}
  {\bibfnamefont {C.}~\bibnamefont {Fu}}, \bibinfo {author} {\bibfnamefont
  {H.}~\bibnamefont {Guo}},\ and\ \bibinfo {author} {\bibfnamefont {B.~Q.}\
  \bibnamefont {Liu}},\ }\href {https://doi.org/10.1103/PhysRevLett.115.182001}
  {\bibfield  {journal} {\bibinfo  {journal} {Physical Review Letters}\
  }\textbf {\bibinfo {volume} {115}},\ \bibinfo {pages} {182001} (\bibinfo
  {year} {2015})}\BibitemShut {NoStop}%
\bibitem [{\citenamefont {Bulatowicz}\ \emph {et~al.}(2013)\citenamefont
  {Bulatowicz}, \citenamefont {Griffith}, \citenamefont {Larsen}, \citenamefont
  {Mirijanian}, \citenamefont {Fu}, \citenamefont {Smith}, \citenamefont
  {Snow}, \citenamefont {Yan},\ and\ \citenamefont {Walker}}]{Bulatowicz2013}%
  \BibitemOpen
  \bibfield  {author} {\bibinfo {author} {\bibfnamefont {M.}~\bibnamefont
  {Bulatowicz}}, \bibinfo {author} {\bibfnamefont {R.}~\bibnamefont
  {Griffith}}, \bibinfo {author} {\bibfnamefont {M.}~\bibnamefont {Larsen}},
  \bibinfo {author} {\bibfnamefont {J.}~\bibnamefont {Mirijanian}}, \bibinfo
  {author} {\bibfnamefont {C.~B.}\ \bibnamefont {Fu}}, \bibinfo {author}
  {\bibfnamefont {E.}~\bibnamefont {Smith}}, \bibinfo {author} {\bibfnamefont
  {W.~M.}\ \bibnamefont {Snow}}, \bibinfo {author} {\bibfnamefont
  {H.}~\bibnamefont {Yan}},\ and\ \bibinfo {author} {\bibfnamefont {T.~G.}\
  \bibnamefont {Walker}},\ }\href
  {https://doi.org/10.1103/PhysRevLett.111.102001} {\bibfield  {journal}
  {\bibinfo  {journal} {Physical Review Letters}\ }\textbf {\bibinfo {volume}
  {111}},\ \bibinfo {pages} {102001} (\bibinfo {year} {2013})}\BibitemShut
  {NoStop}%
\bibitem [{\citenamefont {Terrano}\ and\ \citenamefont
  {Romalis}(2022)}]{Terrano2022}%
  \BibitemOpen
  \bibfield  {author} {\bibinfo {author} {\bibfnamefont {W.~A.}\ \bibnamefont
  {Terrano}}\ and\ \bibinfo {author} {\bibfnamefont {M.~V.}\ \bibnamefont
  {Romalis}},\ }\href {https://doi.org/10.1088/2058-9565/ac1ae0} {\bibfield
  {journal} {\bibinfo  {journal} {Quantum Science and Technology}\ }\textbf
  {\bibinfo {volume} {7}},\ \bibinfo {pages} {014001} (\bibinfo {year}
  {2022})}\BibitemShut {NoStop}%
\bibitem [{\citenamefont {Hahn}(1950)}]{Hahn1950}%
  \BibitemOpen
  \bibfield  {author} {\bibinfo {author} {\bibfnamefont {E.~L.}\ \bibnamefont
  {Hahn}},\ }\href {https://doi.org/10.1103/PhysRev.80.580} {\bibfield
  {journal} {\bibinfo  {journal} {Physical Review}\ }\textbf {\bibinfo {volume}
  {80}},\ \bibinfo {pages} {580} (\bibinfo {year} {1950})}\BibitemShut
  {NoStop}%
\bibitem [{\citenamefont {Carr}\ and\ \citenamefont
  {Purcell}(1954)}]{Carr1954}%
  \BibitemOpen
  \bibfield  {author} {\bibinfo {author} {\bibfnamefont {H.~Y.}\ \bibnamefont
  {Carr}}\ and\ \bibinfo {author} {\bibfnamefont {E.~M.}\ \bibnamefont
  {Purcell}},\ }\href {https://doi.org/10.1103/PhysRev.94.630} {\bibfield
  {journal} {\bibinfo  {journal} {Physical Review}\ }\textbf {\bibinfo {volume}
  {94}},\ \bibinfo {pages} {630} (\bibinfo {year} {1954})}\BibitemShut
  {NoStop}%
\bibitem [{\citenamefont {Torrey}(1956)}]{Torrey1956}%
  \BibitemOpen
  \bibfield  {author} {\bibinfo {author} {\bibfnamefont {H.~C.}\ \bibnamefont
  {Torrey}},\ }\href {https://doi.org/10.1103/PhysRev.104.563} {\bibfield
  {journal} {\bibinfo  {journal} {Physical Review}\ }\textbf {\bibinfo {volume}
  {104}},\ \bibinfo {pages} {563} (\bibinfo {year} {1956})}\BibitemShut
  {NoStop}%
\bibitem [{\citenamefont {Stoller}\ \emph {et~al.}(1991)\citenamefont
  {Stoller}, \citenamefont {Happer},\ and\ \citenamefont
  {Dyson}}]{Stoller1991}%
  \BibitemOpen
  \bibfield  {author} {\bibinfo {author} {\bibfnamefont {S.~D.}\ \bibnamefont
  {Stoller}}, \bibinfo {author} {\bibfnamefont {W.}~\bibnamefont {Happer}},\
  and\ \bibinfo {author} {\bibfnamefont {F.~J.}\ \bibnamefont {Dyson}},\ }\href
  {https://doi.org/10.1103/PhysRevA.44.7459} {\bibfield  {journal} {\bibinfo
  {journal} {Physical Review A}\ }\textbf {\bibinfo {volume} {44}},\ \bibinfo
  {pages} {7459} (\bibinfo {year} {1991})}\BibitemShut {NoStop}%
\bibitem [{\citenamefont {Bender}\ and\ \citenamefont
  {Boettcher}(1998)}]{Bender1998}%
  \BibitemOpen
  \bibfield  {author} {\bibinfo {author} {\bibfnamefont {C.~M.}\ \bibnamefont
  {Bender}}\ and\ \bibinfo {author} {\bibfnamefont {S.}~\bibnamefont
  {Boettcher}},\ }\href {https://doi.org/10.1103/PhysRevLett.80.5243}
  {\bibfield  {journal} {\bibinfo  {journal} {Physical Review Letters}\
  }\textbf {\bibinfo {volume} {80}},\ \bibinfo {pages} {5243} (\bibinfo {year}
  {1998})}\BibitemShut {NoStop}%
\bibitem [{\citenamefont {Bender}(2007)}]{Bender2007}%
  \BibitemOpen
  \bibfield  {author} {\bibinfo {author} {\bibfnamefont {C.~M.}\ \bibnamefont
  {Bender}},\ }\href {https://doi.org/10.1088/0034-4885/70/6/R03} {\bibfield
  {journal} {\bibinfo  {journal} {Reports on Progress in Physics}\ }\textbf
  {\bibinfo {volume} {70}},\ \bibinfo {pages} {947} (\bibinfo {year}
  {2007})}\BibitemShut {NoStop}%
\bibitem [{\citenamefont {Makris}\ \emph {et~al.}(2008)\citenamefont {Makris},
  \citenamefont {El-Ganainy}, \citenamefont {Christodoulides},\ and\
  \citenamefont {Musslimani}}]{Makris2008}%
  \BibitemOpen
  \bibfield  {author} {\bibinfo {author} {\bibfnamefont {K.~G.}\ \bibnamefont
  {Makris}}, \bibinfo {author} {\bibfnamefont {R.}~\bibnamefont {El-Ganainy}},
  \bibinfo {author} {\bibfnamefont {D.~N.}\ \bibnamefont {Christodoulides}},\
  and\ \bibinfo {author} {\bibfnamefont {Z.~H.}\ \bibnamefont {Musslimani}},\
  }\href {https://doi.org/10.1103/PhysRevLett.100.103904} {\bibfield  {journal}
  {\bibinfo  {journal} {Physical Review Letters}\ }\textbf {\bibinfo {volume}
  {100}},\ \bibinfo {pages} {103904} (\bibinfo {year} {2008})}\BibitemShut
  {NoStop}%
\bibitem [{\citenamefont {R{\"{u}}ter}\ \emph {et~al.}(2010)\citenamefont
  {R{\"{u}}ter}, \citenamefont {Makris}, \citenamefont {El-Ganainy},
  \citenamefont {Christodoulides}, \citenamefont {Segev},\ and\ \citenamefont
  {Kip}}]{Ruter2010}%
  \BibitemOpen
  \bibfield  {author} {\bibinfo {author} {\bibfnamefont {C.~E.}\ \bibnamefont
  {R{\"{u}}ter}}, \bibinfo {author} {\bibfnamefont {K.~G.}\ \bibnamefont
  {Makris}}, \bibinfo {author} {\bibfnamefont {R.}~\bibnamefont {El-Ganainy}},
  \bibinfo {author} {\bibfnamefont {D.~N.}\ \bibnamefont {Christodoulides}},
  \bibinfo {author} {\bibfnamefont {M.}~\bibnamefont {Segev}},\ and\ \bibinfo
  {author} {\bibfnamefont {D.}~\bibnamefont {Kip}},\ }\href
  {https://doi.org/10.1038/nphys1515} {\bibfield  {journal} {\bibinfo
  {journal} {Nature Physics}\ }\textbf {\bibinfo {volume} {6}},\ \bibinfo
  {pages} {192} (\bibinfo {year} {2010})}\BibitemShut {NoStop}%
\bibitem [{\citenamefont {Chang}\ \emph {et~al.}(2014)\citenamefont {Chang},
  \citenamefont {Jiang}, \citenamefont {Hua}, \citenamefont {Yang},
  \citenamefont {Wen}, \citenamefont {Jiang}, \citenamefont {Li}, \citenamefont
  {Wang},\ and\ \citenamefont {Xiao}}]{Chang2014}%
  \BibitemOpen
  \bibfield  {author} {\bibinfo {author} {\bibfnamefont {L.}~\bibnamefont
  {Chang}}, \bibinfo {author} {\bibfnamefont {X.}~\bibnamefont {Jiang}},
  \bibinfo {author} {\bibfnamefont {S.}~\bibnamefont {Hua}}, \bibinfo {author}
  {\bibfnamefont {C.}~\bibnamefont {Yang}}, \bibinfo {author} {\bibfnamefont
  {J.}~\bibnamefont {Wen}}, \bibinfo {author} {\bibfnamefont {L.}~\bibnamefont
  {Jiang}}, \bibinfo {author} {\bibfnamefont {G.}~\bibnamefont {Li}}, \bibinfo
  {author} {\bibfnamefont {G.}~\bibnamefont {Wang}},\ and\ \bibinfo {author}
  {\bibfnamefont {M.}~\bibnamefont {Xiao}},\ }\href
  {https://doi.org/10.1038/nphoton.2014.133} {\bibfield  {journal} {\bibinfo
  {journal} {Nature Photonics}\ }\textbf {\bibinfo {volume} {8}},\ \bibinfo
  {pages} {524} (\bibinfo {year} {2014})}\BibitemShut {NoStop}%
\bibitem [{\citenamefont {Peng}\ \emph {et~al.}(2014)\citenamefont {Peng},
  \citenamefont {{\"{O}}zdemir}, \citenamefont {Lei}, \citenamefont {Monifi},
  \citenamefont {Gianfreda}, \citenamefont {Long}, \citenamefont {Fan},
  \citenamefont {Nori}, \citenamefont {Bender},\ and\ \citenamefont
  {Yang}}]{Peng2014}%
  \BibitemOpen
  \bibfield  {author} {\bibinfo {author} {\bibfnamefont {B.}~\bibnamefont
  {Peng}}, \bibinfo {author} {\bibfnamefont {S.~K.}\ \bibnamefont
  {{\"{O}}zdemir}}, \bibinfo {author} {\bibfnamefont {F.}~\bibnamefont {Lei}},
  \bibinfo {author} {\bibfnamefont {F.}~\bibnamefont {Monifi}}, \bibinfo
  {author} {\bibfnamefont {M.}~\bibnamefont {Gianfreda}}, \bibinfo {author}
  {\bibfnamefont {G.~L.}\ \bibnamefont {Long}}, \bibinfo {author}
  {\bibfnamefont {S.}~\bibnamefont {Fan}}, \bibinfo {author} {\bibfnamefont
  {F.}~\bibnamefont {Nori}}, \bibinfo {author} {\bibfnamefont {C.~M.}\
  \bibnamefont {Bender}},\ and\ \bibinfo {author} {\bibfnamefont
  {L.}~\bibnamefont {Yang}},\ }\href {https://doi.org/10.1038/nphys2927}
  {\bibfield  {journal} {\bibinfo  {journal} {Nature Physics}\ }\textbf
  {\bibinfo {volume} {10}},\ \bibinfo {pages} {394} (\bibinfo {year}
  {2014})}\BibitemShut {NoStop}%
\bibitem [{\citenamefont {Feng}\ \emph {et~al.}(2014)\citenamefont {Feng},
  \citenamefont {Wong}, \citenamefont {Ma}, \citenamefont {Wang},\ and\
  \citenamefont {Zhang}}]{Feng2014}%
  \BibitemOpen
  \bibfield  {author} {\bibinfo {author} {\bibfnamefont {L.}~\bibnamefont
  {Feng}}, \bibinfo {author} {\bibfnamefont {Z.~J.}\ \bibnamefont {Wong}},
  \bibinfo {author} {\bibfnamefont {R.-M.}\ \bibnamefont {Ma}}, \bibinfo
  {author} {\bibfnamefont {Y.}~\bibnamefont {Wang}},\ and\ \bibinfo {author}
  {\bibfnamefont {X.}~\bibnamefont {Zhang}},\ }\href
  {https://doi.org/10.1126/science.1258479} {\bibfield  {journal} {\bibinfo
  {journal} {Science}\ }\textbf {\bibinfo {volume} {346}},\ \bibinfo {pages}
  {972} (\bibinfo {year} {2014})}\BibitemShut {NoStop}%
\bibitem [{\citenamefont {Zhen}\ \emph {et~al.}(2015)\citenamefont {Zhen},
  \citenamefont {Hsu}, \citenamefont {Igarashi}, \citenamefont {Lu},
  \citenamefont {Kaminer}, \citenamefont {Pick}, \citenamefont {Chua},
  \citenamefont {Joannopoulos},\ and\ \citenamefont
  {Solja{\v{c}}i{\'{c}}}}]{Zhen2015}%
  \BibitemOpen
  \bibfield  {author} {\bibinfo {author} {\bibfnamefont {B.}~\bibnamefont
  {Zhen}}, \bibinfo {author} {\bibfnamefont {C.~W.}\ \bibnamefont {Hsu}},
  \bibinfo {author} {\bibfnamefont {Y.}~\bibnamefont {Igarashi}}, \bibinfo
  {author} {\bibfnamefont {L.}~\bibnamefont {Lu}}, \bibinfo {author}
  {\bibfnamefont {I.}~\bibnamefont {Kaminer}}, \bibinfo {author} {\bibfnamefont
  {A.}~\bibnamefont {Pick}}, \bibinfo {author} {\bibfnamefont {S.~L.}\
  \bibnamefont {Chua}}, \bibinfo {author} {\bibfnamefont {J.~D.}\ \bibnamefont
  {Joannopoulos}},\ and\ \bibinfo {author} {\bibfnamefont {M.}~\bibnamefont
  {Solja{\v{c}}i{\'{c}}}},\ }\href {https://doi.org/10.1038/nature14889}
  {\bibfield  {journal} {\bibinfo  {journal} {Nature}\ }\textbf {\bibinfo
  {volume} {525}},\ \bibinfo {pages} {354} (\bibinfo {year}
  {2015})}\BibitemShut {NoStop}%
\bibitem [{\citenamefont {Doppler}\ \emph {et~al.}(2016)\citenamefont
  {Doppler}, \citenamefont {Mailybaev}, \citenamefont {B{\"{o}}hm},
  \citenamefont {Kuhl}, \citenamefont {Girschik}, \citenamefont {Libisch},
  \citenamefont {Milburn}, \citenamefont {Rabl}, \citenamefont {Moiseyev},\
  and\ \citenamefont {Rotter}}]{Doppler2016}%
  \BibitemOpen
  \bibfield  {author} {\bibinfo {author} {\bibfnamefont {J.}~\bibnamefont
  {Doppler}}, \bibinfo {author} {\bibfnamefont {A.~A.}\ \bibnamefont
  {Mailybaev}}, \bibinfo {author} {\bibfnamefont {J.}~\bibnamefont
  {B{\"{o}}hm}}, \bibinfo {author} {\bibfnamefont {U.}~\bibnamefont {Kuhl}},
  \bibinfo {author} {\bibfnamefont {A.}~\bibnamefont {Girschik}}, \bibinfo
  {author} {\bibfnamefont {F.}~\bibnamefont {Libisch}}, \bibinfo {author}
  {\bibfnamefont {T.~J.}\ \bibnamefont {Milburn}}, \bibinfo {author}
  {\bibfnamefont {P.}~\bibnamefont {Rabl}}, \bibinfo {author} {\bibfnamefont
  {N.}~\bibnamefont {Moiseyev}},\ and\ \bibinfo {author} {\bibfnamefont
  {S.}~\bibnamefont {Rotter}},\ }\href {https://doi.org/10.1038/nature18605}
  {\bibfield  {journal} {\bibinfo  {journal} {Nature}\ }\textbf {\bibinfo
  {volume} {537}},\ \bibinfo {pages} {76} (\bibinfo {year} {2016})}\BibitemShut
  {NoStop}%
\bibitem [{\citenamefont {Xu}\ \emph {et~al.}(2016)\citenamefont {Xu},
  \citenamefont {Mason}, \citenamefont {Jiang},\ and\ \citenamefont
  {Harris}}]{Xu2016}%
  \BibitemOpen
  \bibfield  {author} {\bibinfo {author} {\bibfnamefont {H.}~\bibnamefont
  {Xu}}, \bibinfo {author} {\bibfnamefont {D.}~\bibnamefont {Mason}}, \bibinfo
  {author} {\bibfnamefont {L.}~\bibnamefont {Jiang}},\ and\ \bibinfo {author}
  {\bibfnamefont {J.~G.}\ \bibnamefont {Harris}},\ }\href
  {https://doi.org/10.1038/nature18604} {\bibfield  {journal} {\bibinfo
  {journal} {Nature}\ }\textbf {\bibinfo {volume} {537}},\ \bibinfo {pages}
  {80} (\bibinfo {year} {2016})}\BibitemShut {NoStop}%
\bibitem [{\citenamefont {Zhang}\ \emph {et~al.}(2016)\citenamefont {Zhang},
  \citenamefont {Zhang}, \citenamefont {Sheng}, \citenamefont {Yang},
  \citenamefont {Miri}, \citenamefont {Christodoulides}, \citenamefont {He},
  \citenamefont {Zhang},\ and\ \citenamefont {Xiao}}]{Zhang2016}%
  \BibitemOpen
  \bibfield  {author} {\bibinfo {author} {\bibfnamefont {Z.}~\bibnamefont
  {Zhang}}, \bibinfo {author} {\bibfnamefont {Y.}~\bibnamefont {Zhang}},
  \bibinfo {author} {\bibfnamefont {J.}~\bibnamefont {Sheng}}, \bibinfo
  {author} {\bibfnamefont {L.}~\bibnamefont {Yang}}, \bibinfo {author}
  {\bibfnamefont {M.~A.}\ \bibnamefont {Miri}}, \bibinfo {author}
  {\bibfnamefont {D.~N.}\ \bibnamefont {Christodoulides}}, \bibinfo {author}
  {\bibfnamefont {B.}~\bibnamefont {He}}, \bibinfo {author} {\bibfnamefont
  {Y.}~\bibnamefont {Zhang}},\ and\ \bibinfo {author} {\bibfnamefont
  {M.}~\bibnamefont {Xiao}},\ }\href
  {https://doi.org/10.1103/PhysRevLett.117.123601} {\bibfield  {journal}
  {\bibinfo  {journal} {Physical Review Letters}\ }\textbf {\bibinfo {volume}
  {117}},\ \bibinfo {pages} {123601} (\bibinfo {year} {2016})}\BibitemShut
  {NoStop}%
\bibitem [{\citenamefont {Li}\ \emph {et~al.}(2019)\citenamefont {Li},
  \citenamefont {Peng}, \citenamefont {Han}, \citenamefont {Miri},
  \citenamefont {Li}, \citenamefont {Xiao}, \citenamefont {Zhu}, \citenamefont
  {Zhao}, \citenamefont {Al{\`{u}}}, \citenamefont {Fan},\ and\ \citenamefont
  {Qiu}}]{Li2019}%
  \BibitemOpen
  \bibfield  {author} {\bibinfo {author} {\bibfnamefont {Y.}~\bibnamefont
  {Li}}, \bibinfo {author} {\bibfnamefont {Y.~G.}\ \bibnamefont {Peng}},
  \bibinfo {author} {\bibfnamefont {L.}~\bibnamefont {Han}}, \bibinfo {author}
  {\bibfnamefont {M.~A.}\ \bibnamefont {Miri}}, \bibinfo {author}
  {\bibfnamefont {W.}~\bibnamefont {Li}}, \bibinfo {author} {\bibfnamefont
  {M.}~\bibnamefont {Xiao}}, \bibinfo {author} {\bibfnamefont {X.~F.}\
  \bibnamefont {Zhu}}, \bibinfo {author} {\bibfnamefont {J.}~\bibnamefont
  {Zhao}}, \bibinfo {author} {\bibfnamefont {A.}~\bibnamefont {Al{\`{u}}}},
  \bibinfo {author} {\bibfnamefont {S.}~\bibnamefont {Fan}},\ and\ \bibinfo
  {author} {\bibfnamefont {C.~W.}\ \bibnamefont {Qiu}},\ }\href
  {https://doi.org/10.1126/science.aaw6259} {\bibfield  {journal} {\bibinfo
  {journal} {Science}\ }\textbf {\bibinfo {volume} {364}},\ \bibinfo {pages}
  {170} (\bibinfo {year} {2019})}\BibitemShut {NoStop}%
\bibitem [{\citenamefont {Schaden}\ \emph {et~al.}(2007)\citenamefont
  {Schaden}, \citenamefont {Zhao},\ and\ \citenamefont {Wu}}]{Schaden2007}%
  \BibitemOpen
  \bibfield  {author} {\bibinfo {author} {\bibfnamefont {M.}~\bibnamefont
  {Schaden}}, \bibinfo {author} {\bibfnamefont {K.~F.}\ \bibnamefont {Zhao}},\
  and\ \bibinfo {author} {\bibfnamefont {Z.}~\bibnamefont {Wu}},\ }\href
  {https://doi.org/10.1103/PhysRevA.76.062502} {\bibfield  {journal} {\bibinfo
  {journal} {Physical Review A}\ }\textbf {\bibinfo {volume} {76}},\ \bibinfo
  {pages} {062502} (\bibinfo {year} {2007})}\BibitemShut {NoStop}%
\bibitem [{\citenamefont {Zhao}\ \emph
  {et~al.}(2008{\natexlab{a}})\citenamefont {Zhao}, \citenamefont {Schaden},\
  and\ \citenamefont {Wu}}]{Zhao2008b}%
  \BibitemOpen
  \bibfield  {author} {\bibinfo {author} {\bibfnamefont {K.~F.}\ \bibnamefont
  {Zhao}}, \bibinfo {author} {\bibfnamefont {M.}~\bibnamefont {Schaden}},\ and\
  \bibinfo {author} {\bibfnamefont {Z.}~\bibnamefont {Wu}},\ }\href
  {https://doi.org/10.1103/PhysRevA.78.013418} {\bibfield  {journal} {\bibinfo
  {journal} {Physical Review A}\ }\textbf {\bibinfo {volume} {78}},\ \bibinfo
  {pages} {013418} (\bibinfo {year} {2008}{\natexlab{a}})}\BibitemShut
  {NoStop}%
\bibitem [{\citenamefont {Zhao}\ \emph
  {et~al.}(2008{\natexlab{b}})\citenamefont {Zhao}, \citenamefont {Schaden},\
  and\ \citenamefont {Wu}}]{Zhao2008a}%
  \BibitemOpen
  \bibfield  {author} {\bibinfo {author} {\bibfnamefont {K.~F.}\ \bibnamefont
  {Zhao}}, \bibinfo {author} {\bibfnamefont {M.}~\bibnamefont {Schaden}},\ and\
  \bibinfo {author} {\bibfnamefont {Z.}~\bibnamefont {Wu}},\ }\href
  {https://doi.org/10.1103/PhysRevA.78.034901} {\bibfield  {journal} {\bibinfo
  {journal} {Physical Review A}\ }\textbf {\bibinfo {volume} {78}},\ \bibinfo
  {pages} {034901} (\bibinfo {year} {2008}{\natexlab{b}})}\BibitemShut
  {NoStop}%
\bibitem [{\citenamefont {Zhao}\ \emph {et~al.}(2010)\citenamefont {Zhao},
  \citenamefont {Schaden},\ and\ \citenamefont {Wu}}]{Zhao2010}%
  \BibitemOpen
  \bibfield  {author} {\bibinfo {author} {\bibfnamefont {K.~F.}\ \bibnamefont
  {Zhao}}, \bibinfo {author} {\bibfnamefont {M.}~\bibnamefont {Schaden}},\ and\
  \bibinfo {author} {\bibfnamefont {Z.}~\bibnamefont {Wu}},\ }\href
  {https://doi.org/10.1103/PhysRevA.81.042903} {\bibfield  {journal} {\bibinfo
  {journal} {Physical Review A}\ }\textbf {\bibinfo {volume} {81}},\ \bibinfo
  {pages} {042903} (\bibinfo {year} {2010})}\BibitemShut {NoStop}%
\bibitem [{\citenamefont {Wu}(2021)}]{Wu2021}%
  \BibitemOpen
  \bibfield  {author} {\bibinfo {author} {\bibfnamefont {Z.}~\bibnamefont
  {Wu}},\ }\href {https://doi.org/10.1103/RevModPhys.93.035006} {\bibfield
  {journal} {\bibinfo  {journal} {Reviews of Modern Physics}\ }\textbf
  {\bibinfo {volume} {93}},\ \bibinfo {pages} {035006} (\bibinfo {year}
  {2021})}\BibitemShut {NoStop}%
\bibitem [{\citenamefont {Saam}\ \emph {et~al.}(1996)\citenamefont {Saam},
  \citenamefont {Drukker},\ and\ \citenamefont {Happer}}]{Saam1996}%
  \BibitemOpen
  \bibfield  {author} {\bibinfo {author} {\bibfnamefont {B.}~\bibnamefont
  {Saam}}, \bibinfo {author} {\bibfnamefont {N.}~\bibnamefont {Drukker}},\ and\
  \bibinfo {author} {\bibfnamefont {W.}~\bibnamefont {Happer}},\ }\href
  {https://doi.org/10.1016/S0009-2614(96)01238-9} {\bibfield  {journal}
  {\bibinfo  {journal} {Chemical Physics Letters}\ }\textbf {\bibinfo {volume}
  {263}},\ \bibinfo {pages} {481} (\bibinfo {year} {1996})}\BibitemShut
  {NoStop}%
\bibitem [{\citenamefont {{Cohen-Tannoudji, C.}}\ \emph
  {et~al.}(1970)\citenamefont {{Cohen-Tannoudji, C.}}, \citenamefont
  {{Dupont-Roc, J.}}, \citenamefont {{Haroche, S.}},\ and\ \citenamefont
  {{Lalo\"e, F.}}}]{Cohen-Tannoudji1970}%
  \BibitemOpen
  \bibfield  {author} {\bibinfo {author} {\bibnamefont {{Cohen-Tannoudji,
  C.}}}, \bibinfo {author} {\bibnamefont {{Dupont-Roc, J.}}}, \bibinfo {author}
  {\bibnamefont {{Haroche, S.}}},\ and\ \bibinfo {author} {\bibnamefont
  {{Lalo\"e, F.}}},\ }\href {https://doi.org/10.1051/rphysap:019700050109500}
  {\bibfield  {journal} {\bibinfo  {journal} {Rev. Phys. Appl. (Paris)}\
  }\textbf {\bibinfo {volume} {5}},\ \bibinfo {pages} {95} (\bibinfo {year}
  {1970})}\BibitemShut {NoStop}%
\bibitem [{\citenamefont {Eklund}(2008)}]{Eklund2008}%
  \BibitemOpen
  \bibfield  {author} {\bibinfo {author} {\bibfnamefont {E.~J.}\ \bibnamefont
  {Eklund}},\ }\emph {\bibinfo {title} {{Microgyroscope Based on Spin-Polarized
  Nuclei}}},\ \href@noop {} {Ph.D. thesis},\ \bibinfo  {school} {University of
  California, Irvine} (\bibinfo {year} {2008})\BibitemShut {NoStop}%
\bibitem [{\citenamefont {Tang}\ \emph {et~al.}(2019)\citenamefont {Tang},
  \citenamefont {Li}, \citenamefont {Zhang}, \citenamefont {Wang},\ and\
  \citenamefont {Zhao}}]{Tang2019}%
  \BibitemOpen
  \bibfield  {author} {\bibinfo {author} {\bibfnamefont {F.}~\bibnamefont
  {Tang}}, \bibinfo {author} {\bibfnamefont {A.-x.}\ \bibnamefont {Li}},
  \bibinfo {author} {\bibfnamefont {K.}~\bibnamefont {Zhang}}, \bibinfo
  {author} {\bibfnamefont {Y.}~\bibnamefont {Wang}},\ and\ \bibinfo {author}
  {\bibfnamefont {N.}~\bibnamefont {Zhao}},\ }\href
  {https://doi.org/10.1088/1361-6455/ab2a99} {\bibfield  {journal} {\bibinfo
  {journal} {Journal of Physics B: Atomic, Molecular and Optical Physics}\
  }\textbf {\bibinfo {volume} {52}},\ \bibinfo {pages} {205001} (\bibinfo
  {year} {2019})}\BibitemShut {NoStop}%
\bibitem [{\citenamefont {Zhang}\ \emph
  {et~al.}(2020{\natexlab{a}})\citenamefont {Zhang}, \citenamefont {Luo},
  \citenamefont {Tang}, \citenamefont {Zhao},\ and\ \citenamefont
  {Wang}}]{Zhang2020}%
  \BibitemOpen
  \bibfield  {author} {\bibinfo {author} {\bibfnamefont {K.}~\bibnamefont
  {Zhang}}, \bibinfo {author} {\bibfnamefont {Z.}~\bibnamefont {Luo}}, \bibinfo
  {author} {\bibfnamefont {F.}~\bibnamefont {Tang}}, \bibinfo {author}
  {\bibfnamefont {N.}~\bibnamefont {Zhao}},\ and\ \bibinfo {author}
  {\bibfnamefont {Y.}~\bibnamefont {Wang}},\ }\href
  {https://doi.org/10.35848/1347-4065/ab6caf} {\bibfield  {journal} {\bibinfo
  {journal} {Japanese Journal of Applied Physics}\ }\textbf {\bibinfo {volume}
  {59}},\ \bibinfo {pages} {030907} (\bibinfo {year}
  {2020}{\natexlab{a}})}\BibitemShut {NoStop}%
\bibitem [{\citenamefont {Song}\ \emph {et~al.}(2021)\citenamefont {Song},
  \citenamefont {Wang},\ and\ \citenamefont {Zhao}}]{Song2021}%
  \BibitemOpen
  \bibfield  {author} {\bibinfo {author} {\bibfnamefont {B.}~\bibnamefont
  {Song}}, \bibinfo {author} {\bibfnamefont {Y.}~\bibnamefont {Wang}},\ and\
  \bibinfo {author} {\bibfnamefont {N.}~\bibnamefont {Zhao}},\ }\href
  {https://doi.org/10.1103/PhysRevA.104.023105} {\bibfield  {journal} {\bibinfo
   {journal} {Physical Review A}\ }\textbf {\bibinfo {volume} {104}},\ \bibinfo
  {pages} {023105} (\bibinfo {year} {2021})}\BibitemShut {NoStop}%
\bibitem [{\citenamefont {Wiersig}(2014)}]{Wiersig2014}%
  \BibitemOpen
  \bibfield  {author} {\bibinfo {author} {\bibfnamefont {J.}~\bibnamefont
  {Wiersig}},\ }\href {https://doi.org/10.1103/PhysRevLett.112.203901}
  {\bibfield  {journal} {\bibinfo  {journal} {Physical Review Letters}\
  }\textbf {\bibinfo {volume} {112}},\ \bibinfo {pages} {203901} (\bibinfo
  {year} {2014})}\BibitemShut {NoStop}%
\bibitem [{\citenamefont {Wiersig}(2016)}]{Wiersig2016}%
  \BibitemOpen
  \bibfield  {author} {\bibinfo {author} {\bibfnamefont {J.}~\bibnamefont
  {Wiersig}},\ }\href {https://doi.org/10.1103/PhysRevA.93.033809} {\bibfield
  {journal} {\bibinfo  {journal} {Physical Review A}\ }\textbf {\bibinfo
  {volume} {93}},\ \bibinfo {pages} {033809} (\bibinfo {year}
  {2016})}\BibitemShut {NoStop}%
\bibitem [{\citenamefont {Liu}\ \emph {et~al.}(2016)\citenamefont {Liu},
  \citenamefont {Zhang}, \citenamefont {{\"O}zdemir}, \citenamefont {Peng},
  \citenamefont {Jing}, \citenamefont {L{\"{u}}}, \citenamefont {Li},
  \citenamefont {Yang}, \citenamefont {Nori},\ and\ \citenamefont
  {Liu}}]{Liu2016a}%
  \BibitemOpen
  \bibfield  {author} {\bibinfo {author} {\bibfnamefont {Z.~P.}\ \bibnamefont
  {Liu}}, \bibinfo {author} {\bibfnamefont {J.}~\bibnamefont {Zhang}}, \bibinfo
  {author} {\bibfnamefont {{\c{S}}.~K.}\ \bibnamefont {{\"O}zdemir}}, \bibinfo
  {author} {\bibfnamefont {B.}~\bibnamefont {Peng}}, \bibinfo {author}
  {\bibfnamefont {H.}~\bibnamefont {Jing}}, \bibinfo {author} {\bibfnamefont
  {X.~Y.}\ \bibnamefont {L{\"{u}}}}, \bibinfo {author} {\bibfnamefont {C.~W.}\
  \bibnamefont {Li}}, \bibinfo {author} {\bibfnamefont {L.}~\bibnamefont
  {Yang}}, \bibinfo {author} {\bibfnamefont {F.}~\bibnamefont {Nori}},\ and\
  \bibinfo {author} {\bibfnamefont {Y.~X.}\ \bibnamefont {Liu}},\ }\href
  {https://doi.org/10.1103/PhysRevLett.117.110802} {\bibfield  {journal}
  {\bibinfo  {journal} {Physical Review Letters}\ }\textbf {\bibinfo {volume}
  {117}},\ \bibinfo {pages} {110802} (\bibinfo {year} {2016})}\BibitemShut
  {NoStop}%
\bibitem [{\citenamefont {Chen}\ \emph {et~al.}(2017)\citenamefont {Chen},
  \citenamefont {Kaya~{\"O}zdemir}, \citenamefont {Zhao}, \citenamefont
  {Wiersig},\ and\ \citenamefont {Yang}}]{Chen2017}%
  \BibitemOpen
  \bibfield  {author} {\bibinfo {author} {\bibfnamefont {W.}~\bibnamefont
  {Chen}}, \bibinfo {author} {\bibfnamefont {{\c{S}}.}~\bibnamefont
  {Kaya~{\"O}zdemir}}, \bibinfo {author} {\bibfnamefont {G.}~\bibnamefont
  {Zhao}}, \bibinfo {author} {\bibfnamefont {J.}~\bibnamefont {Wiersig}},\ and\
  \bibinfo {author} {\bibfnamefont {L.}~\bibnamefont {Yang}},\ }\href
  {https://doi.org/10.1038/nature23281} {\bibfield  {journal} {\bibinfo
  {journal} {Nature}\ }\textbf {\bibinfo {volume} {548}},\ \bibinfo {pages}
  {192} (\bibinfo {year} {2017})}\BibitemShut {NoStop}%
\bibitem [{\citenamefont {Hodaei}\ \emph {et~al.}(2017)\citenamefont {Hodaei},
  \citenamefont {Hassan}, \citenamefont {Wittek}, \citenamefont
  {Garcia-Gracia}, \citenamefont {El-Ganainy}, \citenamefont
  {Christodoulides},\ and\ \citenamefont {Khajavikhan}}]{Hodaei2017}%
  \BibitemOpen
  \bibfield  {author} {\bibinfo {author} {\bibfnamefont {H.}~\bibnamefont
  {Hodaei}}, \bibinfo {author} {\bibfnamefont {A.~U.}\ \bibnamefont {Hassan}},
  \bibinfo {author} {\bibfnamefont {S.}~\bibnamefont {Wittek}}, \bibinfo
  {author} {\bibfnamefont {H.}~\bibnamefont {Garcia-Gracia}}, \bibinfo {author}
  {\bibfnamefont {R.}~\bibnamefont {El-Ganainy}}, \bibinfo {author}
  {\bibfnamefont {D.~N.}\ \bibnamefont {Christodoulides}},\ and\ \bibinfo
  {author} {\bibfnamefont {M.}~\bibnamefont {Khajavikhan}},\ }\href
  {https://doi.org/10.1038/nature23280} {\bibfield  {journal} {\bibinfo
  {journal} {Nature}\ }\textbf {\bibinfo {volume} {548}},\ \bibinfo {pages}
  {187} (\bibinfo {year} {2017})}\BibitemShut {NoStop}%
\bibitem [{\citenamefont {Lai}\ \emph {et~al.}(2019)\citenamefont {Lai},
  \citenamefont {Lu}, \citenamefont {Suh}, \citenamefont {Yuan},\ and\
  \citenamefont {Vahala}}]{Lai2019}%
  \BibitemOpen
  \bibfield  {author} {\bibinfo {author} {\bibfnamefont {Y.~H.}\ \bibnamefont
  {Lai}}, \bibinfo {author} {\bibfnamefont {Y.~K.}\ \bibnamefont {Lu}},
  \bibinfo {author} {\bibfnamefont {M.~G.}\ \bibnamefont {Suh}}, \bibinfo
  {author} {\bibfnamefont {Z.}~\bibnamefont {Yuan}},\ and\ \bibinfo {author}
  {\bibfnamefont {K.}~\bibnamefont {Vahala}},\ }\href
  {https://doi.org/10.1038/s41586-019-1777-z} {\bibfield  {journal} {\bibinfo
  {journal} {Nature}\ }\textbf {\bibinfo {volume} {576}},\ \bibinfo {pages}
  {65} (\bibinfo {year} {2019})}\BibitemShut {NoStop}%
\bibitem [{\citenamefont {Hokmabadi}\ \emph {et~al.}(2019)\citenamefont
  {Hokmabadi}, \citenamefont {Schumer}, \citenamefont {Christodoulides},\ and\
  \citenamefont {Khajavikhan}}]{Hokmabadi2019}%
  \BibitemOpen
  \bibfield  {author} {\bibinfo {author} {\bibfnamefont {M.~P.}\ \bibnamefont
  {Hokmabadi}}, \bibinfo {author} {\bibfnamefont {A.}~\bibnamefont {Schumer}},
  \bibinfo {author} {\bibfnamefont {D.~N.}\ \bibnamefont {Christodoulides}},\
  and\ \bibinfo {author} {\bibfnamefont {M.}~\bibnamefont {Khajavikhan}},\
  }\href {https://doi.org/10.1038/s41586-019-1780-4} {\bibfield  {journal}
  {\bibinfo  {journal} {Nature}\ }\textbf {\bibinfo {volume} {576}},\ \bibinfo
  {pages} {70} (\bibinfo {year} {2019})}\BibitemShut {NoStop}%
\bibitem [{\citenamefont {Makulski}(2015)}]{Makulski2015}%
  \BibitemOpen
  \bibfield  {author} {\bibinfo {author} {\bibfnamefont {W.}~\bibnamefont
  {Makulski}},\ }\href {https://doi.org/10.1002/mrc.4191} {\bibfield  {journal}
  {\bibinfo  {journal} {Magnetic Resonance in Chemistry}\ }\textbf {\bibinfo
  {volume} {53}},\ \bibinfo {pages} {273} (\bibinfo {year} {2015})}\BibitemShut
  {NoStop}%
\bibitem [{\citenamefont {Sheng}\ \emph {et~al.}(2014)\citenamefont {Sheng},
  \citenamefont {Kabcenell},\ and\ \citenamefont {Romalis}}]{Sheng2014}%
  \BibitemOpen
  \bibfield  {author} {\bibinfo {author} {\bibfnamefont {D.}~\bibnamefont
  {Sheng}}, \bibinfo {author} {\bibfnamefont {A.}~\bibnamefont {Kabcenell}},\
  and\ \bibinfo {author} {\bibfnamefont {M.~V.}\ \bibnamefont {Romalis}},\
  }\href {https://doi.org/10.1103/PhysRevLett.113.163002} {\bibfield  {journal}
  {\bibinfo  {journal} {Physical Review Letters}\ }\textbf {\bibinfo {volume}
  {113}},\ \bibinfo {pages} {163002} (\bibinfo {year} {2014})}\BibitemShut
  {NoStop}%
\bibitem [{\citenamefont {Limes}\ \emph {et~al.}(2018)\citenamefont {Limes},
  \citenamefont {Sheng},\ and\ \citenamefont {Romalis}}]{Limes2018}%
  \BibitemOpen
  \bibfield  {author} {\bibinfo {author} {\bibfnamefont {M.~E.}\ \bibnamefont
  {Limes}}, \bibinfo {author} {\bibfnamefont {D.}~\bibnamefont {Sheng}},\ and\
  \bibinfo {author} {\bibfnamefont {M.~V.}\ \bibnamefont {Romalis}},\ }\href
  {https://doi.org/10.1103/PhysRevLett.120.033401} {\bibfield  {journal}
  {\bibinfo  {journal} {Physical Review Letters}\ }\textbf {\bibinfo {volume}
  {120}},\ \bibinfo {pages} {033401} (\bibinfo {year} {2018})}\BibitemShut
  {NoStop}%
\bibitem [{\citenamefont {Korver}\ \emph {et~al.}(2015)\citenamefont {Korver},
  \citenamefont {Thrasher}, \citenamefont {Bulatowicz},\ and\ \citenamefont
  {Walker}}]{Korver2015}%
  \BibitemOpen
  \bibfield  {author} {\bibinfo {author} {\bibfnamefont {A.}~\bibnamefont
  {Korver}}, \bibinfo {author} {\bibfnamefont {D.}~\bibnamefont {Thrasher}},
  \bibinfo {author} {\bibfnamefont {M.}~\bibnamefont {Bulatowicz}},\ and\
  \bibinfo {author} {\bibfnamefont {T.~G.}\ \bibnamefont {Walker}},\ }\href
  {https://doi.org/10.1103/PhysRevLett.115.253001} {\bibfield  {journal}
  {\bibinfo  {journal} {Physical Review Letters}\ }\textbf {\bibinfo {volume}
  {115}},\ \bibinfo {pages} {253001} (\bibinfo {year} {2015})}\BibitemShut
  {NoStop}%
\bibitem [{\citenamefont {Thrasher}\ \emph {et~al.}(2019)\citenamefont
  {Thrasher}, \citenamefont {Sorensen}, \citenamefont {Weber}, \citenamefont
  {Bulatowicz}, \citenamefont {Korver}, \citenamefont {Larsen},\ and\
  \citenamefont {Walker}}]{Thrasher2019}%
  \BibitemOpen
  \bibfield  {author} {\bibinfo {author} {\bibfnamefont {D.~A.}\ \bibnamefont
  {Thrasher}}, \bibinfo {author} {\bibfnamefont {S.~S.}\ \bibnamefont
  {Sorensen}}, \bibinfo {author} {\bibfnamefont {J.}~\bibnamefont {Weber}},
  \bibinfo {author} {\bibfnamefont {M.}~\bibnamefont {Bulatowicz}}, \bibinfo
  {author} {\bibfnamefont {A.}~\bibnamefont {Korver}}, \bibinfo {author}
  {\bibfnamefont {M.}~\bibnamefont {Larsen}},\ and\ \bibinfo {author}
  {\bibfnamefont {T.~G.}\ \bibnamefont {Walker}},\ }\href
  {https://doi.org/10.1103/PhysRevA.100.061403} {\bibfield  {journal} {\bibinfo
   {journal} {Physical Review A}\ }\textbf {\bibinfo {volume} {100}},\ \bibinfo
  {pages} {061403(R)} (\bibinfo {year} {2019})}\BibitemShut {NoStop}%
\bibitem [{\citenamefont {Sheng}\ \emph {et~al.}(2017)\citenamefont {Sheng},
  \citenamefont {Perry}, \citenamefont {Krzyzewski}, \citenamefont {Geller},
  \citenamefont {Kitching},\ and\ \citenamefont {Knappe}}]{Sheng2017}%
  \BibitemOpen
  \bibfield  {author} {\bibinfo {author} {\bibfnamefont {D.}~\bibnamefont
  {Sheng}}, \bibinfo {author} {\bibfnamefont {A.~R.}\ \bibnamefont {Perry}},
  \bibinfo {author} {\bibfnamefont {S.~P.}\ \bibnamefont {Krzyzewski}},
  \bibinfo {author} {\bibfnamefont {S.}~\bibnamefont {Geller}}, \bibinfo
  {author} {\bibfnamefont {J.}~\bibnamefont {Kitching}},\ and\ \bibinfo
  {author} {\bibfnamefont {S.}~\bibnamefont {Knappe}},\ }\href
  {https://doi.org/10.1063/1.4974349} {\bibfield  {journal} {\bibinfo
  {journal} {Appl Phys Lett}\ }\textbf {\bibinfo {volume} {110}},\ \bibinfo
  {pages} {31106} (\bibinfo {year} {2017})}\BibitemShut {NoStop}%
\bibitem [{\citenamefont {Zhang}\ \emph
  {et~al.}(2020{\natexlab{b}})\citenamefont {Zhang}, \citenamefont {Xiao},
  \citenamefont {Ding}, \citenamefont {Feng}, \citenamefont {Peng},
  \citenamefont {Shen}, \citenamefont {Sun}, \citenamefont {Wu}, \citenamefont
  {Wu}, \citenamefont {Yang}, \citenamefont {Zheng}, \citenamefont {Zhang},
  \citenamefont {Chen},\ and\ \citenamefont {Guo}}]{Zhang2020a}%
  \BibitemOpen
  \bibfield  {author} {\bibinfo {author} {\bibfnamefont {R.}~\bibnamefont
  {Zhang}}, \bibinfo {author} {\bibfnamefont {W.}~\bibnamefont {Xiao}},
  \bibinfo {author} {\bibfnamefont {Y.}~\bibnamefont {Ding}}, \bibinfo {author}
  {\bibfnamefont {Y.}~\bibnamefont {Feng}}, \bibinfo {author} {\bibfnamefont
  {X.}~\bibnamefont {Peng}}, \bibinfo {author} {\bibfnamefont {L.}~\bibnamefont
  {Shen}}, \bibinfo {author} {\bibfnamefont {C.}~\bibnamefont {Sun}}, \bibinfo
  {author} {\bibfnamefont {T.}~\bibnamefont {Wu}}, \bibinfo {author}
  {\bibfnamefont {Y.}~\bibnamefont {Wu}}, \bibinfo {author} {\bibfnamefont
  {Y.}~\bibnamefont {Yang}}, \bibinfo {author} {\bibfnamefont {Z.}~\bibnamefont
  {Zheng}}, \bibinfo {author} {\bibfnamefont {X.}~\bibnamefont {Zhang}},
  \bibinfo {author} {\bibfnamefont {J.}~\bibnamefont {Chen}},\ and\ \bibinfo
  {author} {\bibfnamefont {H.}~\bibnamefont {Guo}},\ }\href
  {https://doi.org/10.1126/sciadv.aba8792} {\bibfield  {journal} {\bibinfo
  {journal} {Science Advances}\ }\textbf {\bibinfo {volume} {6}},\ \bibinfo
  {pages} {eaba8792} (\bibinfo {year} {2020}{\natexlab{b}})}\BibitemShut
  {NoStop}%
\bibitem [{\citenamefont {Langbein}(2018)}]{Langbein2018}%
  \BibitemOpen
  \bibfield  {author} {\bibinfo {author} {\bibfnamefont {W.}~\bibnamefont
  {Langbein}},\ }\href {https://doi.org/10.1103/PhysRevA.98.023805} {\bibfield
  {journal} {\bibinfo  {journal} {Physical Review A}\ }\textbf {\bibinfo
  {volume} {98}},\ \bibinfo {pages} {023805} (\bibinfo {year}
  {2018})}\BibitemShut {NoStop}%
\bibitem [{\citenamefont {Zhang}\ \emph {et~al.}(2019)\citenamefont {Zhang},
  \citenamefont {Sweeney}, \citenamefont {Hsu}, \citenamefont {Yang},
  \citenamefont {Stone},\ and\ \citenamefont {Jiang}}]{Zhang2019}%
  \BibitemOpen
  \bibfield  {author} {\bibinfo {author} {\bibfnamefont {M.}~\bibnamefont
  {Zhang}}, \bibinfo {author} {\bibfnamefont {W.}~\bibnamefont {Sweeney}},
  \bibinfo {author} {\bibfnamefont {C.~W.}\ \bibnamefont {Hsu}}, \bibinfo
  {author} {\bibfnamefont {L.}~\bibnamefont {Yang}}, \bibinfo {author}
  {\bibfnamefont {A.~D.}\ \bibnamefont {Stone}},\ and\ \bibinfo {author}
  {\bibfnamefont {L.}~\bibnamefont {Jiang}},\ }\href
  {https://doi.org/10.1103/PhysRevLett.123.180501} {\bibfield  {journal}
  {\bibinfo  {journal} {Physical Review Letters}\ }\textbf {\bibinfo {volume}
  {123}},\ \bibinfo {pages} {180501} (\bibinfo {year} {2019})}\BibitemShut
  {NoStop}%
\bibitem [{\citenamefont {Lau}\ and\ \citenamefont {Clerk}(2018)}]{Lau2018}%
  \BibitemOpen
  \bibfield  {author} {\bibinfo {author} {\bibfnamefont {H.~K.}\ \bibnamefont
  {Lau}}\ and\ \bibinfo {author} {\bibfnamefont {A.~A.}\ \bibnamefont
  {Clerk}},\ }\href {https://doi.org/10.1038/s41467-018-06477-7} {\bibfield
  {journal} {\bibinfo  {journal} {Nature Communications}\ }\textbf {\bibinfo
  {volume} {9}},\ \bibinfo {pages} {4320} (\bibinfo {year} {2018})}\BibitemShut
  {NoStop}%
\bibitem [{\citenamefont {Chen}\ \emph {et~al.}(2019)\citenamefont {Chen},
  \citenamefont {Jin},\ and\ \citenamefont {Liu}}]{Chen2019}%
  \BibitemOpen
  \bibfield  {author} {\bibinfo {author} {\bibfnamefont {C.}~\bibnamefont
  {Chen}}, \bibinfo {author} {\bibfnamefont {L.}~\bibnamefont {Jin}},\ and\
  \bibinfo {author} {\bibfnamefont {R.~B.}\ \bibnamefont {Liu}},\ }\href
  {https://doi.org/10.1088/1367-2630/ab32ab} {\bibfield  {journal} {\bibinfo
  {journal} {New Journal of Physics}\ }\textbf {\bibinfo {volume} {21}},\
  \bibinfo {pages} {083002} (\bibinfo {year} {2019})}\BibitemShut {NoStop}%
\bibitem [{\citenamefont {Happer}(1972)}]{Happer1972}%
  \BibitemOpen
  \bibfield  {author} {\bibinfo {author} {\bibfnamefont {W.}~\bibnamefont
  {Happer}},\ }\href {https://doi.org/10.1103/RevModPhys.44.169} {\bibfield
  {journal} {\bibinfo  {journal} {Reviews of Modern Physics}\ }\textbf
  {\bibinfo {volume} {44}},\ \bibinfo {pages} {169} (\bibinfo {year}
  {1972})}\BibitemShut {NoStop}%
\bibitem [{\citenamefont {Walker}\ and\ \citenamefont
  {Happer}(1997)}]{Walker1997}%
  \BibitemOpen
  \bibfield  {author} {\bibinfo {author} {\bibfnamefont {T.~G.}\ \bibnamefont
  {Walker}}\ and\ \bibinfo {author} {\bibfnamefont {W.}~\bibnamefont
  {Happer}},\ }\href {https://doi.org/10.1103/revmodphys.69.629} {\bibfield
  {journal} {\bibinfo  {journal} {Reviews of Modern Physics}\ }\textbf
  {\bibinfo {volume} {69}},\ \bibinfo {pages} {629} (\bibinfo {year}
  {1997})}\BibitemShut {NoStop}%
\bibitem [{\citenamefont {Seltzer}(2008)}]{Seltzer2008}%
  \BibitemOpen
  \bibfield  {author} {\bibinfo {author} {\bibfnamefont {S.~J.}\ \bibnamefont
  {Seltzer}},\ }\emph {\bibinfo {title} {{Developments in Alkali-Metal Atomic
  Magnetometry}}},\ \href@noop {} {Ph.D. thesis},\ \bibinfo  {school}
  {Princeton University} (\bibinfo {year} {2008})\BibitemShut {NoStop}%
\bibitem [{\citenamefont {Grover}(1978)}]{Grover1978}%
  \BibitemOpen
  \bibfield  {author} {\bibinfo {author} {\bibfnamefont {B.~C.}\ \bibnamefont
  {Grover}},\ }\href {https://doi.org/10.1103/PhysRevLett.40.391} {\bibfield
  {journal} {\bibinfo  {journal} {Physical Review Letters}\ }\textbf {\bibinfo
  {volume} {40}},\ \bibinfo {pages} {391} (\bibinfo {year} {1978})}\BibitemShut
  {NoStop}%
\bibitem [{\citenamefont {Zheng}\ \emph {et~al.}(2011)\citenamefont {Zheng},
  \citenamefont {Gao}, \citenamefont {Liu}, \citenamefont {Zhang},
  \citenamefont {Ye},\ and\ \citenamefont {Swank}}]{Zheng2011}%
  \BibitemOpen
  \bibfield  {author} {\bibinfo {author} {\bibfnamefont {W.}~\bibnamefont
  {Zheng}}, \bibinfo {author} {\bibfnamefont {H.}~\bibnamefont {Gao}}, \bibinfo
  {author} {\bibfnamefont {J.~G.}\ \bibnamefont {Liu}}, \bibinfo {author}
  {\bibfnamefont {Y.}~\bibnamefont {Zhang}}, \bibinfo {author} {\bibfnamefont
  {Q.}~\bibnamefont {Ye}},\ and\ \bibinfo {author} {\bibfnamefont
  {C.}~\bibnamefont {Swank}},\ }\href
  {https://doi.org/10.1103/PhysRevA.84.053411} {\bibfield  {journal} {\bibinfo
  {journal} {Physical Review A}\ }\textbf {\bibinfo {volume} {84}},\ \bibinfo
  {pages} {053411} (\bibinfo {year} {2011})}\BibitemShut {NoStop}%
\bibitem [{\citenamefont {Poling}\ \emph {et~al.}(2001)\citenamefont {Poling},
  \citenamefont {Prausnitz},\ and\ \citenamefont {O'Connell}}]{Poling2001}%
  \BibitemOpen
  \bibfield  {author} {\bibinfo {author} {\bibfnamefont {B.~E.}\ \bibnamefont
  {Poling}}, \bibinfo {author} {\bibfnamefont {J.~M.}\ \bibnamefont
  {Prausnitz}},\ and\ \bibinfo {author} {\bibfnamefont {J.~P.}\ \bibnamefont
  {O'Connell}},\ }\href
  {https://www.accessengineeringlibrary.com/content/book/9780070116825} {\emph
  {\bibinfo {title} {{The Properties of Gases and Liquids}}}},\ \bibinfo
  {edition} {5th}\ ed.\ (\bibinfo  {publisher} {McGraw-Hill Education},\
  \bibinfo {address} {New York},\ \bibinfo {year} {2001})\BibitemShut {NoStop}%
\bibitem [{\citenamefont {Feng}\ \emph {et~al.}(2020)\citenamefont {Feng},
  \citenamefont {Zhang}, \citenamefont {Lu},\ and\ \citenamefont
  {Sheng}}]{Feng2020}%
  \BibitemOpen
  \bibfield  {author} {\bibinfo {author} {\bibfnamefont {Y.~K.}\ \bibnamefont
  {Feng}}, \bibinfo {author} {\bibfnamefont {S.~B.}\ \bibnamefont {Zhang}},
  \bibinfo {author} {\bibfnamefont {Z.~T.}\ \bibnamefont {Lu}},\ and\ \bibinfo
  {author} {\bibfnamefont {D.}~\bibnamefont {Sheng}},\ }\href
  {https://doi.org/10.1103/PhysRevA.102.043109} {\bibfield  {journal} {\bibinfo
   {journal} {Physical Review A}\ }\textbf {\bibinfo {volume} {102}},\ \bibinfo
  {pages} {043109} (\bibinfo {year} {2020})}\BibitemShut {NoStop}%
\end{thebibliography}%


\providecommand{\noopsort}[1]{}\providecommand{\singleletter}[1]{#1}%
\begin{thebibliography}{14}%
\makeatletter
\providecommand \@ifxundefined [1]{%
 \@ifx{#1\undefined}
}%
\providecommand \@ifnum [1]{%
 \ifnum #1\expandafter \@firstoftwo
 \else \expandafter \@secondoftwo
 \fi
}%
\providecommand \@ifx [1]{%
 \ifx #1\expandafter \@firstoftwo
 \else \expandafter \@secondoftwo
 \fi
}%
\providecommand \natexlab [1]{#1}%
\providecommand \enquote  [1]{``#1''}%
\providecommand \bibnamefont  [1]{#1}%
\providecommand \bibfnamefont [1]{#1}%
\providecommand \citenamefont [1]{#1}%
\providecommand \href@noop [0]{\@secondoftwo}%
\providecommand \href [0]{\begingroup \@sanitize@url \@href}%
\providecommand \@href[1]{\@@startlink{#1}\@@href}%
\providecommand \@@href[1]{\endgroup#1\@@endlink}%
\providecommand \@sanitize@url [0]{\catcode `\\12\catcode `\$12\catcode
  `\&12\catcode `\#12\catcode `\^12\catcode `\_12\catcode `\%12\relax}%
\providecommand \@@startlink[1]{}%
\providecommand \@@endlink[0]{}%
\providecommand \url  [0]{\begingroup\@sanitize@url \@url }%
\providecommand \@url [1]{\endgroup\@href {#1}{\urlprefix }}%
\providecommand \urlprefix  [0]{URL }%
\providecommand \Eprint [0]{\href }%
\providecommand \doibase [0]{https://doi.org/}%
\providecommand \selectlanguage [0]{\@gobble}%
\providecommand \bibinfo  [0]{\@secondoftwo}%
\providecommand \bibfield  [0]{\@secondoftwo}%
\providecommand \translation [1]{[#1]}%
\providecommand \BibitemOpen [0]{}%
\providecommand \bibitemStop [0]{}%
\providecommand \bibitemNoStop [0]{.\EOS\space}%
\providecommand \EOS [0]{\spacefactor3000\relax}%
\providecommand \BibitemShut  [1]{\csname bibitem#1\endcsname}%
\let\auto@bib@innerbib\@empty
\bibitem [{\citenamefont {Happer}(1972)}]{Happer1972}%
  \BibitemOpen
  \bibfield  {author} {\bibinfo {author} {\bibfnamefont {W.}~\bibnamefont
  {Happer}},\ }\href {https://doi.org/10.1103/RevModPhys.44.169} {\bibfield
  {journal} {\bibinfo  {journal} {Reviews of Modern Physics}\ }\textbf
  {\bibinfo {volume} {44}},\ \bibinfo {pages} {169} (\bibinfo {year}
  {1972})}\BibitemShut {NoStop}%
\bibitem [{\citenamefont {Walker}\ and\ \citenamefont
  {Happer}(1997)}]{Walker1997}%
  \BibitemOpen
  \bibfield  {author} {\bibinfo {author} {\bibfnamefont {T.~G.}\ \bibnamefont
  {Walker}}\ and\ \bibinfo {author} {\bibfnamefont {W.}~\bibnamefont
  {Happer}},\ }\href {https://doi.org/10.1103/revmodphys.69.629} {\bibfield
  {journal} {\bibinfo  {journal} {Reviews of Modern Physics}\ }\textbf
  {\bibinfo {volume} {69}},\ \bibinfo {pages} {629} (\bibinfo {year}
  {1997})}\BibitemShut {NoStop}%
\bibitem [{\citenamefont {Seltzer}(2008)}]{Seltzer2008}%
  \BibitemOpen
  \bibfield  {author} {\bibinfo {author} {\bibfnamefont {S.~J.}\ \bibnamefont
  {Seltzer}},\ }\emph {\bibinfo {title} {{Developments in Alkali-Metal Atomic
  Magnetometry}}},\ \href@noop {} {Ph.D. thesis},\ \bibinfo  {school}
  {Princeton University} (\bibinfo {year} {2008})\BibitemShut {NoStop}%
\bibitem [{\citenamefont {Grover}(1978)}]{Grover1978}%
  \BibitemOpen
  \bibfield  {author} {\bibinfo {author} {\bibfnamefont {B.~C.}\ \bibnamefont
  {Grover}},\ }\href {https://doi.org/10.1103/PhysRevLett.40.391} {\bibfield
  {journal} {\bibinfo  {journal} {Physical Review Letters}\ }\textbf {\bibinfo
  {volume} {40}},\ \bibinfo {pages} {391} (\bibinfo {year} {1978})}\BibitemShut
  {NoStop}%
\bibitem [{\citenamefont {{Cohen-Tannoudji, C.}}\ \emph
  {et~al.}(1970)\citenamefont {{Cohen-Tannoudji, C.}}, \citenamefont
  {{Dupont-Roc, J.}}, \citenamefont {{Haroche, S.}},\ and\ \citenamefont
  {{Lalo\"e, F.}}}]{Cohen-Tannoudji1970}%
  \BibitemOpen
  \bibfield  {author} {\bibinfo {author} {\bibnamefont {{Cohen-Tannoudji,
  C.}}}, \bibinfo {author} {\bibnamefont {{Dupont-Roc, J.}}}, \bibinfo {author}
  {\bibnamefont {{Haroche, S.}}},\ and\ \bibinfo {author} {\bibnamefont
  {{Lalo\"e, F.}}},\ }\href {https://doi.org/10.1051/rphysap:019700050109500}
  {\bibfield  {journal} {\bibinfo  {journal} {Rev. Phys. Appl. (Paris)}\
  }\textbf {\bibinfo {volume} {5}},\ \bibinfo {pages} {95} (\bibinfo {year}
  {1970})}\BibitemShut {NoStop}%
\bibitem [{\citenamefont {Eklund}(2008)}]{Eklund2008}%
  \BibitemOpen
  \bibfield  {author} {\bibinfo {author} {\bibfnamefont {E.~J.}\ \bibnamefont
  {Eklund}},\ }\emph {\bibinfo {title} {{Microgyroscope Based on Spin-Polarized
  Nuclei}}},\ \href@noop {} {Ph.D. thesis},\ \bibinfo  {school} {University of
  California, Irvine} (\bibinfo {year} {2008})\BibitemShut {NoStop}%
\bibitem [{\citenamefont {Tang}\ \emph {et~al.}(2019)\citenamefont {Tang},
  \citenamefont {Li}, \citenamefont {Zhang}, \citenamefont {Wang},\ and\
  \citenamefont {Zhao}}]{Tang2019}%
  \BibitemOpen
  \bibfield  {author} {\bibinfo {author} {\bibfnamefont {F.}~\bibnamefont
  {Tang}}, \bibinfo {author} {\bibfnamefont {A.-x.}\ \bibnamefont {Li}},
  \bibinfo {author} {\bibfnamefont {K.}~\bibnamefont {Zhang}}, \bibinfo
  {author} {\bibfnamefont {Y.}~\bibnamefont {Wang}},\ and\ \bibinfo {author}
  {\bibfnamefont {N.}~\bibnamefont {Zhao}},\ }\href
  {https://doi.org/10.1088/1361-6455/ab2a99} {\bibfield  {journal} {\bibinfo
  {journal} {Journal of Physics B: Atomic, Molecular and Optical Physics}\
  }\textbf {\bibinfo {volume} {52}},\ \bibinfo {pages} {205001} (\bibinfo
  {year} {2019})}\BibitemShut {NoStop}%
\bibitem [{\citenamefont {Zhang}\ \emph {et~al.}(2020)\citenamefont {Zhang},
  \citenamefont {Luo}, \citenamefont {Tang}, \citenamefont {Zhao},\ and\
  \citenamefont {Wang}}]{Zhang2020}%
  \BibitemOpen
  \bibfield  {author} {\bibinfo {author} {\bibfnamefont {K.}~\bibnamefont
  {Zhang}}, \bibinfo {author} {\bibfnamefont {Z.}~\bibnamefont {Luo}}, \bibinfo
  {author} {\bibfnamefont {F.}~\bibnamefont {Tang}}, \bibinfo {author}
  {\bibfnamefont {N.}~\bibnamefont {Zhao}},\ and\ \bibinfo {author}
  {\bibfnamefont {Y.}~\bibnamefont {Wang}},\ }\href
  {https://doi.org/10.35848/1347-4065/ab6caf} {\bibfield  {journal} {\bibinfo
  {journal} {Japanese Journal of Applied Physics}\ }\textbf {\bibinfo {volume}
  {59}},\ \bibinfo {pages} {030907} (\bibinfo {year} {2020})}\BibitemShut
  {NoStop}%
\bibitem [{\citenamefont {Song}\ \emph {et~al.}(2021)\citenamefont {Song},
  \citenamefont {Wang},\ and\ \citenamefont {Zhao}}]{Song2021}%
  \BibitemOpen
  \bibfield  {author} {\bibinfo {author} {\bibfnamefont {B.}~\bibnamefont
  {Song}}, \bibinfo {author} {\bibfnamefont {Y.}~\bibnamefont {Wang}},\ and\
  \bibinfo {author} {\bibfnamefont {N.}~\bibnamefont {Zhao}},\ }\href
  {https://doi.org/10.1103/PhysRevA.104.023105} {\bibfield  {journal} {\bibinfo
   {journal} {Physical Review A}\ }\textbf {\bibinfo {volume} {104}},\ \bibinfo
  {pages} {023105} (\bibinfo {year} {2021})}\BibitemShut {NoStop}%
\bibitem [{\citenamefont {Torrey}(1956)}]{Torrey1956}%
  \BibitemOpen
  \bibfield  {author} {\bibinfo {author} {\bibfnamefont {H.~C.}\ \bibnamefont
  {Torrey}},\ }\href {https://doi.org/10.1103/PhysRev.104.563} {\bibfield
  {journal} {\bibinfo  {journal} {Physical Review}\ }\textbf {\bibinfo {volume}
  {104}},\ \bibinfo {pages} {563} (\bibinfo {year} {1956})}\BibitemShut
  {NoStop}%
\bibitem [{\citenamefont {Zheng}\ \emph {et~al.}(2011)\citenamefont {Zheng},
  \citenamefont {Gao}, \citenamefont {Liu}, \citenamefont {Zhang},
  \citenamefont {Ye},\ and\ \citenamefont {Swank}}]{Zheng2011}%
  \BibitemOpen
  \bibfield  {author} {\bibinfo {author} {\bibfnamefont {W.}~\bibnamefont
  {Zheng}}, \bibinfo {author} {\bibfnamefont {H.}~\bibnamefont {Gao}}, \bibinfo
  {author} {\bibfnamefont {J.~G.}\ \bibnamefont {Liu}}, \bibinfo {author}
  {\bibfnamefont {Y.}~\bibnamefont {Zhang}}, \bibinfo {author} {\bibfnamefont
  {Q.}~\bibnamefont {Ye}},\ and\ \bibinfo {author} {\bibfnamefont
  {C.}~\bibnamefont {Swank}},\ }\href
  {https://doi.org/10.1103/PhysRevA.84.053411} {\bibfield  {journal} {\bibinfo
  {journal} {Physical Review A}\ }\textbf {\bibinfo {volume} {84}},\ \bibinfo
  {pages} {053411} (\bibinfo {year} {2011})}\BibitemShut {NoStop}%
\bibitem [{\citenamefont {Poling}\ \emph {et~al.}(2001)\citenamefont {Poling},
  \citenamefont {Prausnitz},\ and\ \citenamefont {O'Connell}}]{Poling2001}%
  \BibitemOpen
  \bibfield  {author} {\bibinfo {author} {\bibfnamefont {B.~E.}\ \bibnamefont
  {Poling}}, \bibinfo {author} {\bibfnamefont {J.~M.}\ \bibnamefont
  {Prausnitz}},\ and\ \bibinfo {author} {\bibfnamefont {J.~P.}\ \bibnamefont
  {O'Connell}},\ }\href
  {https://www.accessengineeringlibrary.com/content/book/9780070116825} {\emph
  {\bibinfo {title} {{The Properties of Gases and Liquids}}}},\ \bibinfo
  {edition} {5th}\ ed.\ (\bibinfo  {publisher} {McGraw-Hill Education},\
  \bibinfo {address} {New York},\ \bibinfo {year} {2001})\BibitemShut {NoStop}%
\bibitem [{\citenamefont {Stoller}\ \emph {et~al.}(1991)\citenamefont
  {Stoller}, \citenamefont {Happer},\ and\ \citenamefont
  {Dyson}}]{Stoller1991}%
  \BibitemOpen
  \bibfield  {author} {\bibinfo {author} {\bibfnamefont {S.~D.}\ \bibnamefont
  {Stoller}}, \bibinfo {author} {\bibfnamefont {W.}~\bibnamefont {Happer}},\
  and\ \bibinfo {author} {\bibfnamefont {F.~J.}\ \bibnamefont {Dyson}},\ }\href
  {https://doi.org/10.1103/PhysRevA.44.7459} {\bibfield  {journal} {\bibinfo
  {journal} {Physical Review A}\ }\textbf {\bibinfo {volume} {44}},\ \bibinfo
  {pages} {7459} (\bibinfo {year} {1991})}\BibitemShut {NoStop}%
\bibitem [{\citenamefont {Feng}\ \emph {et~al.}(2020)\citenamefont {Feng},
  \citenamefont {Zhang}, \citenamefont {Lu},\ and\ \citenamefont
  {Sheng}}]{Feng2020}%
  \BibitemOpen
  \bibfield  {author} {\bibinfo {author} {\bibfnamefont {Y.~K.}\ \bibnamefont
  {Feng}}, \bibinfo {author} {\bibfnamefont {S.~B.}\ \bibnamefont {Zhang}},
  \bibinfo {author} {\bibfnamefont {Z.~T.}\ \bibnamefont {Lu}},\ and\ \bibinfo
  {author} {\bibfnamefont {D.}~\bibnamefont {Sheng}},\ }\href
  {https://doi.org/10.1103/PhysRevA.102.043109} {\bibfield  {journal} {\bibinfo
   {journal} {Physical Review A}\ }\textbf {\bibinfo {volume} {102}},\ \bibinfo
  {pages} {043109} (\bibinfo {year} {2020})}\BibitemShut {NoStop}%
\end{thebibliography}%

\end{document}


\title{Supplemental Material \\for \\``Stable Atomic Magnetometer in Parity-Time Symmetry Broken Phase"}
\author{Xiangdong Zhang}
\author{Jinbo Hu}
\author{Nan Zhao}
\email{nzhao@csrc.ac.cn}
\affiliation{ Beijing Computational Science Research Center }

\date{\today}

\maketitle

\tableofcontents

\renewcommand\thefigure{S\arabic{figure}} 
\renewcommand\theequation{S\arabic{equation}} 
\setcounter{figure}{0}
\setcounter{equation}{0}

\section{Experimental Setup and Measurement Details\label{SMsec:expSetup}}
\subsection{Experimental Setup}
(See Fig.~\ref{fig:expSetup} of main text.) The core component of our experimental setup is a cubic 
cell (labeled as CG2 in our laboratory) made of Pyrex glass, with $\rm ^{129}Xe$ (4 Torr), 
$\rm ^{131}Xe$ (14 Torr), $\rm N_2$ (450 Torr) gas and natural abundance Rb metal 
sealed inside. $\rm N_2$ serves as buffer gas to ensure the diffusive motion of Xe 
nuclear spins. The outer side length of the cell is 1.0 cm and inner side length is 
0.8 cm. The cell is placed in an oven made of boron nitride ceramics, which is 
electrically heated by high-frequency (300 kHz) sinusoidal AC current. 
Temperature of the cell is monitored by four resistance temperature detectors (RTDs) 
distributed inside the oven and stabilized by a proportional-integral-derivative (PID) 
controller. The temperature fluctuation measured by each RTD is in a range of 
$\pm 0.05~{\rm ^\circ C}$ and no long-term drift is observed in the timescale of 
48 hours. A small temperature non-uniformity of $\sim 1~{\rm ^\circ C}$ is observed 
across the oven.

The homogeneous magnetic field along $z$ direction and the transverse field 
along $x$ and $y$ directions are generated by Helmholtz coil and saddle coils, 
respectively. The linear gradient field along $z$ direction is generated by 
gradient Maxwell coil. The coils are driven by a group of current sources. 
The oven and the coils are firmly located in a magnetic shielding consisting 
of five permalloy layers.

The electron spin of Rb atoms is polarized via optical pumping with a $\sigma^+$ 
pump laser beam tuned to Rb $\rm D_1$ transition along $z$ direction~\cite{Happer1972}. 
Then, the Xe nuclear spins are polarized along the $z$ direction through the 
spin-exchange collisions with the Rb atoms~\cite{Walker1997}. The intensity 
distribution of the pump beam is approximately Gaussian with a diameter 
$d_{\rm 4\sigma}=4.96\pm0.44~{\rm mm}$, measured by a beam profiler. A linearly 
polarized probe laser beam along $x$ direction, which is blue detuned by 14GHz 
from the Rb $\rm D_2$ transition, probes the $x$ component of Rb electron spin via the 
Faraday rotation effect~\cite{[][{, Section 2.4.}]Seltzer2008}. The power of pump and probe beam is 
controlled and stabilized by acousto-optic modulators (AOMs). The rotation of the 
probe beam polarization plane is detected by a balanced 
photodetector (BPD), whose output signal is demodulated by a lock-in amplifier and 
then recorded to a PC.

\subsection{Detection of Nuclear Spin Polarization}
The polarized Xe nuclear spins create an effective magnetic field~\cite{Grover1978} 
$\mathbf{B}_K= \chi \mathbf{K}(\mathbf{r})$ in the order of 100 nT, where $\mathbf{K}(\mathbf{r})$ 
is the nuclear spin magnetic moment of Xe atoms and $\chi$ is a proportional 
constant. The effective magnetic field $\mathbf{B}_K$ is detected by the Rb atoms, 
which serves as an in-situ atomic magnetometer. A parametric modulation of the 
Rb spins~\cite{Cohen-Tannoudji1970, [][{, Section 3.1.}]Eklund2008, Tang2019} is applied to establish 
a parametric magnetometer (PM) which can sense the transverse components of the 
nuclear spin polarization $\mathbf{K}$. The demodulation phase is tuned so that 
the PM output is only sensitive to the $K_y$ component~\cite{Tang2019, Zhang2020, Song2021}. 
The typical sensitivity of our PM is $7~{\rm pT/\sqrt{\rm Hz}}$ and 
the bandwidth is above 2 kHz.

\begin{figure*}[hbpt]
  \includegraphics[]{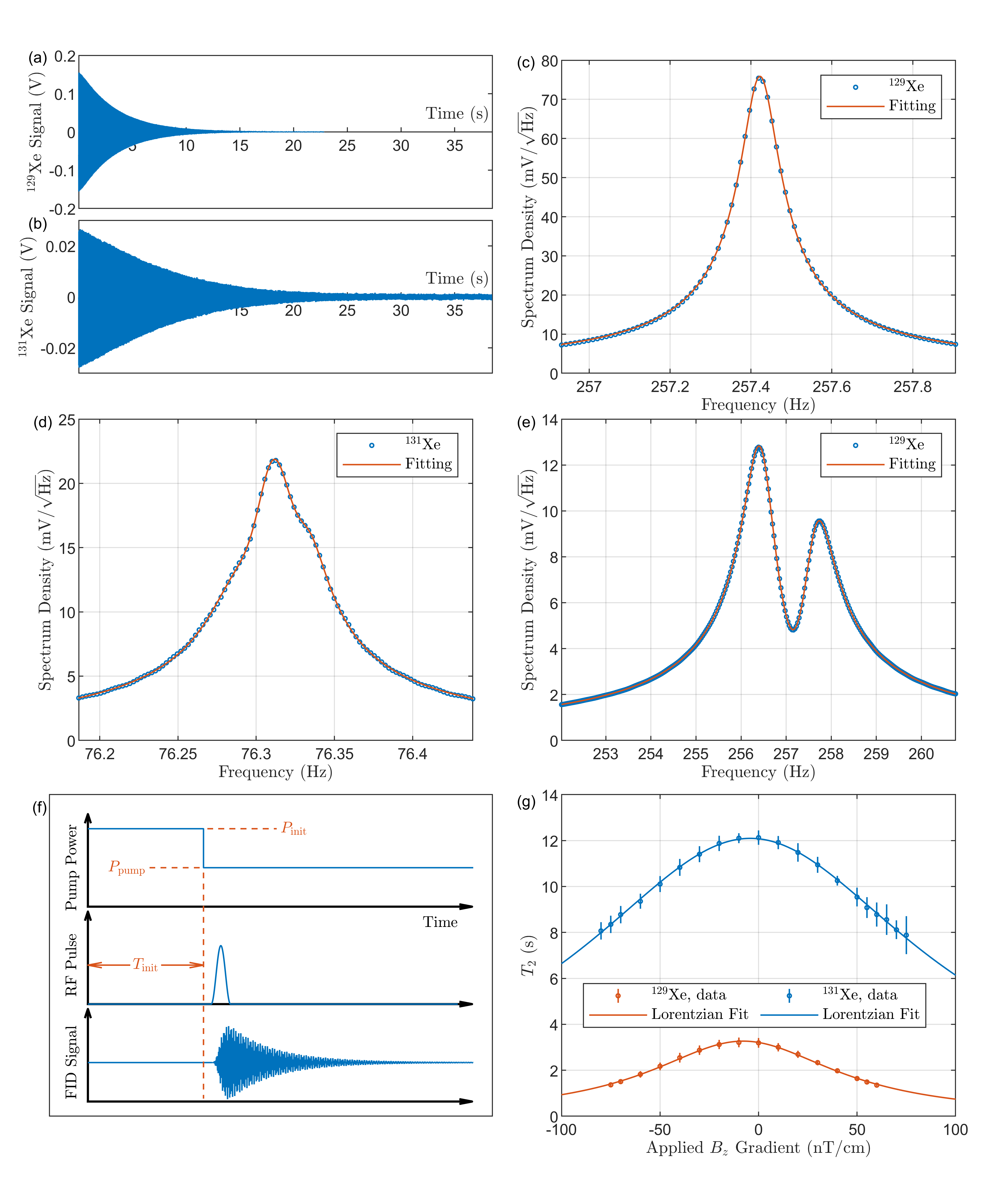}
  \caption{\label{fig:sigDemo_and_sweepG} 
  {FID signal and the measurement of diffusion constant.}
  (a) and (b), An example of the measured time domain FID signal of 
  $\rm ^{129}Xe$ and $\rm ^{131}Xe$ at zero $B_z$ gradient.
  (c) and (d), The FFT spectrum of (a) and (b), together 
  with Lorentzian fitting curve. SLP for (c) and TLP for (d). 
  In (d), the spectrum consists of three Lorentzian peaks where the 
  splitting between two adjacent peaks is approximately 22~mHz and the $T_2$ 
  of all three peaks are around 10 seconds.
  (e), The FFT spectrum of $\rm ^{129}Xe$ at $|G|=250~{\rm nT/cm}$, 
  together with fitting curve using DLP model.
  (f), The time sequence of an FID measurement.
  (g), Experimentally measured $T_2(G)$ curve together with fitting 
  curve using Eq.~\eqref{Eq:T2_grad_dep}. $P_{\rm init}=P_{\rm pump}=250~{\rm mW}$. 
  Data points of $\rm ^{129}Xe$ are extracted from the FFT spectrum by 
  SLP fitting, while points of $\rm ^{131}Xe$ are using TLP fitting. 
  The error bars in (g) represent the $95\%$ confident intervals 
  and data points are the mean value of five repeated measurements.
  }
\end{figure*}

\subsection{FID Measurement}
An FID measurement has three stages (see Fig.~\ref{fig:sigDemo_and_sweepG}(f)). 
In the initialization stage, the power of pump beam is set to $P_{\rm init}$ 
for a period $T_{\rm init}$ (typically 60 s) to polarize the Xe nuclear spins 
along $z$ direction via the Rb-Xe spins-exchange pumping. At the same time, 
the PM is initialized. In the second stage, the pump power is reduced to a 
lower value $P_{\rm pump}$, after which a radiofrequency (RF) pulse is applied 
via the $B_x$ coil to resonantly tilt the $\rm ^{129}Xe$ and $\rm ^{131}Xe$ 
polarization away from the $z$ axis simultaneously. In the recording stage, 
the time domain FID signal of Xe spins is acquired by the output of the PM for 
a period at least 5 times longer than the $T_2$ time of the Xe spins. 
(See Figs.~\ref{fig:sigDemo_and_sweepG}(a) and~\ref{fig:sigDemo_and_sweepG}(b) 
for examples of time domain FID signal.) The pump power for Fig.~\ref{fig:lambdaSpliting} 
and Fig.~\ref{fig:eigenModeDist} in main text is $P_{\rm pump}=P_{\rm init}=250~{\rm mW}$. 
The $P_{\rm init}$ values for Fig.~\ref{fig:stableMag} are always 60 mW, 
and the $P_{\rm pump}$ values are labeled in the figure and caption.

\section{Theoretical Model\label{SMsec:theoreticalModel}}
\subsection{Torrey Equation of Xe Spins}
The dynamics of Xe spins is governed by the general Torrey equation~\cite{Torrey1956} 
\begin{equation}
  \frac{\partial \mathbf{K}}{\partial t} = D \nabla^2 \mathbf{K} - \gamma \mathbf{B} \times \mathbf{K}
   - \mathbf{\Gamma}_{\rm c} \cdot \mathbf{K} + \mathbf{R}_{\rm SE},
  \label{Eq:torreyEqVector}
\end{equation}
where $D$ is the diffusion constant of Xe atoms in the cell, 
$\mathbf{K}=\mathbf{K}(\mathbf{r},t)$ is the spin polarization vector of Xe atoms, 
$\mathbf{B}=\mathbf{B}(\mathbf{r},t)$ is the magnetic field felt by the Xe spins, 
$\mathbf{\Gamma}_{\rm c}=\Gamma_{\rm 1c} \hat{\mathbf{z}}\hat{\mathbf{z}} + \Gamma_{\rm 2c} ( \hat{\mathbf{x}}\hat{\mathbf{x}} + \hat{\mathbf{y}}\hat{\mathbf{y}} )$ 
is spin relaxation rate induced by atomic collisions with the longitudinal and 
transverse relaxation rates $\Gamma_{\rm 1c}$ and $\Gamma_{\rm 2c}$ 
($\hat{\mathbf{x}}$, $\hat{\mathbf{y}}$ and $\hat{\mathbf{z}}$ are the unit vectors 
of $x$, $y$ and $z$ directions respectively), and $\mathbf{R}_{\rm SE}$ is the 
spin-exchange pumping rate due to collisions with Rb atoms.

Two conditions in our experiment allows us to simplify the general Torrey 
Eq.~\eqref{Eq:torreyEqVector}. Firstly, a strong bias magnetic field $|B_0| \sim 22~{\rm \mu T}$ 
is always applied along the $z$ direction. Compared to the bias field $B_0$, 
the transverse components are so weak ($|B_{x,y}|<10~{\rm nT}$) that neglecting 
the weak static transverse field in Eq.~\eqref{Eq:torreyEqVector} is a good 
approximation. Secondly, the vector of spin-exchange pumping rate 
$\mathbf{R}_{\rm SE}=\Gamma_{\rm SE}\mathbf{S}$, 
which parallels to the Rb spin polarization $\mathbf{S}$, is along the $z$ direction. 
The spin-exchange process only changes the longitudinal component $K_z$ of the Xe 
spin polarization. With these conditions, the longitudinal component $K_z$ and the 
components $K_{x,y}$ of the Xe spin polarization are decoupled. The equation of 
motion of the complex transverse component $K_+ \equiv K_x + i K_y$ can be written as
\begin{equation}
  \frac{\partial K_+ (\mathbf{r},t)}{\partial t} = D \nabla^2 K_+ (\mathbf{r},t) 
  - (i \gamma B_z + \Gamma_{\rm 2c} ) K_+ (\mathbf{r},t),
  \label{Eq:torreyEq_transverse}
\end{equation}
which is the Eq.~\eqref{Eq:BlochTorrey} of the main text.

\subsection{Boundary Condition}
Spin relaxation occurs when Xe atoms colliding on the cell wall (the boundary of 
the cubic domain $\Omega$, denoted by $\partial \Omega$). The wall-relaxation is 
usually described by the boundary condition of the spin polarization 
$\mathbf{K}(\mathbf{r},t)$. We assume the transverse component spin polarization 
satisfies the third-type Robin boundary condition
\begin{equation}
  \left.
  \left(  \frac{\lambda}{L} K_+ (\mathbf{r},t) + 
  \hat{\mathbf{n}} \cdot \nabla K_+ (\mathbf{r},t) \right) 
  \right|_{\mathbf{r} \in \partial \Omega}
  =0,
  \label{Eq:boundaryCond}
\end{equation}
where $\Omega = \left\{  (x,y,z) \middle| -L/2 \le x,y,z \le +L/2 \right\}$ 
is the cubic cell domain, $\hat{\mathbf{n}}$ is the unit normal vector of the cell 
walls (pointing from the cell center to the wall), $L$ is the inner side length of 
glass cell and $\lambda$ is a real, dimensionless factor characterizing the 
wall-relaxation strength. In the limit of $\lambda \to +\infty$, the spin polarization 
$K_+ \to 0$ on the cell wall, which means the nuclear spins are fully depolarized 
near the wall. The opposite limit $\lambda \to 0$ implies an ideal surface with no 
wall-relaxation.

The experimental value of parameter $\lambda$ for cell CG2 is in the order of $10^{-2}$ 
and the wall-relaxation rate $\Gamma_{\rm wall}$ is in the order of 
$10^{-2}~{\rm s^{-1}}$ for both $\rm ^{129}Xe$ and $\rm ^{131}Xe$. 
In most cases, $\Gamma_{\rm wall}$ is much smaller than the relaxation rate 
($\sim 10^0~{\rm s^{-1}}$) induced by the inhomogeneous magnetic field, particularly 
in the $PT$-broken phase with large gradient strength (see Fig.~\ref{fig:lambdaSpliting} 
of the main text). So, the non-relaxation boundary condition 
$\hat{\mathbf{n}} \cdot \nabla K_+ (\mathbf{r}) = 0$ is a good approximation in 
dealing with $PT$ transition.

\subsection{Perturbative Solution of Torrey Equation with Robin Boundary and Non-uniform Magnetic Field}
Consider a simplified case with uniform magnetic field $B_z(\mathbf{r})=B_0$ and 
constant decay rate $\Gamma_{\rm 2c}$. The eigenvalue problem corresponding 
to Eq.~\eqref{Eq:torreyEq_transverse} is
\begin{equation}
  D \nabla^2 \phi_{mnp} - (i \gamma B_0 + \Gamma_{\rm 2c}) \phi_{mnp}
  = - s_{mnp}^{(0)} \phi_{mnp},
  \label{Eq:eigenEq_const}
\end{equation}
where $-s_{mnp}^{(0)}$ is the eigenvalue of the eigenmode $\phi_{mnp}(\mathbf{r})$. 
To count the effect of wall relaxation, we consider the third-type Robin boundary 
condition Eq.~\eqref{Eq:boundaryCond}. In the derivation below, we will always 
assume $\lambda \ll 1$.

The eigenmodes $\phi_{mnp}(\mathbf{r})$ of Eq.~\eqref{Eq:eigenEq_const} 
can be written in a factorized form of
\begin{equation}
  \phi_{mnp}(\mathbf{r}) = \phi_m(x)\phi_n(y)\phi_p(z),
  \label{Eq:eigenMode_sepForm}
\end{equation}
where $m$, $n$ and $p$ are non-negative integers labelling the eigenmodes 
in the $x$, $y$ and $z$ directions, respectively. Since $x$, $y$, $z$ directions 
are equivalent, we use $z$ direction as an example below. We can first assume 
$\phi_p(z)$ to be the form
\begin{equation}
  \phi_p(z) = \frac{1}{\mathcal{N}_p} \sin{\left(\kappa_p z + \delta_p\right)},
  \label{Eq:phiz_ansatz}
\end{equation}
where $\mathcal{N}_p$ is normalization factor, and $\kappa_p$ and $\delta_p$ are 
real numbers determined by the boundary condition Eq.~\eqref{Eq:boundaryCond}. 
It's easy to check that solutions \eqref{Eq:eigenMode_sepForm} and \eqref{Eq:phiz_ansatz} 
satisfy the eigen-equation \eqref{Eq:eigenEq_const} with the eigenvalues shown in 
Eq.~\eqref{Eq:eigenvalue_0th}. 
By substituting Eqs.~\eqref{Eq:eigenMode_sepForm} and \eqref{Eq:phiz_ansatz} into 
the boundary condition Eq.~\eqref{Eq:boundaryCond}, we found that the 
wave numbers $\kappa_p$ are the solutions of the transcendental equation
\begin{equation}
  \tan{\left(\kappa_p L\right)} = \frac{2 \lambda \kappa_p L}{\kappa_p^2 L^2 - \lambda^2},
  \label{Eq:kappa_eq}
\end{equation}
and the phase shifts are determined by
\begin{equation}
  \tan{\left(\delta_p - \frac{\kappa_p L}{2}\right)} = \frac{\kappa_p L}{\lambda}.
  \label{Eq:delta_eq}
\end{equation}
With the normalization condition $\int{\phi_p^2(z) {\rm d}z} = 1$, 
the normalization factor is
\begin{equation}
  \mathcal{N}_p = \sqrt{ \frac{L}{2} + \frac{\lambda L}{\kappa_p^2 L^2 + \lambda^2} }.
  \label{Eq:normalizationFactor}
\end{equation}
The eigenvalues corresponding to $\phi_{mnp}(\mathbf{r})$ are 
\begin{equation}
  s^{(0)}_{mnp} = D \kappa_{mnp}^2 + \Gamma_{\rm 2c} + i \gamma B_0,
  \label{Eq:eigenvalue_0th}
\end{equation}
where $\kappa_{mnp}^2 \equiv \kappa_m^2+\kappa_n^2+\kappa_p^2$. The real part of the 
eigenvalue, $\Gamma_2=D \kappa_{mnp}^2+\Gamma_{\rm 2c}$, is the spin decay rate of the 
eigenmode $\phi_{mnp}(\mathbf{r})$, and the imaginary part, $\gamma B_0$, is the spin 
precession frequency in the uniform magnetic field $B_0$. The superscript of 
$s_{mnp}^{(0)}$ represents that this is the $\rm 0^{th}$ order correction of the perturbation 
solution in the following.

To reveal the underlying physics of the wall-relaxation, we expand the tangent 
function in Eq.~\eqref{Eq:kappa_eq} in the neighborhood of $\kappa_p L=p \pi$, 
and find the solution of wave number in the limit of $\lambda \ll 1$ to be
\begin{align}
  \kappa_p \approx
  \left\{ 
    \begin{aligned}
      \frac{ \sqrt{2\lambda + \lambda^2} }{L} &\approx \frac{\sqrt{2\lambda}}{L}&,& p=0 \\
      \frac{p \pi}{L} + \frac{2 \lambda}{p \pi L} &\approx \frac{p \pi}{L}&,& p=1,2,\dots
    \end{aligned} 
  \right. .
  \label{Eq:kappa_solution}
\end{align}
For the fundamental mode ($m=n=p=0$), the wall-relaxation rate is
\begin{equation}
  \Gamma_{000} \equiv D \kappa_{000}^2 \approx 6 \lambda \frac{D}{L^2}.
\end{equation}
For the excited modes ($[m,n,p] \neq [0,0,0]$), the relaxation rate due to spin 
diffusion is
\begin{equation}
  \Gamma_{mnp} \equiv D \kappa_{mnp}^2 \approx (m^2+n^2+p^2) \pi^2 \frac{D}{L^2}.
\end{equation}
As the parameter $\lambda$ is in the order of $10^{-2}$, even for the lowest excited 
modes (with $m^2+n^2+p^2=1$), the decay is much faster than the fundamental mode, 
i.e., $\Gamma_{mnp} \gg \Gamma_{000}$. 

The eigenvalues above is valid only for a uniform magnetic field $B_z({\bf r}) = B_0$. 
To see the effect of a small non-uniform magnetic field, i.e. $B_z({\bf r}) = B_0 + B_1({\bf r})$, 
we rewrite the eigen equation~\eqref{Eq:eigenEq_const} as
\begin{equation}
  \left[ \hat{H}_0 + \hat{H}_1 (\mathbf{r})  \right] \Psi_{mnp} (\mathbf{r})
  = - s_{mnp} \Psi_{mnp} (\mathbf{r}),
  \label{Eq:eigenEq}
\end{equation}
where $\hat{H}_0 \equiv D \nabla^2 - (i \gamma B_0 +\Gamma_{\rm 2c} )$ and 
$\hat{H}_1 \equiv -i \gamma B_1 (\mathbf{r})$. 
$B_0$ is a homogeneous magnetic field while $B_1 (\mathbf{r})$ is spatially non-uniform. 

Then, we calculate the matrix form of $\hat{H}_0$ and $\hat{H}_1$ under the basis 
$\{ \phi_{mnp} (\mathbf{r}) \}$ and apply the results of standard perturbation theory, 
only to remember that $\hat{H}_0$ and $\hat{H}_1$ are non-Hermitian. 

$H_0$ is diagonal, with the eigenvalues $-s_{mnp}^{(0)}$ in Eq.~\eqref{Eq:eigenvalue_0th}. 
The matrix elements of $H_1$ are $(H_1)_{\alpha \beta}=-i \gamma b_{\alpha \beta}$ 
with $b_{\alpha \beta}$ being the $B_1 (\mathbf{r})$ induced coupling between eigenmodes:
\begin{equation}
  b_{\alpha \beta} \equiv \int_\Omega{\phi_\alpha (\mathbf{r}) B_1 (\mathbf{r}) 
                                      \phi_\beta (\mathbf{r}) {\rm d^3}\mathbf{r}}.
\end{equation}
Above, we have simplified the indices of the eigenmode by a single Greek letter, 
e.g., $\alpha = [ mnp ]$, and $\alpha=0$ ($\alpha>0$) stands for the fundamental 
(excited) mode.

Particularly, we are interested in the perturbative correction of the eigenvalue 
$s_0$ of the fundamental mode, which can be directly observed via the spin precession 
frequency shift and the spin decay rate. Up to the $\rm 2^{nd}$ order correction, 
the eigenvalue of the fundamental mode is
\begin{equation}
\begin{aligned}
  s_0 &= \underbrace{D \kappa_0^2 + \Gamma_{\rm 2c} + i \gamma B_0}_{\rm 0^{th}\ order}
    + \underbrace{i \gamma b_{00}}_{\rm 1^{st}\ order}
    + \underbrace{\gamma^2 \sum_{\alpha=1}^{\infty}{\frac{b_{0\alpha}^2}{D(\kappa_\alpha^2 - \kappa_0^2)}} }_{\rm 2^{nd}\ order}
    \\ &\equiv \sum_{k=0}^{2}{s_0^{(k)}},
  \label{Eq:s0_pertb}
\end{aligned}
\end{equation}
where $s_0^{(k)}$ is the $k^{\rm th}$ order correction. In Eq.~\eqref{Eq:s0_pertb}, 
the indices of the wave number $\kappa_{mnp}$ is also replaced by a single Greek 
letter, and $b_{00}$ is the mean value of the inhomogeneous field $B_1 (\mathbf{r})$ 
averaged over the fundamental mode $\phi_0 (\mathbf{r})$. As the perturbation matrix 
element $b_{\alpha \beta}$ is real, the $\rm 1^{st}$ order correction are purely 
imaginary, which contribute to the frequency shift. The $\rm 2^{nd}$ order correction 
is real, which modifies the spin decay rate.

Note that the unperturbed eigenmodes
$\phi_\alpha (\mathbf{r})=\phi_\alpha (\mathbf{r};\lambda)$ 
depend on boundary parameter $\lambda$. Since $\rm ^{129}Xe$ and $\rm ^{131}Xe$ 
usually have different $\lambda$ values, a non-uniform magnetic field can contribute 
different $\rm 1^{st}$ order correction $b_{00}$ to the two isotopes, causing bias 
of rotation signal.

The solution of the transverse Torrey Eq.~\eqref{Eq:torreyEq_transverse} can be 
written in the form
\begin{equation}
  K_+({\bf r},t) = \sum_\alpha{a_\alpha {\rm e}^{-s_\alpha t} \Psi_\alpha({\bf r})},
\end{equation}
where $a_\alpha$ are expansion coefficients, $s_\alpha$ and $\Psi_\alpha({\bf r})$ are the eigen 
solutions of Eq.~\eqref{Eq:eigenEq}. This is a superposition of many exponentially decaying 
eigenmodes. When the non-uniform magnetic field $B_1$ is small, due to the fast decay rate 
of excited modes, only the $\alpha=0$ term is experimentally observable, and we have
\begin{equation}
  K_+({\bf r},t) \approx a_0 {\rm e}^{-s_0 t} \Psi_0({\bf r}).
  \label{Eq:timeSolution_mode0}
\end{equation}

\subsection{Measurement of the Diffusion Constant\label{SMsec:measureD}}
To determine the diffusion constants of Xe nuclear spins experimentally, 
we measure the relaxation time $T_2$ via the FID signal of Xe nuclear spins. 
The magnetic field is generated by coils with the form 
$B_z({\bf r}) = B_0 + G \cdot z$. 
For small $G$, the FID signal can be described by the real part of 
Eq.~\eqref{Eq:timeSolution_mode0}. $T_2$ is the reciprocal of $\Re{[s_0]}$. 
From the $\rm 2^{nd}$ order terms in Eq.~\eqref{Eq:s0_pertb}, we can find that 
$\Re{[s_0]}$ has a quadratic dependence on gradient $G$: 

Notice that for linear gradient field, only the excited modes with 
$\alpha_p=[0,0,2p-1]$ for $p=1,2,3,\dots$ have non-zero matrix element
\begin{equation}
  b_{0 \alpha_p}^{\rm (G)} = G 
  \int_\Omega{z \phi_0 (\mathbf{r}) \phi_{\alpha_p} (\mathbf{r}) {\rm d^3}\mathbf{r}}
  = \frac{2\sqrt{2} G L}{(2p-1)^2 \pi^2}.
\end{equation}
So, the gradient induced spin decay rate is
\begin{equation}
  \begin{aligned}
    \Gamma_{\rm G} &= 
    \sum_{\alpha_p}{\frac{\gamma^2 \left( b_{0 \alpha_p}^{\rm (G)} \right)^2}
                        {D \left( \kappa_{\alpha_p}^2 - \kappa_0^2 \right)}} 
    \approx 
    \sum_{\alpha_p}{\frac{\gamma^2 \left( b_{0 \alpha_p}^{\rm (G)} \right)^2}
                        {D \kappa_{\alpha_p}^2 }} 
    \\ &= \frac{8 \gamma^2 G^2 L^4}{\pi^6 D} \sum_{p=1}^{\infty}{\frac{1}{(2p-1)^6}}
    \\ &= \frac{\gamma^2 G^2 L^4}{120D}.
  \end{aligned}
\end{equation}
The gradient dependence of $T_2$ is
\begin{equation}
  T_2^{-1} = \Re{[s_0]} \approx \Gamma_{\rm 2,min} + \frac{\gamma^2 G^2 L^4}{120D},
  \label{Eq:T2_grad_dep}
\end{equation}
where $\Gamma_{\rm 2,min} \equiv D \kappa_0^2+\Gamma_{\rm 2c}$ is the relaxation 
rate when magnetic field is homogeneous. Similar relation is also reported in 
ref.~\cite{Zheng2011} previously. 

Fig.~\ref{fig:sigDemo_and_sweepG}(g) shows 
the measured $T_2 (G)$ curves of cell CG2 at $\rm 110~^\circ C$. The diffusion 
constants estimated by the least squares fitting with Eq.~\eqref{Eq:T2_grad_dep} 
are $D_{129}=(0.21 \pm 0.03)~{\rm cm^2/s}$ for $\rm ^{129}Xe$ and $D_{131}=(0.23 \pm 0.02)~{\rm cm^2/s}$ 
for $\rm ^{131}Xe$. These values are consistent with the theoretical estimation 
$D=0.24~{\rm cm^2/s}$, which usually agrees with experiment data in $10\%\sim20\%$ 
accuracy~\cite{[][{, Chapter 11.}]Poling2001}.

\subsection{Explanation of the $PT$ Transition\label{SMsec:PT_explanation}}
\subsubsection{Exact Solution with Linear Gradient Field\label{SMsec:exactSolution}}
Reference~\cite{Stoller1991} has shown the exact solution of the Torrey 
equation under linear gradient fields. 
The solution can be briefly summarized as follows:

Consider a simplified version of the eigen equation~\eqref{Eq:eigenEq} under linear gradient field 
$B_z({\bf r}) = B_0 + G \cdot z$
\begin{equation}
  \left( D \frac{\rm d^2}{{\rm d}z^2} - i \gamma G \cdot z \right) M_k(z)
  = - s'_{k} M_k (z), \quad
  k=0,1,2,\dots
  \label{Eq:eigenEq2}
\end{equation}
Here, we ignore the $x,y$ directions because $B_z$ is uniform along these directions, 
and the solutions for these directions are trivial. We also drop the $i \gamma B_0$ and 
$\Gamma_{\rm 2c}$ terms because they only contribute a constant shift to all eigenvalues 
and do not affect the distribution of eigenmodes. $s'_k$ is the eigenvalue corresponding 
to the eigenmode $M_k(z)$.

Also, apply the non-depolarization boundary conditions on both ends of the cell
\begin{equation}
  \left. \frac{{\rm d}M_k}{{\rm d}z} \right|_{z=\pm L/2} = 0.
  \label{Eq:boundaryCond2}
\end{equation}

We transform Eq.~\eqref{Eq:eigenEq2} to a dimensionless form by defining new spatial coordinate 
and normalized gradient
\begin{equation}
  \zeta \equiv \frac{2z}{L} \in [-1,+1], \quad g' \equiv \frac{\gamma G L^3}{8 D}.
\end{equation}
The eigen equation and boundary condition becomes
\begin{equation}
  \left( \frac{\rm d^2}{{\rm d}\zeta^2} - i g' \zeta \right) M_k(\zeta)
  = - q_k M_k (\zeta), \quad 
  \left. \frac{{\rm d}M_k}{{\rm d}\zeta} \right|_{\zeta=\pm 1} = 0, 
  \label{Eq:eigenEq_dimensionless}
\end{equation}
where $q_k \equiv s'_{k} L^2/(4D)$ is dimensionless eigenvalues.

It's easy to check that, by introducing an appropriate spatial coordinate $\xi$, 
Eq.~\eqref{Eq:eigenEq_dimensionless} can transform to the form 
$M_k''(\xi) - \xi M_k(\xi) = 0$, 
which is the Airy equation. So, the general solution of Eq.~\eqref{Eq:eigenEq_dimensionless} 
is the superposition of the two Airy functions ${\rm Ai}(x)$ and ${\rm Bi}(x)$, 
\begin{equation}
  M_k(\zeta) = c_1^{(k)} {\rm Ai} \left(  \frac{-q_k + i g' \zeta}{ (i g')^{2/3}}  \right)
             + c_2^{(k)} {\rm Bi} \left(  \frac{-q_k + i g' \zeta}{ (i g')^{2/3}}  \right).
  \label{Eq:generalSol_exact}
\end{equation}
The coefficients $\{ c_i^{(k)}\}$ and eigenvalues $\{ q_k \}$ can be obtained by solving the 
following boundary conditions
\begin{equation}
  \begin{aligned}
    c_1^{(k)} {\rm Ai} \left(  \frac{-q_k + i g' }{ (i g')^{2/3}}  \right)
  + c_2^{(k)} {\rm Bi} \left(  \frac{-q_k + i g' }{ (i g')^{2/3}}  \right) &= 0, \\
    c_1^{(k)} {\rm Ai} \left(  \frac{-q_k - i g' }{ (i g')^{2/3}}  \right)
  + c_2^{(k)} {\rm Bi} \left(  \frac{-q_k - i g' }{ (i g')^{2/3}}  \right) &= 0.
  \end{aligned}
  \label{Eq:generalSol_exact_bCond}
\end{equation}
This is homogeneous linear equations of $c_1^{(k)}$ and $c_2^{(k)}$. To have non-zero solution, 
the determinant of coefficient matrix should be zero, which gives the eigenvalues $\{ q_k \}$. 
For the convenience of reference, we sort the eigenvalues as: 
for $q_k$ and $q_l$, if $\Re{[q_k]} < \Re{[q_l]}$, then $k<l$; 
if $\Re{[q_k]} = \Re{[q_l]}$ and $\Im{[q_k]} < \Im{[q_l]}$, then $k<l$. 
Under this convention, $q_0$ is the eigenvalue with the smallest real part and negative imaginary part.

\begin{figure*}[hbpt]
  \includegraphics[]{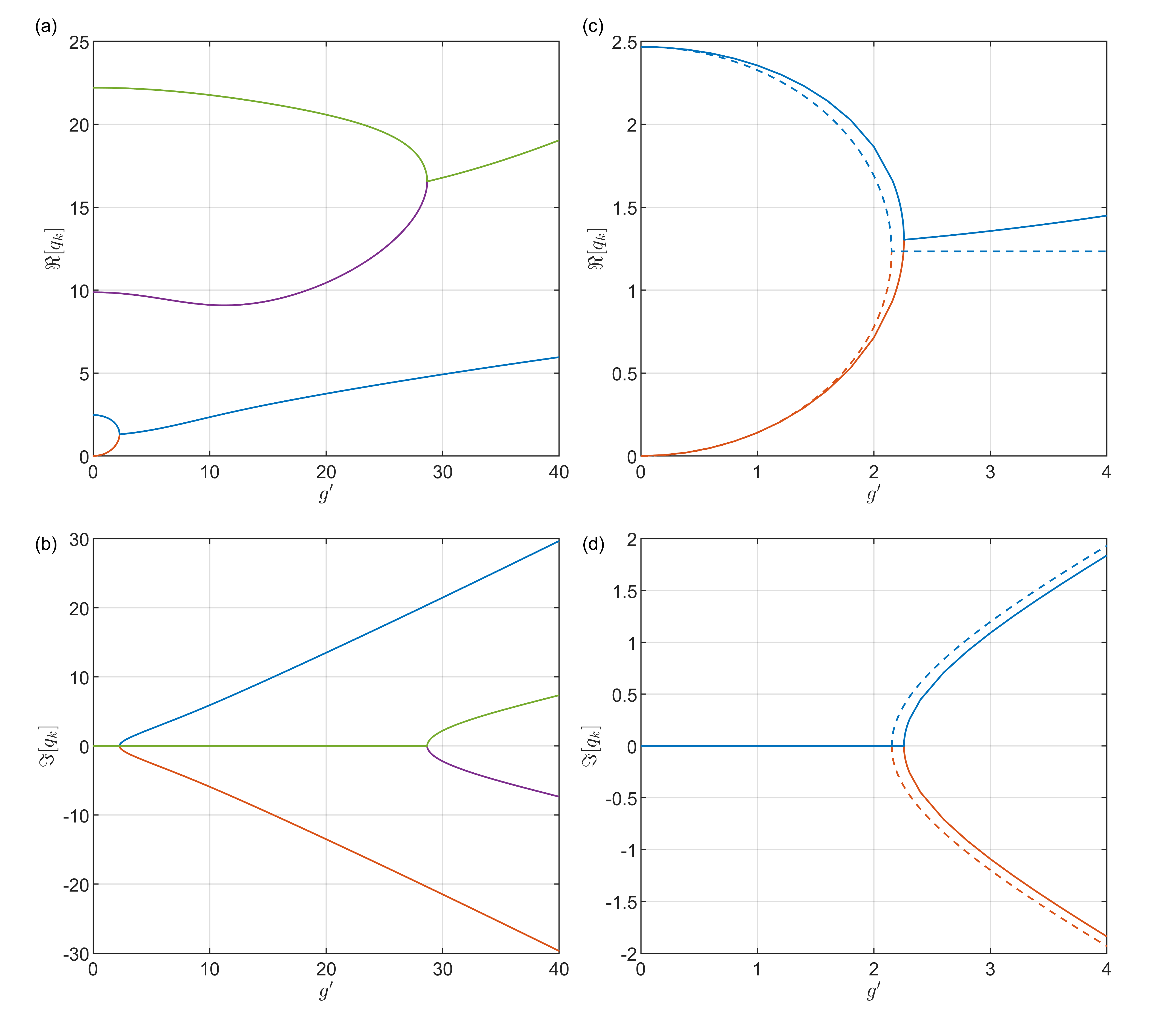}
  \caption{\label{fig:eigenvalue_spectrum} 
  {The eigenvalue spectrum of Torrey equation.}
  (a), the real part of the eigenvalues $q_k$ in Eq.~\eqref{Eq:eigenEq_dimensionless}. 
  Red, blue, purple and green line correspond to $q_0 \sim q_3$, respectively. 
  (b), the imaginary part of $q_k$, similar to (a). 
  (c) and (d), the real and imaginary part of $q_0$ and $q_1$ (solid lines), together with 
  the eigenvalues of Eq.~\eqref{Eq:torreyEq_transverse_2modeApprox} (dashed lines). 
  The value of dashed line is $(\epsilon + \xi_\pm) L^2 / (4D)$, where $\lambda = 0$ is assumed. 
  }
\end{figure*}

The eigenvalues $\{q_0, q_1, q_2, q_3\}$, as a function of $g'$, is shown in Fig.~\ref{fig:eigenvalue_spectrum}(a)(b). 
We can see that, there are some points where two eigenvalues merge together -- so-called "branch point" 
or "exceptional point" (EP). The position of these EPs is given by
\begin{equation}
  g'_p = \frac{27\sqrt{3}}{32} j_p^2, \quad 
  q_{2p-2}(g'_p) = q_{2p-1}(g'_p) = \frac{g'_p}{\sqrt{3}}, \quad p=1,2,\cdots,
  \label{Eq:EP_position}
\end{equation}
where $j_p$ is the $p$-th zero point of the Bessel function $J_{-2/3}(x)$. 

\subsubsection{$PT$ Symmetry Breaking}
The first EP is at $g'=g'_1 \approx 2.258$, where $q_0 = q_1 \approx 1.304 $. 
In the region $g' < g'_1$, we have $ \Re{[q_0]} < \Re{[q_1]}$ and $\Im{[q_0]} = \Im{[q_1]} = 0$, 
meaning that the eigenmodes $M_0(\zeta)$ and $M_1(\zeta)$ have different relaxation rates but 
the same Larmor frequency. In the region $g' > g'_1$, the relaxation rate of these two eigenmodes 
becomes the same, and their Larmor frequency splits. We can define 
the frequency difference of these two eigenmodes as $\Delta \equiv \Im{[q_1]} - \Im{[q_0]}$. 
By numerical analysis, in the right neighborhood of $g'_1$, 
\begin{equation}
  \Delta \approx 2.34 \left(g' - g'_1 \right)^{0.5}, 
  \label{Eq:criticalExpo_num}
\end{equation}
and ${\rm d}\Delta / {\rm d}g'$ diverges as $g' \rightarrow g_1^{'+}$. This is a sign of phase transition. 

\begin{figure*}[hbpt]
  \includegraphics[]{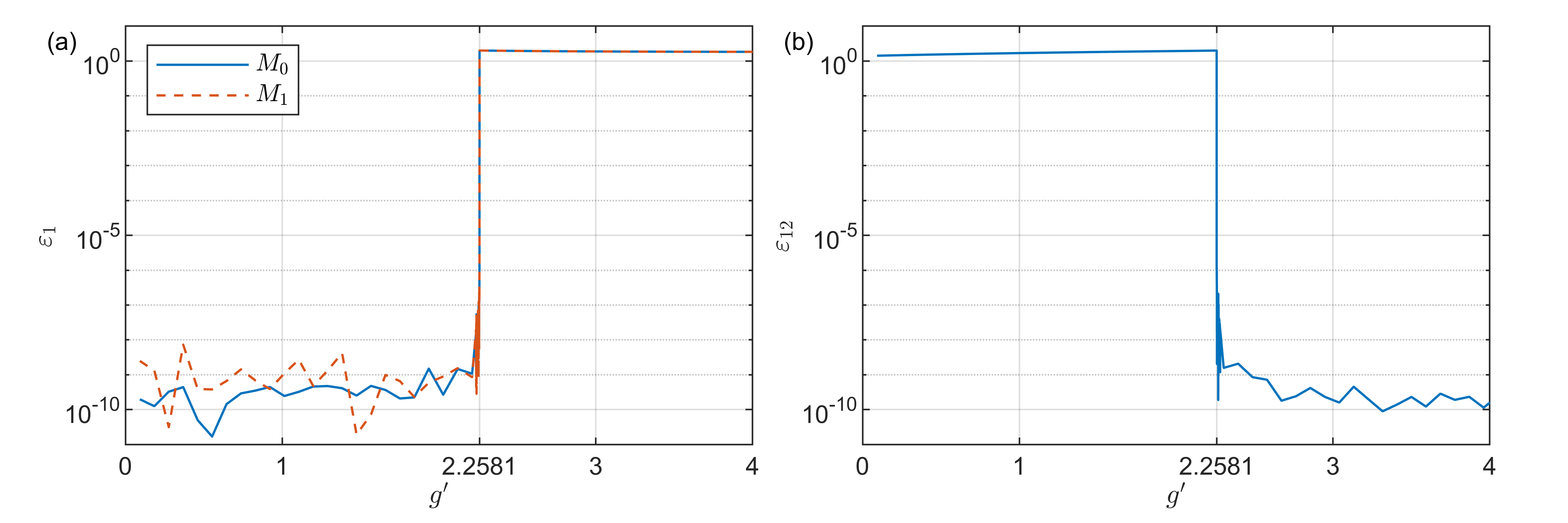}
  \caption{\label{fig:symChk_eigenmode} 
  {The $PT$ symmetry of eigenmodes $M_0$ and $M_1$.}
  (a), Numerical verification of the $PT$ symmetry of eigenmodes $M_0$ and $M_1$. 
  $ \varepsilon_1(M_k) \equiv \left(\int{\left| M_k(z) - M_k^*(-z) \right| {\rm d}z} \right) / \int{\left| M_k(z) \right| {\rm d}z}$. 
  In the region $g'<g'_1$, $\varepsilon_1(M_0) = \varepsilon_1(M_1) = 0$. 
  (b), $PT$ operation transforms $M_0$ into $M_1$ when $g'>g'_1$. 
  $ \varepsilon_{12} \equiv \left(\int{\left| M_0(z) - M_1^*(-z) \right| {\rm d}z} \right) / \int{\left| M_0(z) \right| {\rm d}z}$. 
  }
\end{figure*}

On the other hand, the eigen equation~\eqref{Eq:eigenEq2} is invariant under $PT$ operation, which implies that 
the eigenmodes should have $PT$ symmetry. To demonstrate this, we apply $PT$ operation on Eq.~\eqref{Eq:eigenEq2} 
($P$: replace $z$ with $-z$, $T$: complex conjugate) 
\begin{equation}
  \begin{aligned}
  {\rm LHS:}& \quad PT \left\{  \left( D \frac{\rm d^2}{{\rm d}z^2} - i \gamma G \cdot z \right) M_k(z)  \right\} &=& 
                          \left( D \frac{\rm d^2}{{\rm d}z^2} - i \gamma G \cdot z \right) M_k^*(-z),  \\
  {\rm RHS:}& \quad PT \left\{  -s'_k M_k(z) \right\} &=& -\left(s'_k \right)^* M_k^*(-z).
  \end{aligned}
  \label{Eq:PT_on_eigenEq2}
\end{equation}
This shows that, if $M_k(z)$ is an eigenmode with eigenvalue $s'_k$, then $M_k^*(-z)$ is also an eigenmode, with eigenvalue 
$\left(s'_k \right)^*$. 

If $s'_k$ is purely real and no degeneracy exist, then $M_k(z)$ and $M_k^*(-z)$ should be the same, which means 
the eigenmode $M_k(z)$ itself has $PT$ symmetry, i.e. $PT\{M_k(z)\} = M_k(z)$. This is what happens to 
$M_0(z)$ and $M_1(z)$ in the region $g'<g'_1$. Figure~\ref{fig:eigenModeDist}(a) in the main text shows an example of $M_1(z)$ 
in this region. Figure~\ref{fig:symChk_eigenmode}(a) shows a numerical verification of this $PT$ symmetry for $M_0(z)$ and $M_1(z)$. 

When $g'$ goes across $g'_1$ from left to right, the $PT$ symmetry of eigenmodes $M_0(z)$ and $M_1(z)$ breaks. In the region $g'>g'_1$, 
the eigenvalues $s'_0$ and $s'_1$ are complex. The $PT$ operation transforms $\{s'_0, M_0(z)\}$ into $\{s'_1, M_1^*(-z)\}$, 
and {\it vice versa}. The eigenmode itself do not have $PT$ symmetry anymore. An example of $M_0(z)$ and $M_1(z)$ in this region is 
shown in Fig.~\ref{fig:eigenModeDist}(b)(c) of the main text, with $M_+(z)$ being $M_1(z)$ and $M_-(z)$ being $M_0(z)$. 
Figure~\ref{fig:symChk_eigenmode}(b) shows a numerical verification of this property. 

\emph{
In this sense, we call the $g'<g'_1$ region $PT$-symmetric phase, and the $g'>g'_1$ region $PT$-broken phase.
}

Actually, at each exceptional point $g'_p$, the real part of two eigenvalues, $s'_{2p-2}$ and $s'_{2p-1}$, merges. And the $PT$ symmetry 
of eigenmodes $M_{2p-2}(z)$ and $M_{2p-1}(z)$ breaks. So, the $PT$ symmetry of this diffusive spin system breaks step by step as magnetic 
field gradient goes through each EP. In our experiment, the relaxation rate of eigenmodes $M_k(z), k>1$ is too large that it's impossible 
to see their signals in FID measurement. Therefore, we only focus on the first two eigenmodes. 

In the $PT$-symmetric phase, the $PT$ symmetry requires $M_0$ and $M_1$ to be symmetric against 
the center of the cell. 
In a linear gradient field, since these two eigenmodes are both symmetric, 
the average magnetic field they ``feel'' will always be the same. 
So, the resonance frequency of these two eigenmodes must be the same. 
The gradient only contributes a boradening to the resonance line. 
However, in the $PT$-broken phase, $M_0$ and $M_1$ do not have $PT$ symmetry anymore. 
As the gradient getting larger, they become more and more localized at the opposite ends of the cell. 
Then, the average magnetic field they ``feel'' become more and more different. 
That's why their resonance frequency splits -- 
\emph{it's the breaking of $PT$ symmetry that makes the splitting possible.}

\subsubsection{Two Mode Approximation}
The exact solution using Airy functions is complicated and abstract. In this section, we try to solve the eigenmodes $M_0(z)$ and $M_1(z)$ 
in a mathematically simpler way by using perturbation and two mode approximation, 
and explain the theoretical lines in Fig.~\ref{fig:eigenModeDist} of main text. 

Due to the fast relaxation rate of high excitation modes, 
we can approximately expand $K_+ (\mathbf{r},t)$ on the lowest two unperturbed 
eigenmodes as $K_+ (\mathbf{r},t) \approx c_0 (t) \phi_0 (\mathbf{r})+c_1 (t) \phi_1 (\mathbf{r})$, 
where $\phi_1 (\mathbf{r})$ stands for 
$\phi_{001} (\mathbf{r})=\phi_0 (x) \phi_0 (y) \phi_1 (z)$. Considering a linear 
gradient field $B_z (\mathbf{r})=B_0+G \cdot z$, Torrey equation~\eqref{Eq:torreyEq_transverse} 
becomes
\begin{equation}
  \frac{\rm d}{{\rm d}t} \begin{pmatrix} c_0 \\ c_1 \end{pmatrix} = - \left[  
    \frac{D}{2} \left( \kappa_0^2 + \kappa_1^2 \right) 
    + \underbrace{ 
      \epsilon
      \begin{pmatrix}
      - 1 & i g \\
      i g & + 1
      \end{pmatrix} 
    }_{H_2}
  \right]
  \begin{pmatrix}
    c_0 \\ c_1
  \end{pmatrix},
  \label{Eq:torreyEq_transverse_2modeApprox}
\end{equation}
where
\begin{equation}
  \epsilon \equiv \frac{D(\kappa_1^2-\kappa_0^2 )}{2}, \quad
  g \equiv \frac{\gamma b_{01}}{\epsilon} = \frac{4\sqrt{2} \gamma G L^3}{\pi^4 D} 
  = \frac{32\sqrt{2}}{\pi^4} g'.
\end{equation}
Both $\epsilon$ and $g$ are purely real. 
Here, we drop the $i \gamma B_0$ and $\Gamma_{\rm 2c}$ terms in 
Eq.~\eqref{Eq:torreyEq_transverse_2modeApprox} for the same reason stated below 
Eq.~\eqref{Eq:eigenEq2}. 
The solution of Eq.~\eqref{Eq:torreyEq_transverse_2modeApprox} are determined by 
the non-Hermitian coefficient matrix $H_2$, whose eigenvalues are 
$\xi_\pm=\pm \epsilon \sqrt{1-g^2}$. We plot the eigenvalues of 
$H_2$ together with the exact solution in Fig.~\ref{fig:eigenvalue_spectrum}(c)(d). 

The value of $\epsilon$ represents the 
``diffusion energy'' while the value of $\gamma b_{01}$ represents the coupling strength 
between eigenmodes $\phi_0 (\mathbf{r})$ and $\phi_1 (\mathbf{r})$ caused by the 
linear gradient field. For weak coupling ($|g|<1$), $\xi_\pm$ are purely real, 
resulting in two oscillations with the same resonance frequency but different decay 
rates. For strong coupling ($|g|>1$), $\xi_\pm$ are purely imaginary, 
leading to frequency splitting of the two oscillations. 
The strong coupling between $\phi_0 (\mathbf{r})$ and $\phi_1 (\mathbf{r})$ 
destroys the $PT$ symmetry of both eigenmodes. The $PT$ transition occurs 
when the coupling strength matches the ``diffusion energy'' ($|g|=1$), giving 
the field gradient of the exceptional point (assume $\lambda = 0$)
\begin{equation}
  g'_{\rm EP} = \frac{\pi^4}{32\sqrt{2}} \approx 2.15, 
\end{equation}
which match with the exact solution $g'_1$ in $5\%$ accuracy. The power law behavior Eq.~\eqref{Eq:criticalExpo_num} 
near EP can be derived from $\xi_\pm$: 
\begin{equation}
  \Delta = \frac{L^2}{4D} \Im{[\xi_+ - \xi_-]} = 8^{1/4} \left( g' - g'_{\rm EP} \right) ^{1/2}, \quad 
  g' \rightarrow (g'_{\rm EP})^+ .
\end{equation}

Let's consider the solution of FID signal in $PT$-broken phase. The two eigen vectors of 
matrix $H_2$, corresponding to eigenvalue $\xi_\pm = \pm i \epsilon \sqrt{g^2-1}$, are 
\begin{equation}
  \nu_+ = \frac{1}{\sqrt{2}}
  \begin{pmatrix} 1 \\ + {\rm e}^{-i \varphi}  \end{pmatrix}, \quad
  \nu_- = \frac{1}{\sqrt{2}}
  \begin{pmatrix} 1 \\ - {\rm e}^{+i \varphi}  \end{pmatrix},
\end{equation}
where
\begin{equation}
  \sin{\varphi} \equiv \frac{1}{g}, \quad
  \cos{\varphi} \equiv \frac{1}{g|g|} \sqrt{g^2 - 1}.
\end{equation}
Note that, $\nu_+$ is not orthogonal to $\nu_-$, because $H_2$ is non-Hermitian. 
The base functions, assuming $\lambda=0$, are
\begin{equation}
  \phi_0({\bf r}) =  \frac{1}{L^{3/2}} , \quad
  \phi_1({\bf r}) = -\frac{\sqrt{2}}{L^{3/2}} \sin{\left( \frac{\pi z}{L} \right)}.
\end{equation}
In our experiment, the initial polarization of an FID measurement is approximately uniform, i.e. 
$K_+({\bf r},0) = |K| \phi_0 ({\bf r})$. In the vector form, 
\begin{equation}
  K_{+,0} = |K| \begin{pmatrix}  1 \\ 0  \end{pmatrix} 
          = \frac{|K|}{\sqrt{2} \cos{\varphi}} \left( {\rm e}^{+i\varphi} \nu_+ 
          +  {\rm e}^{-i\varphi} \nu_- \right).
\end{equation}
Then, the solution of $K_+({\bf r}, t)$ is
\begin{equation}
  K_+({\bf r},t) = \frac{|K|}{\sqrt{2} \cos{\varphi}} 
  \left[ {\rm e}^{+i\varphi} {\rm e}^{-s_+ t} \nu_+({\bf r}) 
  +  {\rm e}^{-i\varphi} {\rm e}^{-s_- t} \nu_-({\bf r}) \right],
\end{equation}
where $s_\pm = \Gamma_2 + i \omega_\pm \equiv \Gamma_{\rm 2c} + \epsilon + i \gamma B_0 \pm i \epsilon \sqrt{g^2-1}$ 
are the eigenvalues of eigenmodes $\nu_\pm$. 
Our probe beam, along $x$ direction, is located at $(y_{\rm p}, z_{\rm p})$ on the $yz$ plane. 
The average spin polarization $K_x$ within the probe beam is
\begin{equation}
  \begin{aligned}
    K_x(t) &= \frac{1}{L} \int_{-\frac{L}{2}}^{+\frac{L}{2}}{ 
                   \Re{[K_+(x,y_{\rm p},z_{\rm p},t)]} {\rm d}x} \\
           &= \frac{|K|}{\sqrt{2} \cos{\varphi}} 
             \Re{\left[ {\rm e}^{+i\varphi} {\rm e}^{-s_+ t} \nu_+(z_{\rm p}) 
             +  {\rm e}^{-i\varphi} {\rm e}^{-s_- t} \nu_-(z_{\rm p}) \right]} \\
           &= \frac{|K| {\rm e}^{-\Gamma_2 t} }{\sqrt{2} \cos{\varphi}} 
           \left\{  a_+ \cos{[\omega_+ t - (\beta_+ + \varphi)]}
                +   a_- \cos{[\omega_- t - (\beta_- - \varphi)]} \right\}, 
  \end{aligned}
  \label{Eq:FIDSignal_twoModeSolution}
\end{equation}
where 
\begin{equation}
  \begin{aligned}
    a_\pm {\rm e}^{i \beta_\pm} &\equiv \nu_\pm (z_{\rm p}) \\
            &= \frac{1}{L^{3/2}}\left\{  
              \left[ \frac{1}{\sqrt{2}} \mp \cos{\varphi} \cdot I_{\rm sin}(z_{\rm p}) \right]  
                            + i \sin{\varphi} \cdot I_{\rm sin}(z_{\rm p}) \right\}, 
           \quad  I_{\rm sin}(z_{\rm p}) \equiv \sin{\frac{\pi z_{\rm p}}{L}}.
  \end{aligned}
\end{equation}
From Eq.~\eqref{Eq:FIDSignal_twoModeSolution}, we see that the FID signal in 
$PT$-broken phase is composed of two exponentially decaying oscillations 
with frequencies $\omega_\pm$. The amplitude ratio of these two oscillations is 
\begin{equation}
  \frac{a_+}{a_-} = \sqrt{\frac{  
    1-2\sqrt{2} \cos{\varphi} \cdot I_{\rm sin} + 2 I_{\rm sin}^2
   }{
    1+2\sqrt{2} \cos{\varphi} \cdot I_{\rm sin} + 2 I_{\rm sin}^2
   }}.
   \label{Eq:amplRatio_anaSol_brk}
\end{equation}
The initial phase difference of these two oscillations is 
\begin{equation}
  \delta \theta (z_{\rm p}) = \pm \left[ 
    (\omega_+ - \omega_-)t_0  - (\beta_+ - \beta_-) - 2 \varphi 
    \right],
   \label{Eq:phaseDiff_anaSol_brk}
\end{equation}
where $t_0$ is related to the time origin of FID measurement. 

The theoretical lines in Figs.~\ref{fig:eigenModeDist}(d) and (e) of main text 
is calculated similarly as stated above, but utilizing the 
exact solution in Section~\ref{SMsec:exactSolution}.

\section{Data Fitting and Analysis\label{SMsec:spectrumFitting}}
\subsection{Fitting Model}
Since the dynamics of Xe spins is governed by Eq.~\eqref{Eq:torreyEqVector}, 
the time domain FID signal of Xe spins should have the following form
\begin{equation}
  y(t) = \sum_{k=1}^{n}{A_k {\rm e}^{-\frac{t}{\tau_k}} \sin{\left( \omega_k t + \varphi_k \right)}}.
  \label{Eq:generalTimeDomainFID}
\end{equation}
The Fourier transform of $y(t)$ is the superposition of $2n$ Lorentzian 
peaks:
\begin{equation}
  Y(\omega) \equiv \frac{1}{\sqrt{2 \pi}} \int_{-\infty}^{+\infty}{\theta(t) y(t) {\rm e}^{-i\omega t} {\rm d}t}
  = \frac{-i}{\sqrt{8 \pi}} \sum_{k=1}^{n}{ A_k \left[
    \frac{{\rm e}^{+ i \varphi_k}}{1/\tau_k + i (\omega - \omega_k)} 
  - \frac{{\rm e}^{- i \varphi_k}}{1/\tau_k + i (\omega + \omega_k)}
  \right]},
  \label{Eq:generalFT}
\end{equation}
where $\theta(t)$ is the Heaviside step function. 

By choosing the value $n$ in Eq.~\eqref{Eq:generalTimeDomainFID}, 
we get three model functions:
\begin{itemize}
\item Single-Lorentzian-peak fitting (SLP, $n=1$) is used to 
analyze the $\rm ^{129}Xe$ FID signal in the $PT$-symmetric phase with zero or weak 
gradient field. As the phase $\varphi_1$ is irrelevant in $|Y_{\rm SLP}|$, three 
fitting parameters $A_1$, $\tau_1$ and $\omega_1$ are extracted. 
The model function is ($b$ represents a noise background)
\begin{equation}
  |Y_{\rm SLP}(\omega;A_1,\tau_1,\omega_1,b)| 
  \equiv \frac{1}{\sqrt{8 \pi}} 
    \frac{A_1}{ | 1/\tau_1 + i (\omega - \omega_1) |} + b.
    \label{Eq:modelFun_SLP}
\end{equation}

\item Double-Lorentzian-peak fitting (DLP, $n=2$) 
is used for the spectrum of $\rm ^{129}Xe$ in large gradient 
field (close to or larger than the EPs). Fitting parameters $A_k$, $\tau_k$, 
$\omega_k$ ($k=1, 2$) and $\Delta \varphi=\varphi_2-\varphi_1$ are extracted. 
The model function is
\begin{equation}
  |Y_{\rm DLP}(\omega;A_1,A_2,\tau_1,\tau_2,\omega_1,\omega_2,\Delta \varphi,b)| 
  \equiv \frac{1}{\sqrt{8 \pi}} \left|
    \frac{A_1}{ 1/\tau_1 + i (\omega - \omega_1) } 
  + \frac{A_2 {\rm e}^{+i\Delta \varphi}}{ 1/\tau_2 + i (\omega - \omega_2) }
    \right| + b.
    \label{Eq:modelFun_DLP}
\end{equation}

\item Triple-Lorentzian-peak fitting (TLP, $n=3$) 
is used to analyze the $\rm ^{131}Xe$ 
FID signal with quadrupole splitting. To reduce the degrees of freedom (DOFs) and 
improve the stability of the fitting process, all the phases $\varphi_k$ 
($k=1, 2, 3$) are set to zero and the frequency splitting is assumed to be 
symmetric ($\omega_1+\omega_3=2\omega_2$)~\cite{Feng2020}. 
The model function is
\begin{equation} 
  \begin{split}
  &|Y_{\rm TLP}(\omega;\{A_k\},\{\tau_k\},\omega_0,\Delta \omega_Q,b)|
  \\ \equiv &\frac{1}{\sqrt{8 \pi}} \left|
    \frac{A_1}{ 1/\tau_1 + i (\omega - \omega_0 + \Delta \omega_Q) } 
  + \frac{A_2}{ 1/\tau_2 + i (\omega - \omega_0) }
  + \frac{A_3}{ 1/\tau_3 + i (\omega - \omega_0 - \Delta \omega_Q) }
    \right| + b.
    \label{Eq:modelFun_TLP}
  \end{split}
\end{equation}
\end{itemize}

\begin{figure*}[hbpt]
  \includegraphics[]{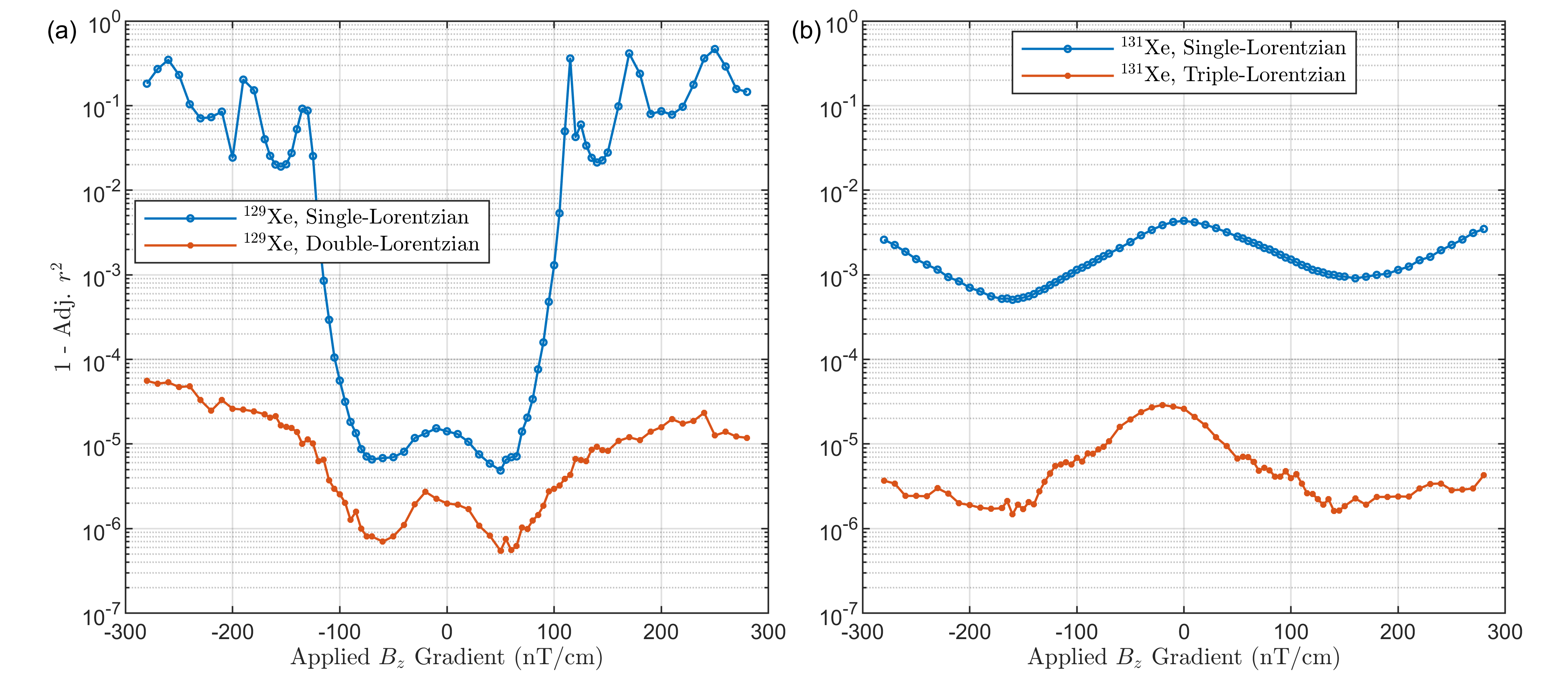}
  \caption{\label{fig:gof_vs_grad} 
  {The goodness of fitting for different models.}
  (a), The Adj. R-Squared value of the Lorentzian fitting on the $\rm ^{129}Xe$ 
  FID data of Fig.~\ref{fig:lambdaSpliting} and Fig.~\ref{fig:eigenModeDist} in the 
  main text. Both SLP and DLP fittings give satisfactory results in small gradient 
  region with $|G| < 70~{\rm nT/cm}$. SLP fitting is significantly worse in the 
  large gradient region because the decay rate of the lowest two eigen modes get 
  close to each other.
  (b), Similar to (a) but for $\rm ^{131}Xe$. Goodness of fitting keeps 
  smooth as gradient changes. The large difference between SLP and TLP fitting 
  performances indicates that the quadrupole splitting is not negligible and SLP 
  model may underestimate the $T_2$ of $\rm ^{131}Xe$.
  }
\end{figure*}

The FID signal of $\rm ^{131}Xe$ can be analyzed by either SLP or TLP model. 
SLP fitting usually overestimates the relaxation rate $T_2^{-1}$, 
particularly when the quadrupole splitting $\Delta \omega_{\rm Q}$ is not negligible 
(in our experiment $\Delta \omega_{\rm Q}/(2\pi) \sim 20~{\rm mHz}$). TLP model 
gives more accurate value of relaxation rate and, accordingly, is used when the 
accurate value of $T_2^{-1}$ is important. On the other hand, both SLP and TLP 
models give similar results for the central frequency of $\rm ^{131}Xe$. The SLP 
model involves much less DOFs, and the fitting process is numerically more stable. 
So, the SLP model is used to fit the resonance frequency of $\rm ^{131}Xe$ in 
Fig.~\ref{fig:stableMag} of the main text. The physical results (e.g., the 
$P_{\rm pump}$ dependence of the rotation rate $\Omega_{\rm rot}$) do not 
essentially rely on the choice of SLP or TLP model. The fitting model used in 
each figure is summarized in Table~\ref{tab:fit_models}.
\begin{table}
\caption{
  A summary of the fitting models used in processing each figure's data.
  \label{tab:fit_models}
}
  \begin{ruledtabular}
    \begin{tabular}{cc}
    \textbf{Figure}&
    \textbf{Fitting Model}\\
    \colrule
    Fig.~\ref{fig:lambdaSpliting} & DLP \\
    Fig.~\ref{fig:eigenModeDist} & DLP \\
    Fig.~\ref{fig:stableMag} & DLP for $\rm ^{129}Xe$, SLP for $\rm ^{131}Xe$ \\
    Fig.~\ref{fig:sigDemo_and_sweepG}(c) & SLP \\
    Fig.~\ref{fig:sigDemo_and_sweepG}(d) & TLP \\
    Fig.~\ref{fig:sigDemo_and_sweepG}(e) & DLP \\
    Fig.~\ref{fig:sigDemo_and_sweepG}(g) & DLP for $\rm ^{129}Xe$, TLP for $\rm ^{131}Xe$ \\
    Fig.~\ref{fig:gof_vs_grad} & (Labeled in figure)
    \end{tabular}
  \end{ruledtabular}
\end{table}

\subsection{Discrete Fourier Spectrum}
The discrete Fourier spectrum of experiment FID signal is a little different from
Eq.~\eqref{Eq:generalFT} due to finite sampling time. 
We use the MATLAB built-in function plomb() to estimate the Power Spectrum 
Density (PSD) of experimental time domain FID signal $y(t)$. As long as the sampling rate 
is high enough compared to $\{ \omega_k \}$ and the sampling time is long 
enough compared to $\{ \tau_k \}$, the obtained PSD should be proportional 
to $|Y(\omega)|^2$.

Consider a discrete sampling of $y(t)$:
\begin{equation}
  y_0 \equiv y(0), \quad y_1 \equiv y(\Delta t), \quad \cdots, \quad
  y_{N-1} \equiv y(T-\Delta t), \quad T \equiv N\Delta t.
\end{equation}
The MATLAB built-in function periodogram() calculates the following PSD:
\begin{equation}
  \hat{P}(\omega) = \frac{\Delta t}{N} \left| 
      \sum_{n=0}^{N-1}{y_n {\rm e}^{-i n \omega \Delta t}} 
  \right|^2, \quad
  -\frac{\pi}{\Delta t} < \omega \le +\frac{\pi}{\Delta t}.
\end{equation}
plomb() returns the estimation of $2 \hat{P}(\omega)$, but allowing 
non-uniform sampling. Note that there is a factor of 2 coming from the 
requirement of total energy conservation, because plomb() only calculates 
single-sided PSD.

If sampling rate is high enough (
$|1/\tau_k + i(\omega - \omega_k )|\Delta t \ll 1, \quad \omega_k \Delta t < \pi$
) and sampling time is long enough ($T \gg \tau_k$), we can show that
\begin{equation}
  \hat{P}(\omega) = \frac{2 \pi}{T} |Y(\omega)|^2.
\end{equation}
Thus, we can use Eq.~\eqref{Eq:generalFT} to perform least squares fitting 
on $\hat{P}(\omega)$ in the neighborhood of $\omega=\omega_k$. 
The normalized spectrum in the waterfall plot of Fig.~\ref{fig:lambdaSpliting}(a) 
of main text is $\sqrt{P(\omega)T/(2\pi)}$, where the $2P(\omega)$ here 
represents the power spectrum returned by plomb() function.

\subsection{Fitting Algorithm}
We use the MATLAB built-in function fit() to perform the nonlinear 
least squares fitting. The data points used for fitting is 
$\sqrt{P(\omega)T/(2\pi)}$, and the model functions are 
Eqs.~\eqref{Eq:modelFun_SLP}-\eqref{Eq:modelFun_TLP}. 
Before being fed to plomb() function, the time domain FID signal is 
automatically trimmed to be shorter than 12 times of $T_2$. This is to 
ensure the Lorentzian peaks not being buried by the white noise. 
An oversampling factor of $8 \sim 12$ is used for plomb(), which makes the 
frequency point in $P(\omega)$ $8 \sim 12$ times denser than a standard FFT 
spectrum. This can help the fitting process easier to converge to a 
physically reasonable result. Typical fitting results 
are shown in Figs.~\ref{fig:sigDemo_and_sweepG}(c)-(e) 
and Fig.~\ref{fig:gof_vs_grad}.

The confidence intervals of the fitting parameters is acquired by calling 
the confint() function with the cfit object returned by the fit() function. 
All the error bars in the figures is calculated based on the 95\% confidence 
bounds returned by confint().

\begin{figure*}[hbpt]
  \includegraphics[]{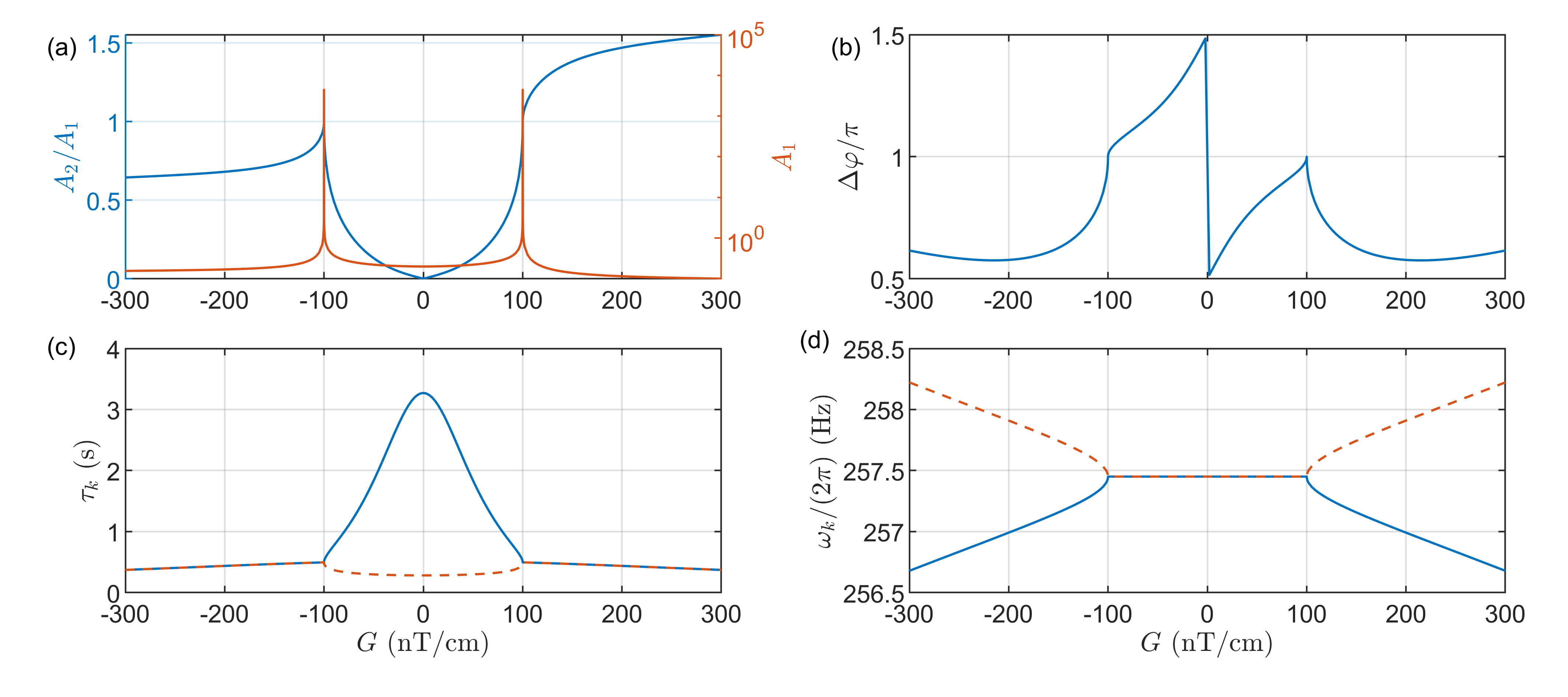}
  \caption{\label{fig:benchmark_parSet} 
  {The parameter set used in the benchmark test.} 
  The parameters in this figure is calculated based on the exact solution demonstrated 
  in Section~\ref{SMsec:exactSolution}, and then fitted to our experiment condition. 
  The exact EP is at $G = \pm 100.051~{\rm nT/cm}$. 
  For the meaning of parameters, see Eq.~\eqref{Eq:benchmark_dataGenerator}. 
  (a), $A_1=0.2$ at $G=0$.
  (c) and (d), blue solid line for $k=1$ and red dashed line for $k=2$. 
  }
\end{figure*}

\subsection{Benchmark of DLP Fitting Algorithm\label{SMsec:fittingBenchmark}}
The DLP fitting above might have intrinsic difficulty in two gradient regions. 
(1) In the zero-gradient region, the FID signal of $\rm ^{129}Xe$ consists of two exponentially decaying 
oscillations with the same frequency. 
One of them have large initial amplitude and decays slow, 
another one have very small initial amplitude and decays very fast. The slowly decaying oscillation make up the main part of the Fourier spectrum, 
while the fast decaying oscillation have very little contribution to the spectrum. 
So, it's very difficult to fit this fast decay rate from the spectrum. (2) In the region 
close to exceptional points, the two oscillations in FID signal have both very 
similar decay rate and very close frequency. The Fourier spectrum includes two Lorentzian 
peaks with similar linewidth and very close center frequency. It seems hard to distinguish 
these two peaks. Thus, a benchmark for the fitting algorithm is necessary to 
ensure that the least squares fitting can correctly extract the 
parameters we want from the spectrum of FID signal. 

The benchmark will focus on two questions: 
\begin{enumerate}
  \item How accurate the fitted parameters can be? Whether the 
  confidence intervals returned by the algorithm correctly represents the error 
  between fitted values and real values? 
  \item Is the fitting result unique? Will the initial values affect the 
  fitting result?
\end{enumerate}

In the benchmark test, a series of artificial signals with the following form are generated 
numerically, and then feed to the fitting algorithm. The fitting output 
is compared with the real value of the parameters to evaluate the accuracy of fitting result. 
\begin{equation}
  y(t) = A_1 {\rm e}^{-t/\tau_1} \sin{(\omega_1 t)}
       + A_2 {\rm e}^{-t/\tau_2} \sin{(\omega_2 t + \Delta \varphi)} + N(t).
  \label{Eq:benchmark_dataGenerator}
\end{equation}
The real value of parameters used in the numerical generation 
is shown in Fig.~\ref{fig:benchmark_parSet}. A small random fluctuation is added 
to the real values for repeated simulation. 
The $N(t)$ above represents the white noise in the experiment signal, and is 
generated using Gaussian distributed random numbers with appropriate variance. 

\begin{figure*}[hbpt]
  \includegraphics[]{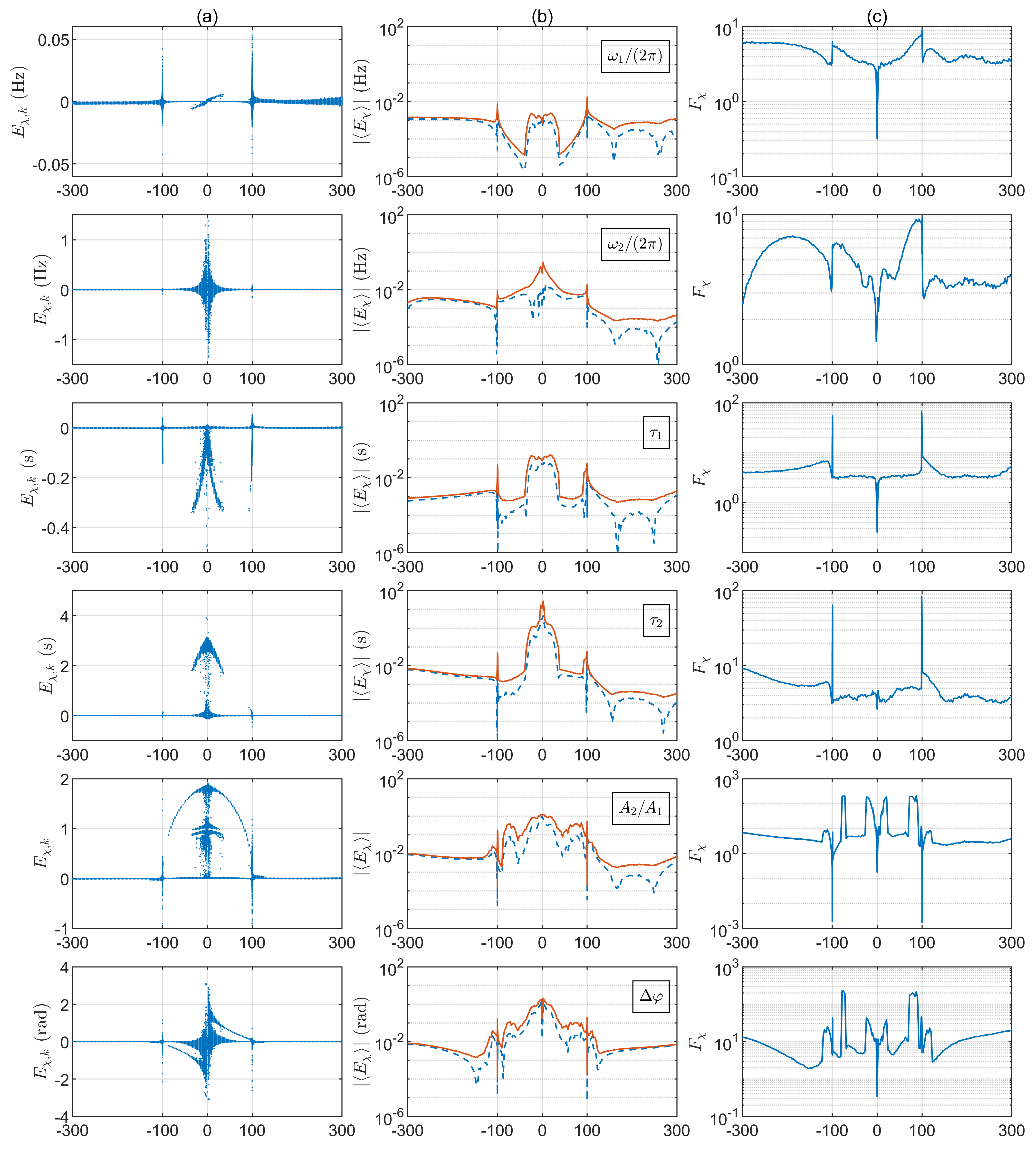}
  \caption{\label{fig:benchmark_fittingAccuracy} 
  {Summary of fitting accuracy.}  
  Each row shows the fitting accuracy of one parameter, which is labeled in 
  the textbox on column (b). The $x$ axis of all subplots are both $G~({\rm nT/cm})$, 
  the magnetic field gradient. 
  Column (a) shows the distribution of actual fitting error $E_\chi$ defined in 
  Eq.~\eqref{Eq:fittingAccuracy_def1}. 
  Column (b) shows the mean value of actual fitting error as defined in 
  Eq.~\eqref{Eq:fittingAccuracy_def2}. 
  The blue dashed lines are the absolute value $|\langle E_\chi \rangle|$, and the 
  red solid lines are the $1\sigma$ upper bound 
  $ \max{\left( |\langle E_\chi \rangle + \sigma_\chi| ,  |\langle E_\chi \rangle - \sigma_\chi | \right)} $. 
  Column (c) shows the U95 factor $F_\chi$ defined in Eq.~\eqref{Eq:fittingAccuracy_def3}. 
  $N=1000$ repeated benchmarks are performed at each gradient. 
  }
\end{figure*}

\subsubsection{Fitting Accuracy}
For a parameter $\chi$, we use $\chi$ to represent its real value. The fitting algorithm gives an 
estimation value of $\chi$, denoted as $\hat{\chi}$, together with a 95\% confidence interval 
$U_\chi$. The fitting error is defined as $E_\chi \equiv \hat{\chi} - \chi$. The relative 
magnitude between the actual fitting error and the estimated confidence interval is 
$\mathcal{R}_\chi \equiv |E_\chi| / U_\chi$. By repeating the benchmark many times, we 
have a set of $E_\chi$ and $\mathcal{R}_\chi$
\begin{equation}
  \{  E_{\chi,k} \}, \quad \{ \mathcal{R}_{\chi,k} \}, \quad k=1,2,\cdots, N
  \label{Eq:fittingAccuracy_def1}
\end{equation}
Then, we can define the mean value and standard derivation of fitting error as 
\begin{equation}
  \langle E_\chi \rangle \equiv \frac{1}{N} \sum_{k=1}^{N}{E_{\chi,k}}, \quad
  \sigma_\chi \equiv \sqrt{ \frac{1}{N-1} 
      \sum_{k=1}^{N}{ \left( E_{\chi,k} - \langle E_\chi \rangle \right)^2 }  }
      \label{Eq:fittingAccuracy_def2}
\end{equation}
According to the meaning of 95\% confidence interval, one may expect that approximately 
95\% of $\{\mathcal{R}_{\chi,k}\}$ should be smaller than one. 
To evaluate the accuracy of confidence interval, we define the U95 factor $F_{\chi}$ as
\begin{equation}
  F_{\chi}: {\rm the~smallest~real~number~such~that~at~least~95\%~of~}
  \{\mathcal{R}_{\chi,k}\} {\rm~are~smaller~than~} F_{\chi}. 
  \label{Eq:fittingAccuracy_def3}
\end{equation}

To verify the fitting accuracy, we generate 1000 artificial signals for each gradient and summarize 
the statistics of fitting accuracy in Fig.~\ref{fig:benchmark_fittingAccuracy}. 

{\bf Resonance frequencies}: from the top two rows of Fig.~\ref{fig:benchmark_fittingAccuracy}, 
we can see that in all gradient region, the U95 factors $F_\chi$ of $\omega_1$ and $\omega_2$ are 
in the range $0 \sim 10$. That means the confidence intervals of these two frequencies 
are slightly under-estimated by the fitting algorithm. We consider this to be acceptable 
for this work. 
From column (b), the mean error has a $1 \sim 10~{\rm mHz}$ bias in most region, 
which means the fitting algorithm is a biased estimator of the two resonance frequencies. 
This small bias is not important in this work, but should be carefully handled when 
the exact resonance frequency is concerned. 
From column (c), the fitting error of $\omega_1/(2\pi)$ can be as large as 50~mHz near EP. 
It's large compared to other gradient region, and is in the same order of magnitude 
with the linewidth $\sim 300~{\rm mHz}$ here. The fitting error of $\omega_2/(2\pi)$ 
can be up to 1~Hz, wihch is due to the very small amplitude and short relaxation time of 
peak \#2. However, the distribution of these fitting errors are symmetric, thus can be 
suppressed by repeating measurement. 

{\bf Relaxation time}: from the middle two rows of Fig.~\ref{fig:benchmark_fittingAccuracy}, 
we can see that except for EP region, the U95 factors $F_\chi$ of $\tau_1$ and $\tau_2$ are 
within the range $0 \sim 10$. This is similar to the case of resonance frequencies 
and is acceptable for this work. 
From column (b), the mean error of $\tau_1$ is always smaller than 0.1~s, which is enough 
for our theory-experiment comparison in Fig.~\ref{fig:lambdaSpliting}(b) of main text. 
The mean error of $\tau_2$ in the zero-gradient region can be as large as $10~{\rm s}$. 
That's why the blue data points in Fig.~\ref{fig:lambdaSpliting}(b) of main text 
deviate from the theoretical line. 
From column (c), we can see that the error distribution near zero-gradient shows an 
asymmetric pattern, meaning that the estimation of $\tau_1$ and $\tau_2$ has a large bias. 
The structure of error distribution also indicates that the fitting may have two local 
minimums, one close to the real values, another far from the real values. For the case of 
$\tau_2$, this is easy to understand -- the fitter just return 
$\hat{\tau}_2 \approx \hat{\tau}_1 \approx \tau_1$ since the peak \#2 is too small to 
recognize. But the reason of multiple solution behavior of $\tau_1$ is unknown to us. 

{\bf Peak amplitude ratio}: from the fifth row of Fig.~\ref{fig:benchmark_fittingAccuracy}, 
we can see that the U95 factor of the peak amplitude ratio $A_2 / A_1$ can be larger 
than $10^2$ in some regions of the $PT$-symmetric phase. Also, the mean error can be 
as large as $10^0$. This means the fitting value $\hat{A}_2 / \hat{A}_1$ is not 
reliable in those regions. However, the experiment data points in 
Fig.~\ref{fig:eigenModeDist}(d) fits so well with the theoretical lines that no bias 
in the order of $10^0$ can be observed. This remains a mystery to us. 
From column (a), the error distribution has a strange pattern, which indicates 
that there may be two to four local minimums. Good news is those local minimums 
separate from the real solution far enough that it's easy to filter them by 
physical criterion. 

{\bf Phase shift}: from the bottomest row of Fig.~\ref{fig:benchmark_fittingAccuracy}, 
we can see that the fitted value of $\Delta \varphi$ is completely unreliable in the 
zero-gradient region since the mean error is in the order of $\pi$. The confidence 
intervals are also severely underestimated in many regions of the $PT$-symmetric phase since 
the U95 factor can be larger than $10^2$. Thus, in Fig.~\ref{fig:eigenModeDist}(e) 
of main text, the data points near zero-gradient region shoud be considered inaccurate. 

As a summary, the benchmark in this section shows that, the fitting accuracy of 
$\omega_k/(2\pi)$ and $\tau_k$ is good enough to support the comparison between 
experiment data and theoretical lines in Fig.~\ref{fig:lambdaSpliting} of main text 
(except the $\tau_2$ in zero-gradient region). The fitting accuracy of $A_2/A_1$ and 
$\Delta \varphi$ is not so good that the comparison in Fig.~\ref{fig:eigenModeDist} 
of main text is not reliable in the $PT$-symmetric phase. As for Fig.~\ref{fig:stableMag} 
of main text, all data are measured at gradient $250~{\rm nT/cm}$, where the fitting accuracy 
is pretty good and thus the results are reliable. 

As mentioned above, the U95 factor varies within $0 \sim 10$ in most of the gradient region. 
When calculating the error bars in the figures of main text, we just omit this U95 factor. 
The error bars are estimated using the original confidence intervals returned by our fitting algorithm 
together with the standard derivation of repeated measurements. This may under-estimate the 
errors but is considered to be acceptable for the demonstration purpose in this work. 

\begin{figure*}[hbpt]
  \includegraphics[]{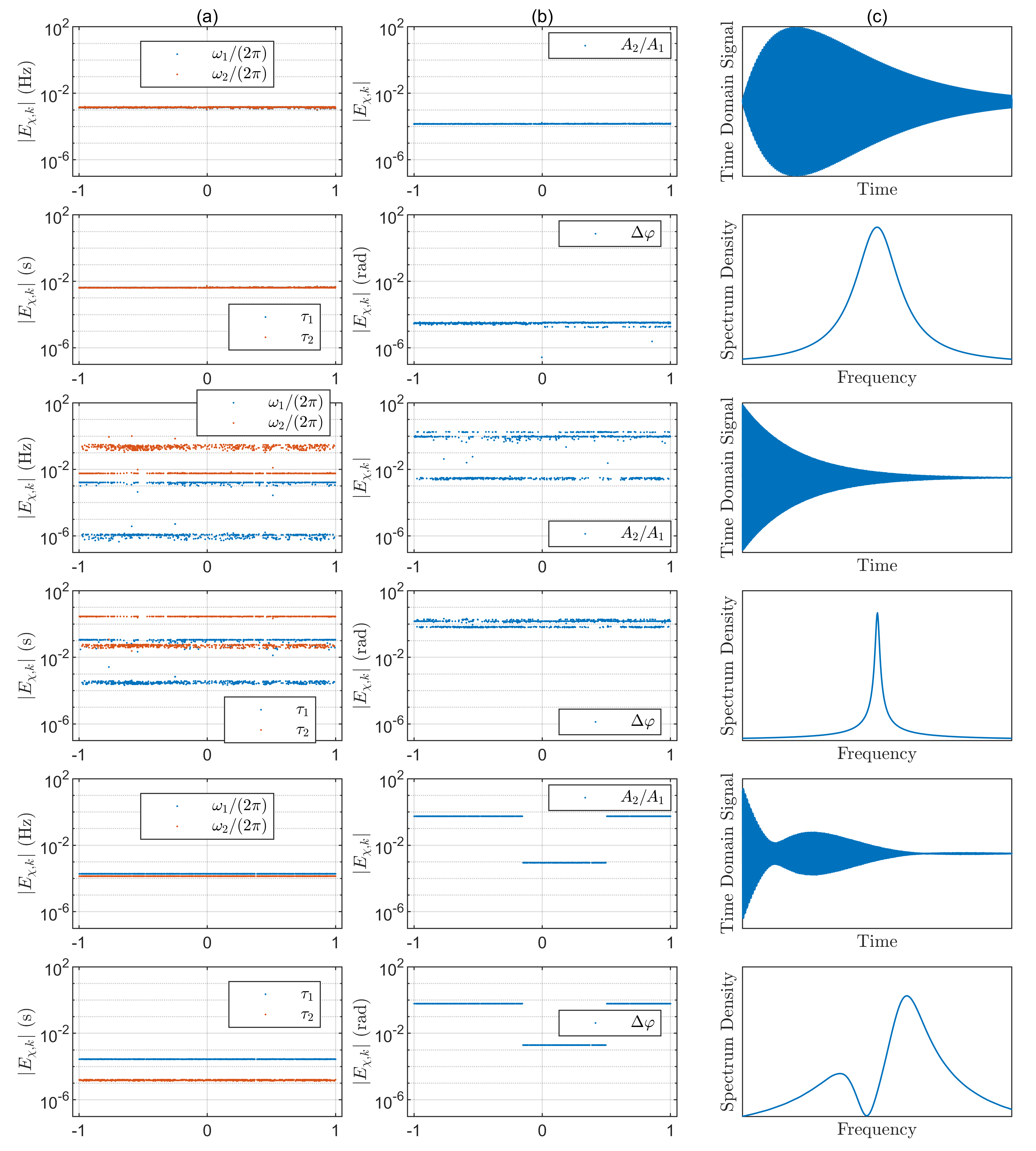}
  \caption{\label{fig:benchmark_phi0Dependence} 
  {Fitting dependence on the initial value of $\Delta\varphi$.}  
  The $x$ axis of subplots in column (a) and (b) are both 
  $(\varphi_0 - \Delta\varphi)/ \pi$, 
  the difference between fitting initial value and real value of $\Delta\varphi$. 
  Column (a) and (b) shows the distribution of actual fitting error $|E_\chi|$ 
  defined in Eq.~\eqref{Eq:fittingAccuracy_def1}. 
  Column (c) shows the FID signal used for test. 
  The top two rows correspond to $G=-100.051~{\rm nT/cm}$, a gradient very close to EP. 
  The middle two rows correspond to $G=+3~{\rm nT/cm}$, near zero-gradient region. 
  The bottom two rows correspond to $G=+177~{\rm nT/cm}$, where the two peaks separate 
  to each other far. 
  $N=1000$ benchmarks are performed at each gradient, with random $\varphi_0$. 
  The time domain signal used for these 1000 benchmarks are exactly the same --  
  both the real value of all parameters and the noise term $N(t)$ are the same. 
  }
\end{figure*}

\subsubsection{Dependence on Initial Values}
The numerical least squares fitting need an initial value for every parameter as the 
start point of the optimization algorithm. It's easy to estimate the value of 
$\omega_k$, $\tau_k$ and $A_k$ in Eq.~\eqref{Eq:modelFun_DLP} from the shape of 
Fourier spectrum. Also, the initial value of $b$ can be safely set to zero as the noise is 
very small. However, it's hard to estimate the value of $\Delta \varphi$ -- can we just 
use a random initial value? 

In the benchmark of the above section, the fitting error distribution indicates that 
the fitting process may have multiple local minimums. That means the choice of initial 
values might affect the result of fitting, and a benchmark is necessary to figure out 
whether the initial value of $\Delta \varphi$ will affect which local minimum 
the fitting algorithm converges to. 

Figure~\ref{fig:benchmark_phi0Dependence} shows the benchmark result. In the gradient 
region close to EP (top two rows), the solution is unique. The initial value $\varphi_0$ 
have no effect on the fitting result. 

In the zero-gradient region (middle two rows), it seems that there are two or three 
different solutions -- the fitting error of $\omega_k$ and $\tau_k$ jumps between 
two distinct values and the fitting error of $A_2/A_1$ jumps between three different 
values. However, we cannot control which solution to converge by choosing the value of 
$\varphi_0$. The fitter just converge to a solution randomly. We have no idea how to 
solve the multi-solution problem in this region so far. 

In the large-gradient region (bottom two rows), the solutions of $\omega_k$ and $\tau_k$ 
are unique, but there are two different solutions for $A_2/A_1$ and $\Delta \varphi$. 
These two solutions are hard to distinguish -- both solutions fit very well with the 
spectrum and their adj. $R^2$ are the same. To get the solution whcih is closer to 
the real parameters, we need to use a $\varphi_0$ that is close to the real value 
of $\Delta\varphi$. But this is impractical since we don't know the real value of 
$\Delta\varphi$ in experiment. 

\begin{figure*}[hbpt]
  \includegraphics[]{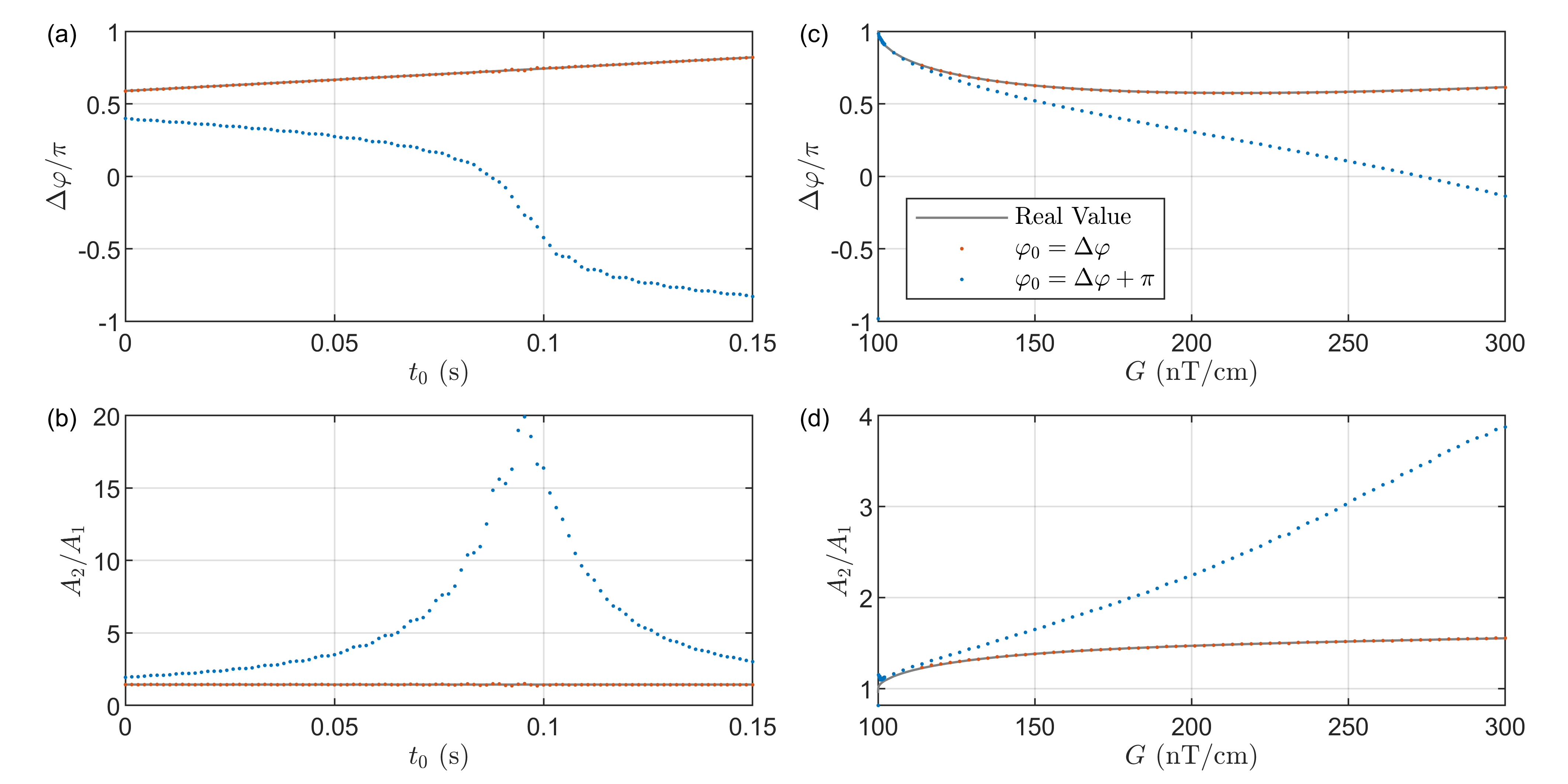}
  \caption{\label{fig:benchmark_distinguish2solutions} 
  {The behavior of the two fitting solutions.} 
  The legend of all subplots is the same -- solid lines represent the real value 
  of the parameter; red dots are the fitting result using the real value of 
  $\Delta\varphi$ as the initial value $\varphi_0$; blue dots are the fitting 
  result using $\varphi_0 = \Delta\varphi + \pi$. 
  Subplots (a) and (b) are generated using the artificial FID signal at 
  $G = +177~{\rm nT/cm}$. The trimming time $t_0$ is defined in 
  Eq.~\eqref{Eq:benchmark_dataGenerator_trimmed}. The real value line is calculated 
  by Eq.~\eqref{Eq:benchmark_phi_t0_dep}. 
  }
\end{figure*}

A practical way to distinguish these two solutions 
is to drop the beginning of the time domain signal and fit again. If we trim a small 
length $t_0$ from the beginning of the time domain signal, it's equivalent to change the 
time origin of Eq.~\eqref{Eq:benchmark_dataGenerator}. The signal after trimming is 
\begin{equation}
  y'(t;t_0) \equiv y(t+t_0) 
       = \left[A_1 {\rm e}^{-t_0/\tau_1}\right] {\rm e}^{-t/\tau_1} \sin{(\omega_1 t + \omega_1 t_0)}
       + \left[A_2 {\rm e}^{-t_0/\tau_2}\right] {\rm e}^{-t/\tau_2} \sin{(\omega_2 t + \omega_2 t_0 + \Delta \varphi)} + N(t+t_0).
  \label{Eq:benchmark_dataGenerator_trimmed}
\end{equation}
Since the frequency of the two peaks are different in this gradient region, the phase 
shift returned by the fitter shoud has the following dependence on the trimming time 
$t_0$:
\begin{equation}
  \widehat{\Delta\varphi} (t_0) = \Delta\varphi + (\omega_2 - \omega_1) t_0.
  \label{Eq:benchmark_phi_t0_dep}
\end{equation}
The relaxation time $\tau_1$ and $\tau_2$ are the same in this gradient region, so the 
amplitude ratio $\hat{A}_2 / \hat{A}_1$ shoud not depend on $t_0$. 

As shown in Figs.~\ref{fig:benchmark_distinguish2solutions}(a)(b), the ``better solution" 
using $\varphi_0 = \Delta\varphi$ do obey these dependences, while the ``worse solution" 
using $\varphi_0 = \Delta\varphi + \pi$ does not. 
From Figs.~\ref{fig:benchmark_distinguish2solutions}(c)(d), we can see that both 
the ``better solution" and the ``worse solution" itself make up a continuous curve. 
So, one should be careful on distinguishing these solutions since the ``worse solution" 
also physically looks good.

\bibliographystyle{apsrev4-2}
\bibliography{../maintext/maintext-APS.bib}